# Bridging Pico-to-Nanonewtons with a Ratiometric Force Probe for Monitoring Nanoscale Polymer Physics Before Damage

Ryota Kotani[1], Soichi Yokoyama[1], Shunpei Nobusue[2], Shigehiro Yamaguchi[3], Atsuhiro Osuka[1], Hiroshi Yabu[4]*, Shohei Saito[1,5]*

**Corresponding Authors:**

Hiroshi Yabu (hiroshi.yabu.d5@tohoku.ac.jp)

Shohei Saito (saito.shohei.4c@kyoto-u.ac.jp)

**Affiliations:**

[1]Graduate School of Science, Kyoto University, Kyoto 606-8502, Japan.

[2]Institute of Advanced Energy, Kyoto University, Uji 611-0011, Japan.

[3]Graduate School of Science, Nagoya University, Nagoya 464-8602, Japan.

[4]WPI-Advanced Institute for Materials Research (AIMR), Tohoku University, Sendai 980-8577, Japan.

[5]PRESTO, Japan Science and Technology Agency, Kyoto 606-8502, Japan.

**Abstract:** Understanding the transmission of nanoscale forces in the pico-to-nanonewton range is important in polymer physics. While physical approaches have limitations in analyzing the local force distribution in condensed environments, chemical analysis using force probes is promising. However, there are stringent requirements for probing the local forces generated before structural damage. The magnitude of those forces corresponds to the range below covalent bond scission (from 200 pN to several nN) and above thermal fluctuation (several pN). Here, we report a conformationally flexible dual-fluorescence force probe with a theoretically estimated threshold of approximately 100 pN. This probe enables ratiometric analysis of the distribution of local forces in a stretched polymer chain network. Without changing the intrinsic properties of the polymer, the force distribution was reversibly monitored in real time. Chemical control of the probe location demonstrated that the local stress concentration is twice as biased at crosslinkers than at main chains, particularly in a strain-hardening region. Due to the high sensitivity, the percentage of stressed force probes was estimated to be more than 1000 times higher than the activation rate of a conventional mechanophore.

**Introduction:**

Understanding the transmission of pico-to-nanonewton forces in complex hierarchical structures is a primary goal in polymer physics[1-3] and mechanobiology[4,5]. Physical approaches such as optical/magnetic tweezers and atomic force microscopy (AFM) are widely used to evaluate piconewton (pN) force, but it is difficult to visualize the distribution of forces in condensed matter or molecular crowding environments with these techniques (Fig. 1a)[6-8]. While there are several instruments, such as surface force apparatuses (SFAs), rheometers, and tensile testing machines, that can analyze physical forces transmitted in meso-to-macroscopic structures, methods to directly quantify the distribution of forces at the molecular level are still being pursued. A promising alternative is chemical doping of mechanoresponsive molecules as a force probe[9-14]. These chromophores for materials and biological systems have been independently developed over the last few decades because the force magnitudes of interest are different in these fields. In the field of mechanobiology, Förster resonance energy transfer (FRET) dyads have been developed to analyze forces ranging from several to tens of pN[15-18]. On the other hand, various mechanophores respond to forces with higher thresholds involving internal covalent bond scission, which have produced a number of stress-responsive materials[19-32]. In particular, spiropyrans bearing different substituents have been reported to isomerize on 100 ms timescales under forces of 200–400 pN,[33,34] while density functional theory (DFT) calculations of most mechanophores have provided force thresholds on the order of a few nanonewtons.[35,36] Reversible responses have also been obtained in several mechanophores, even after covalent bond scission.[37-40] Recently, spiropyrans have also been used as a force probe to address important questions in polymer physics, including those related to load transmission and distribution at the single-chain (segmental) level.[9-14] However, there are still stringent requirements for the design of force probes that can quantitatively monitor local stress concentrations before the polymer chain network is structurally damaged but without changing the intrinsic properties of the polymer by chemical doping. Although sacrificial bond scission of polymer materials has been monitored using turn-on mechanophores[9-11], fully understanding the distribution of nanoscale forces in entangled polymer chains is still challenging, and its importance becomes more obvious in the rational design of uniquely tough materials[41-49]. The target force range for this purpose is approximately 10–100 pN, which is below that for covalent bond scission (200 pN to several nN) and beyond that for spontaneous thermal fluctuation at room temperature ($k_\mathrm{B}T = 4.1$ pN nm)[16,50] (Fig. 1b). Here, we demonstrate that a conformationally flexible flapping force probe enables this study by ratiometric analysis of its bright dual fluorescence (FL), in which the potential energy profiles in the ground state ($S_0$) and the lowest singlet excited state ($S_1$) are suitably designed. This high brightness allows minimal chemical doping to preserve the intrinsic polymer properties. Controlling the chemical location of the probe in crosslinked polymers and observing the dual emission from the stretched polymers in real time provided new quantitative insights into biased nanoscale stress concentrations.

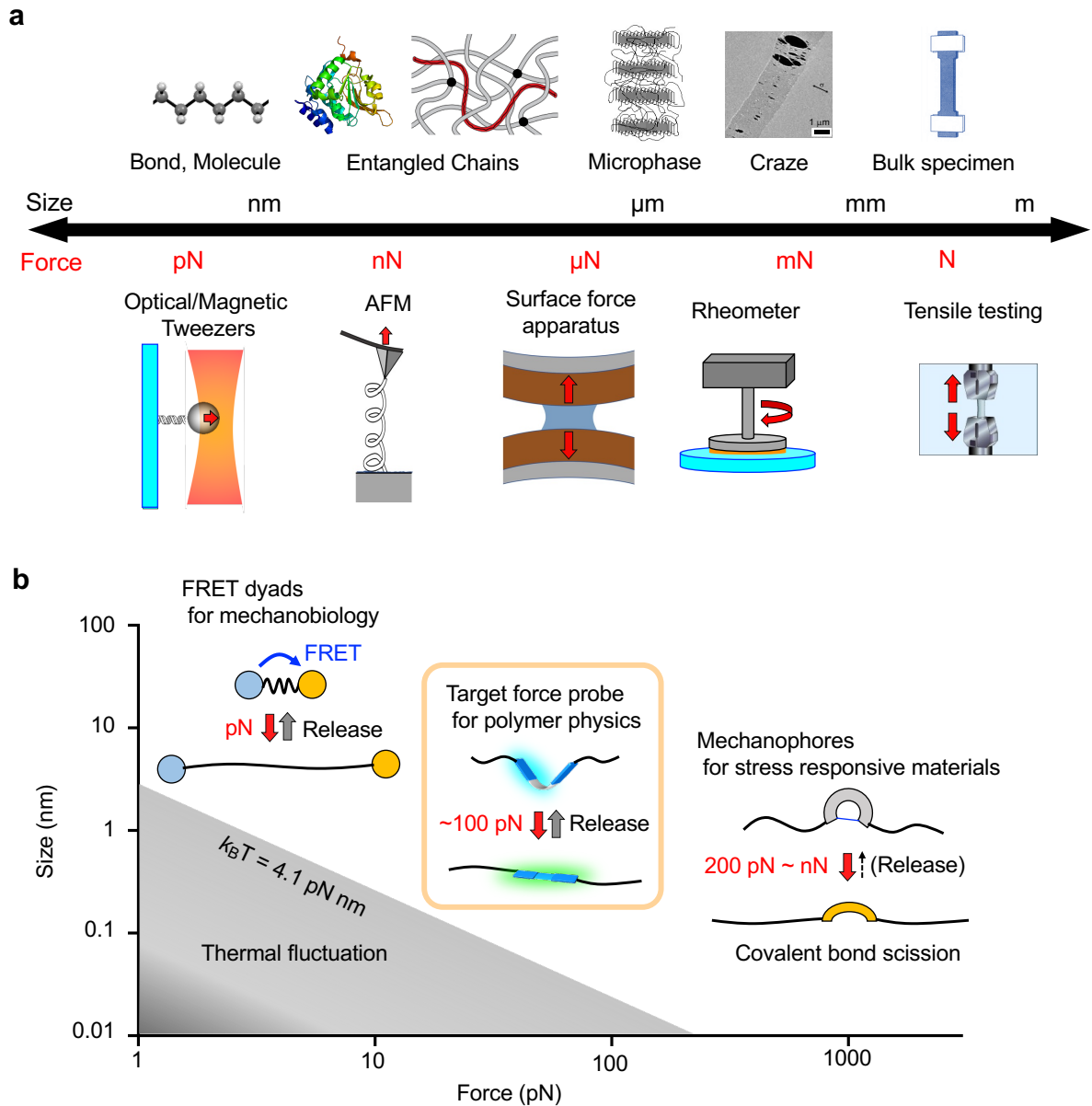

**Fig. 1.** (a) Hierarchical structures of materials and physical measurement methods of mechanical forces at different scales. (b) Target force range for the nanoscale study of polymer physics.

## Results and Discussion:
### subsection 1)
**Dual-fluorescence flapping molecular probe**

A flapping molecule (FLAP), composed of two rigid anthraceneimide wings fused with a conformationally flexible eight-membered ring (cyclooctatetraene, COT), shows dual FL properties (Fig. 2a,b)[51]. In $S_0$, a bent form is most stable, while a planar form has higher energy due to the strain effects of the flat COT ring[52]. On the other hand, FLAP has two energy minima in $S_1$ with bent and planar geometries. Emissions from these minima are both allowed, exhibiting blue FL at approximately 460 nm and green FL at approximately 520 nm. Since the energy barrier is low and the planar geometry is slightly more stable in $S_1$, FLAP changes its conformation from bent to planar on a subnanosecond timescale after ultraviolet (UV) excitation in solution phase[53], emitting predominantly green FL with a high FL quantum yield ($\Phi_F \approx 0.3$). The population of excited species at these $S_1$ minima is largely dependent on the local viscosity in solution, while the polarity dependence can be ignored due to the non-charge-transfer character. Whereas the viscosity-probing function is active in the presence of solvent molecules, the planarization dynamics across the barrier in $S_1$ are suppressed in a polymer matrix, such that blue FL is mainly observed ($\Phi_F \approx 0.3$). On the other hand, planarization in $S_0$ can be mechanically induced by an external force, as we previously demonstrated the mechanophore function in a crystal phase transition of FLAP[54]. Among conformationally flexible mechanophores that can show detectable signals without covalent bond scission[55–62], FLAP is unique in its dual FL properties, which enables ratiometric FL analysis (Fig. 2b). Although a variety of single-component dual-fluorescence chromophores have been reported in photochemistry[63,64], force probe application has not been reported. By ratiometric FL analysis, quantitative evaluation can be realized even when the molecular force probes are distributed heterogeneously and the concentration of the probe changes by polymer deformation.

It is also meaningful to compare the working mechanism of the flapping force probe with that of the reported rigidochromic probes (categorized into fluorogenic molecular rotors), which can provide important microscopic insights into the friction force on a real contact area[65,66]. The similarity between the two probes can be seen in the mechanical control of the distribution of different excited-state species. On the other hand, the difference is the direction of the mechanically induced dynamics; namely, in the rigidochromic probe, the excited-state dynamics from the nearly Franck-Condon emissive geometry to another nonemissive geometry are suppressed in the confined space under mechanical compression. In contrast, the flapping force probe is initially confined with the bent form in a narrow free volume of the polymer materials,[67] but the population of the planarized species increases due to mechanical tension in the stretched polymer chain network (see Fig. 4e for further discussion).

**subsection 2)**
**Force threshold for fluorescence switching before mechanical bond scission**

Flapping molecules bearing OH groups were prepared for covalent doping in polycarbonate (PC) and polyurethane (PU). Terminal bulky substituents were introduced to suppress aggregation of the FLAP probes[68]. Due to the high brightness, the doping ratio could be minimized to less than $10^{-4}$ equivalent to that of other monomer components. The original mechanical properties of the host polymers were preserved even after covalent doping (Supplementary Fig. 20 for PC and Supplementary Fig. 27 for PU), which is an essential requisite for a force probe. The average distance between the doped FLAPs was estimated to be ≈20 nm. Since the distance is much farther than that for FRET (< 10 nm), the FRET-induced FL perturbation can be ignored. Negative control experiments by physical (noncovalent) doping of FLAP confirmed almost no mechanical response in terms of FL (Supplementary Fig. 38). In polymeric environments with chemical (covalent) doping, the sensitive mechanical response of this probe was confirmed. A linear PC sample prepared from the components in Fig. 2C showed glassy (nonelastomeric) behavior, with a glass transition temperature $T_g$ of 153 °C (Supplementary Fig. 19). When the notched PC film was slowly stretched under an inverted microscope equipped with a small tensile tester and a hyperspectral camera, the rapid growth of the stressed area was visualized on a micro- to millimeter scale based on the resulting FL, which changed from a bluish to greenish color under UV light. Two-dimensional imaging as a function of the FL ratio at 525- and 470-nm intensities ($FL_{525}/FL_{470}$) clarified the situation. The spectral response started much earlier than crack propagation from the notch (Fig. 2d). This result strongly suggested that the conformational planarization of FLAP can be specifically induced prior to covalent bond scission of the stressed polymer chains. In addition, compared with that in the crack tip front (region C), the FL ratio in a backward region of the crack (region B) decreased. This observation indicates that the FLAP force probe can trace the spatially transmitted relaxation process, while mechanically irreversible chromophores have a limitation in tracing relaxation after activation[69].

To estimate the threshold of the FL switch at the single-molecule level, (time-dependent) DFT calculations were performed in $S_0$ and $S_1$ on a polymer substructure near FLAP (Fig. 3). The energy profile in $S_0$ was obtained by plotting the relative potential energy of the optimized geometries with a fixed distance between terminal carbon atoms. At the most relaxed structure with a minimum energy, the corresponding distance was 40.4 Å (where extension = 0 Å), and the COT bending angle was 40.2°. Scanning the extension until 15 Å indicated that, in the early stage, the planarization of FLAP occurs specifically without C–C bond elongation in the terminal carbon chains. The planarization completed by 12.0-Å extension is an energetically uphill process, but it requires only 6.7 kcal/mol. The steepness of the early-stage slope, which is the required force for conformational planarization, is several tens of pN. Further extension results in a rapid energy increase over a 200-pN slope. Eventually, the side chain is broken without bond scission in the FLAP skeleton. To determine the threshold for FL switching, the corresponding $S_1$ energy profile must also be considered because a small conformational relaxation in $S_1$ should be expected, even in the polymeric environment that suppresses a large structural change in principle.[66] Therefore, additional optimizations in $S_1$ were conducted with the same extension value at each geometry as obtained in $S_0$. As a result, $S_1$ optimization from half-stretched structures below and above the 10.5-Å extension converged into two specific conformations of the FLAP skeleton, in which the COT bending angles were 28° (bent) and 0° (planar), respectively. The calculation result is consistent with the experimental results. The FL spectra of the stretched FLAP-doped PUs (see Fig. 4) showed a spectral change in only the relative intensity of the two FL bands at 474 and 525 nm, without a gradual redshift in wavelength. The observed unshifted FL bands indicate that the population of the excited species basically accumulates at the two $S_1$ minima. Since the $S_1$ relaxation of the FLAP skeleton proceeds on an ultrafast timescale[53], polymer chains around the FLAP force probe cannot largely move at this instant. Based on these results, FL switching of the single FLAP molecule in the polymer substructure can be expected around the 10.5-Å extension, at which the threshold energy is calculated to be 4.5 kcal/mol with the required force of approximately 100 pN.

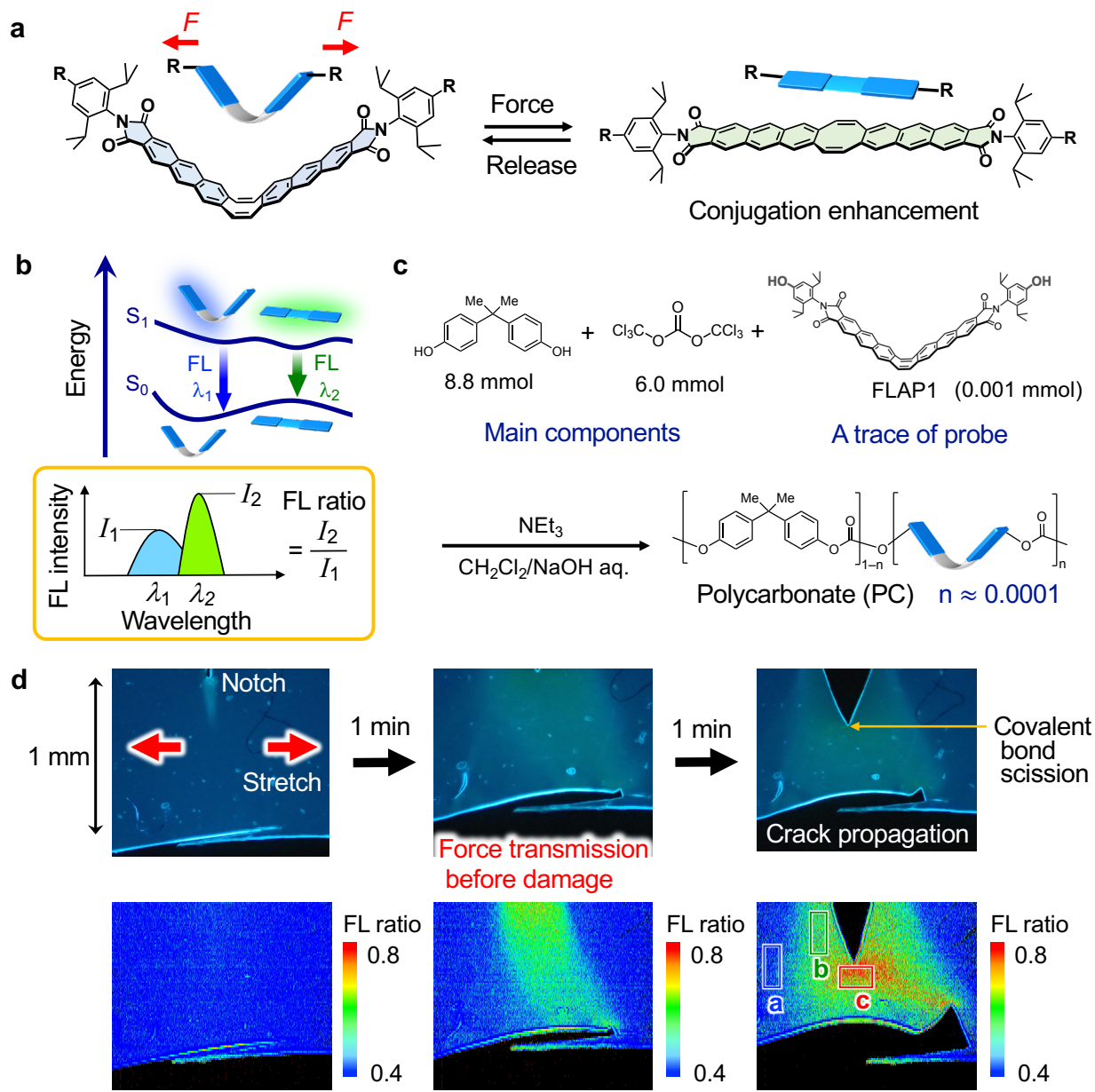

**Fig. 2.** (a) Flapping molecular force probe (FLAP) that enables (b) quantitative ratiometric analysis based on dual fluorescence (FL). (c) Preparation of polycarbonate (PC) from bisphenol A and triphosgene chemically doped with trace amounts of **FLAP1** (0.05 wt%). (d) FL microscopy images of the notched PC specimen under 365-nm UV irradiation. Force transmission induced before crack propagation by tensile testing (strain rate of $2.0 \times 10^{-3}$ s$^{-1}$) (top). Corresponding two-dimensional ratiometric FL images (FL$_{525}$/FL$_{470}$) obtained by a hyperspectral camera (bottom). In the last image, an unstretched region (A), a backward region of the crack (B), and the crack tip front (C) are indicated.

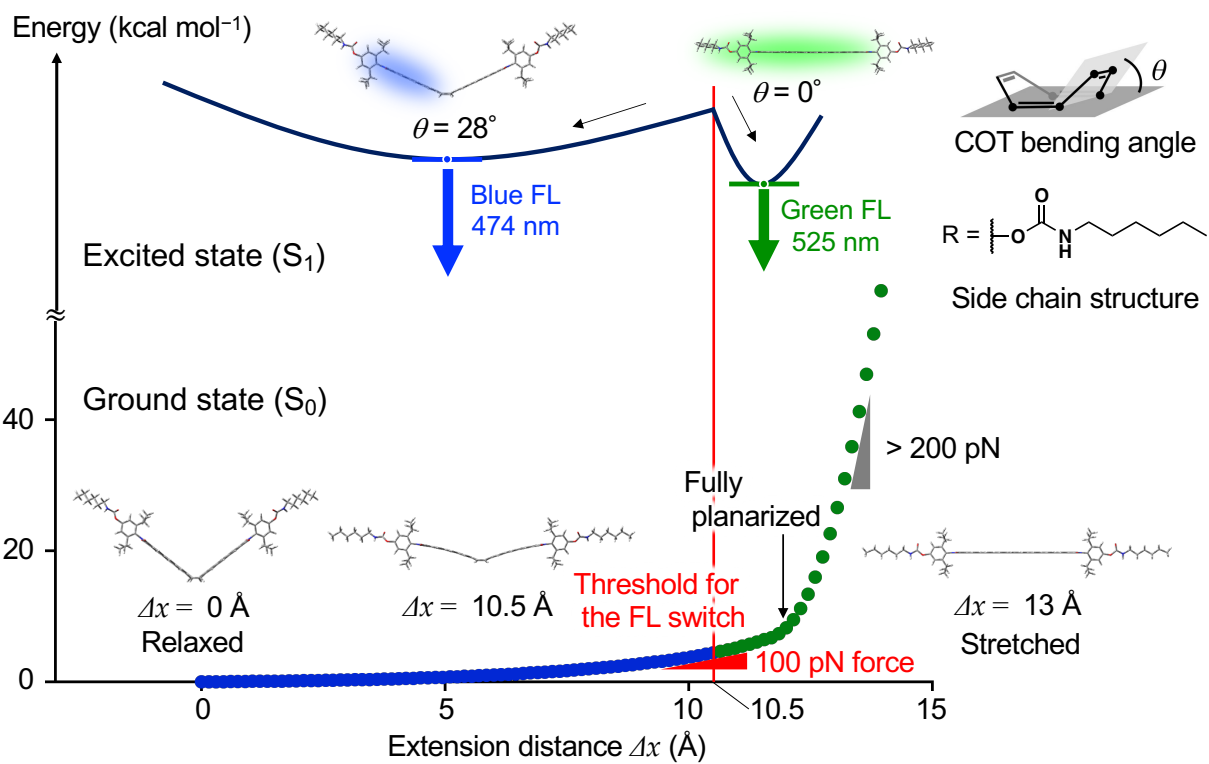

**Fig. 3.** DFT-calculated energy profile of a polymer-chain substructure in the vicinity of FLAP (R = *N*-hexylcarbamate in Fig 2a). Constrained geometry optimization was performed in $S_0$ at the PBE0/6-31G(d) level (bottom), in which the distance between terminal carbon atoms was fixed. From the most relaxed geometry, the relative potential energy was scanned as a function of extension, $\Delta x$, by 0.15 Å. Excited-state geometry optimization in $S_1$ at the TD-PBE0/6-31G(d) level was also performed (top) for each geometry obtained in $S_0$, where the extension was fixed at the same length. The presence of two relaxed conformers of the FLAP skeleton was indicated in $S_1$, whose COT bending angles were 28° (bent) and 0° (planar). Note that the conformational coordinates along the abscissa in $S_1$ (top) are different from those of $S_0$ (bottom) in this figure.

subsection 3)

**Controlling the chemical location of the force probe in crosslinked polyurethanes**

By controlling the chemical location of the FLAP force probe, we gained new quantitative insights into the biased local stress concentration of crosslinked PUs. Two different PUs, **PU1** and **PU2**, were synthesized by chemical doping with **FLAP1** and **FLAP2**, respectively (Fig. 4a). The photophysical properties of **FLAP1** and **FLAP2** were identical in solution. Due to the different numbers and positions of the reactive OH groups, **FLAP1** is incorporated into PU main chains, while **FLAP2** is incorporated into crosslinking points. Except for the difference in the probe locations, the following aspects of these PUs are all the same: stoichiometric amounts of monomer components, polymerization protocols using a tin catalyst, and preparation methods for dumbbell-shaped test specimens. Indeed, **PU1** and **PU2** as well as undoped PU (**PU0**) reproducibly showed almost identical mechanical and thermal properties (Fig. 4B and Supplementary Figs. 25–27). As shown Fig. 4, the crosslinking density, namely, the molar fraction of triethanolamine (TEA) to the sum of the monomers, was set to 0.10, while the molar ratio of **FLAP1** and **FLAP2** was small enough to be ignored, that is, 0.0001 equivalent to the poly(tetrahydrofuran) (PTHF) monomer (0.02 wt% to the obtained PUs). The stress–strain curve of the PU specimens indicated highly elastomeric properties (Fig. 4b). Typical rubbery deformations with Young's modulus values of 1.0–1.1 MPa were confirmed for these PUs. The yield point was not clearly observed for the rubbery PU. After a typical plateau-like region at approximately 1 MPa, strain-hardening behavior started at approximately 500% strain, and rupture eventually occurred at approximately 750% strain at ≈6 MPa. It is worth noting that the rubbery PUs were gradually transformed into semicrystalline PUs within several days (Supplementary Figs. 39–41); however, we focused on rubbery PUs in this study.

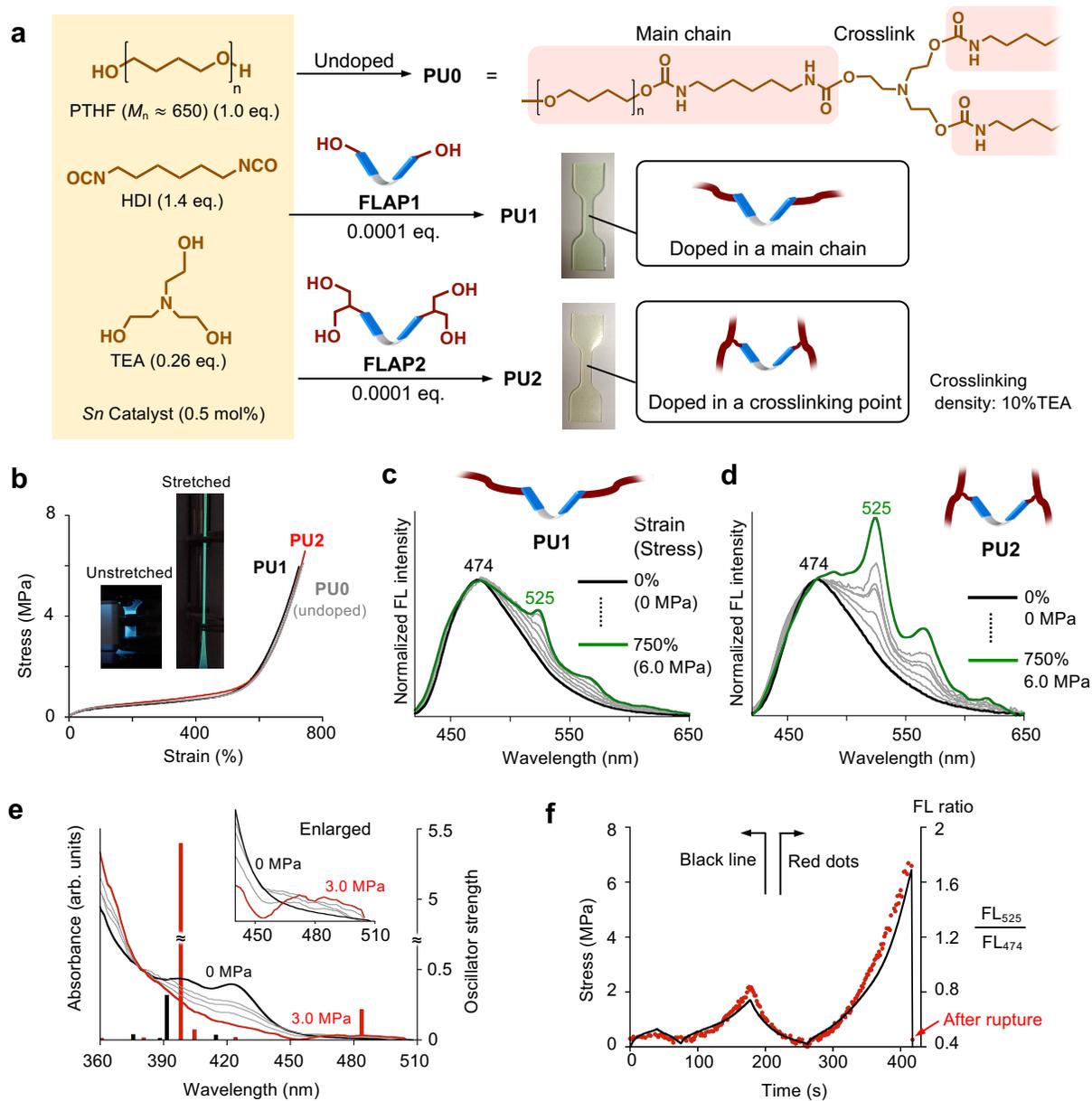

**Fig. 4.** (a) Controlled chemical doping of **FLAP1** and **FLAP2** in crosslinked polyurethanes, **PU1** and **PU2**, respectively. PTHF: poly(tetrahydrofuran); HDI: hexamethylenediisocyanate; TEA: triethanolamine; Sn catalyst: dibutyltin dilaurate. The crosslinking density can be tuned by the molar ratio of TEA to the other monomer components. (b) Engineering stress–strain curves of rubbery polyurethanes, **PU1**, **PU2** and undoped **PU0** with a crosslinking density of 10% TEA. Inset images: unstretched and stretched fluorescent specimens of **PU2** under UV light. (c, d) FL spectral changes during the tensile testing of **PU1** and **PU2**. FL intensity was normalized at 474 nm. Excitation wavelength: 365 nm. (e) An absorption spectral change of **PU2** by stretching the specimen. Oscillator strengths were calculated at the unstretched and fully stretched geometries (Supplementary Tables 12 and 13) at the TD PBE0/6-31+G(d) level. (f) Real-time and reversible FL response during the loading–unloading cycle of **PU2**. Time-dependent stress (black line) and FL ratio plots (red dots).

**subsection 4)**
**Fluorescence spectroscopy simultaneously conducted with tensile testing**

During mechanical testing, the FL spectra of the PU specimens were measured in real time (Fig. 4c and 4d). Under continuous UV irradiation (365 nm, ≈100 mW cm$^{-2}$), the FL emission was collected by an optical fiber connected to a multichannel photodetector. Before stretching, both **PU1** and **PU2** showed the same blue FL band with a single peak at 474 nm, corresponding to emission from the bent FLAP conformation in $S_1$. As the stress increased, a green FL band at 525 nm appeared along with vibronic subbands at 565 and 618 nm emitted from the planarized conformation. On the other hand, the blue FL band did not disappear in the late stage of stretching, meaning that considerable numbers of force probes were still relaxed even just before the specimens ruptured. Namely, the macroscopic stress was not ideally distributed over the whole polymer network. The nanoscale stress concentration cannot be discussed by using "turn-on" mechanophores that show no characteristic signal in their unactivated forms. More importantly, the relative degree of green FL enhancement was remarkably larger in **PU2** than in **PU1**, clearly indicating biased nanoscale stress concentrations at the crosslinkers compared with the PU main chains. A stretch-induced absorption spectral change was also confirmed for **PU2** (Fig. 4e). The appearance of a detectable absorption band in the long-wavelength region reaching 500 nm provided strong evidence for the compulsory conformational change of FLAP in the ground state. This weak absorption band was assigned to a partially allowed excitation in the visible range (Supplementary Table 13) for the effectively π-conjugated system of the fully planarized FLAP conformation in $S_0$. A loading–unloading cycle test was also carried out to confirm the repeated and real-time response of the mechanically reversible force probe (Fig. 4f and Supplementary Fig. 36). The FL ratio excellently traced the time-dependent stress macroscopically loaded on the PU specimen. After rupture, the FL ratio immediately returned to the extent of the unstretched initial phase, which confirmed the reversible response of the FLAP force probe and the negligible influence of photodegradation.

**subsection 5)**

**Quantifying biased nanoscale stress concentrations by ratiometric fluorescence analysis**

The biased stress concentration at crosslinking points was demonstrated in every PU with different crosslinking densities of 8.7%, 10% and 13% TEA (Fig. 5). The engineering stress–strain curves of these PUs are shown in Fig. 5a. The rupture stress and rupture strain decreased with increasing crosslinking density. The strain-hardening region of the highly crosslinked PU (13% TEA) started earlier at ≈400% strain, while that of the other PUs started at ≈500% strain. The FL ratios ($FL_{525}/FL_{474}$) were plotted as a function of macroscopic strain and stress (Fig. 5b, c), respectively. Here, we obtained an important insight. For all **PU2** specimens containing FLAP at crosslinking points, the FL ratio remarkably increased in the strain-hardening region. This behavior is in sharp contrast to the moderately increased FL ratio in the **PU1** specimens with FLAP at the main chains. In the plateau-like region of the stress–strain curve (observed below 1 MPa stress), the FL ratios were not largely different between **PU1** and **PU2**, indicating that the local stress concentrations were observed to the same degree at the main chains and crosslinkers. However, the biased stress concentration at the crosslinkers became pronounced during the strain-hardening process above 1 MPa stress. With decreasing crosslinking density, the difference in the FL ratios between **PU1** and **PU2** at the rupture point increased. On the other hand, when we focused on the specific strain level in a strain-hardening region (for example, 600% strain), larger macroscopic stress was generated for more highly crosslinked polymers, and accordingly, the corresponding FL ratios were relatively larger at the same strain level. Based on spectral separation analysis of the dual FL signals, we estimated the corresponding percentage of the number of stressed FLAP probes over the threshold for FL switching (Fig. 5d and Supplementary Fig. 37), where the efficiencies of UV absorption and FL emission of the bent and planar conformations were theoretically considered. At a 10% TEA crosslinking density, the estimated percentage of stressed probes at the rupture point reached 18 ± 2% (**PU2**) among the crosslinking FLAPs, whereas that in the main chains was approximately 9 ± 1% (**PU1**). Note that the low threshold of the FLAP force probe (theoretically predicted to be ≈100 pN) made the stressed probe percentage more than 1000 times larger than that of a reported mechanophore (diarylbibenzofuranone) that requires covalent bond dissociations[70], demonstrating the suitability of FLAP for polymer physics studies.

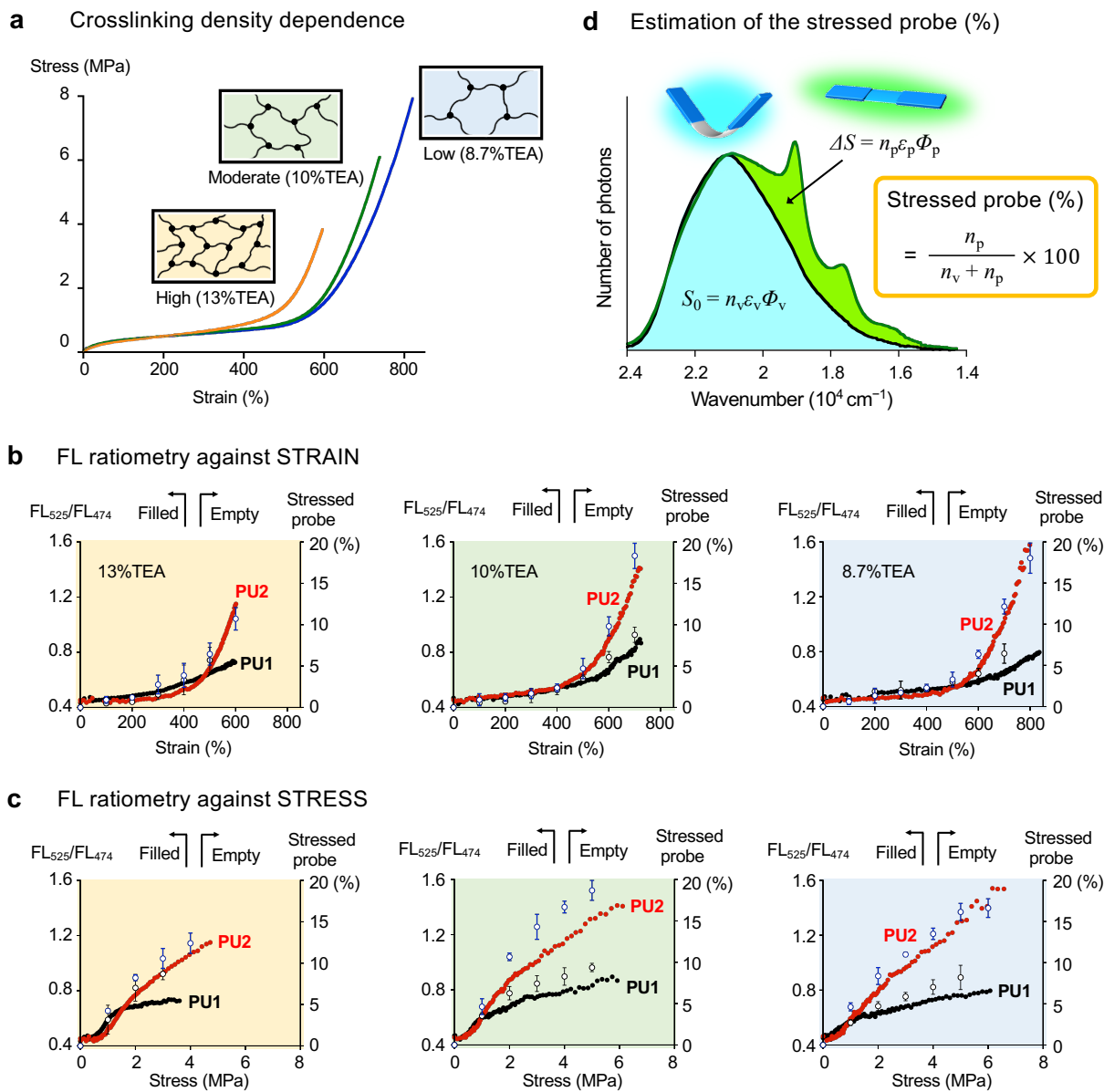

**Fig. 5.** (a) Stress–strain curves of undoped **PU0** with different crosslinking densities. Almost the same curves were obtained for **PU1** and **PU2** at each crosslinking density (Supplementary Fig. 27). (b,c) FL ratio (filled) and percentages of the stressed FLAP probe (empty) as a function of macroscopic strain (b) and stress (c) based on tensile testing. Error bars in the stressed probe (%) indicate the standard deviation obtained from three tensile tests. (d) Estimation of the stressed probe (%) over the threshold force of approximately 100 pN based on the observed FL spectrum. $n$: number of FLAP molecules, $\varepsilon$: molar absorption coefficient at the excitation wavelength (365 nm), and $\Phi$: FL quantum yield, where the subscripts of "v" and "p" mean V-shaped (bent) and planar conformers, respectively. See the details in Supplementary Fig. 37.

## Methods:

**Synthesis and purification**

All reagents and solvents were of commercial grade and were used without further purification unless noted. Tetrahydrofuran (THF) was dried using a glass contour solvent purification system. Superdehydrated dimethylformamide (DMF) was purchased from Wako Chemicals. Poly(tetrahydrofuran) ($M_n \sim 650$) was dried under vacuum at 70 °C for 2 h before use. Thin-layer chromatography (TLC) was performed with silica gel 60 $F_{254}$ (Merck). Column chromatography was performed using Wako gel C-300.

**Measurements**

$^1$H and $^{13}$C nuclear magnetic resonance (NMR) spectra were recorded on a JEOL ECA-600 (600 MHz for $^1$H and 151 MHz for $^{13}$C NMR) spectrometer. Chemical shifts were expressed as $\delta$ in ppm relative to the internal standards $CHCl_3$ ($\delta$ = 7.26 ppm for $^1$H and $\delta$ = 77.16 ppm for $^{13}$C) and DMSO ($\delta$ = 2.50 ppm for $^1$H and $\delta$ = 39.52 ppm for $^{13}$C). High-resolution atmospheric-pressure chemical ionization time-of-flight mass spectrometry (HR-APCI-TOF-MS) was performed on a BRUKER micrOTOF system in positive mode. Ultraviolet (UV)–visible absorption spectra were recorded on a Shimadzu UV-3600 spectrometer. Fluorescence (FL) spectra were recorded on a JASCO FP-8500 spectrofluorometer. Absolute FL quantum yields were determined on HAMAMATSU C9920-02S system. The FL lifetime was recorded on a Hamamatsu Photonics Quantaurus-Tau C11367 spectrometer. Tensile tests were carried out using a SHIMADZU AUTOGRAPH AGS-X tester with a 1-kN load cell at the crosshead. Differential scanning calorimetry (DSC) was performed on a Hitachi High-Tech TA7000 system. Rheological measurements were conducted on an Anton Paar MCR702 rheometer. Real-time FL measurements of PU specimens were performed using an Otsuka Electronics MCPD-6800 multichannel photodetector. Crossed Nicols images were obtained by a Leica DM2500 P optical microscope equipped with a Linkam LTS420E temperature control system. Photographs and movies of specimens were taken using an OLYMPUS Tough TG-6 digital camera.

**Fluorescence microscopy imaging of the stretched polycarbonate film**

This experiment was conducted by the combinational use of an inverted microscope (OLYMPUS IX83 inverted research microscope equipped with a U-HGLGPS mercury lamp), a tensile testing machine (AcroEdge OZ911 tensile testing machine) and a hyperspectral camera (EBAJAPAN NH-8 hyperspectral camera). A polycarbonate (PC) specimen (width ≈ 1 mm and thickness ≈ 0.03 mm) with a small notch was used for tensile testing. This specimen was held with an initial distance of 25 mm and extended from both sides at a speed of 0.05 mm s$^{-1}$. Because of symmetrical stretching, the growing notch fits within the field of view of the microscope. At the same time, FL spectra at each pixel were collected with the hyperspectral camera. The data were acquired with an exposure time of 500 ms at a wavelength interval of 5 nm. FL optical micrographs (Fig. 2d, top) were taken by a digital camera (OLYMPUS TG-6), while the mapping images of the FL ratio (Fig. 2d, bottom) were obtained by analytical software (EBAJAPAN HSAnalyzer) based on the collected FL spectra.

**Theoretical study**

Theoretical calculations were performed using the Gaussian 16 program.

**Acknowledgments:** We thank Dr Hiroya Abe and Dr Yuta Saito (Tohoku Univ.) for their help in the early stage of the preliminary study. **Funding:** JST PRESTO (FRONTIER) and JST FOREST, Grant numbers JPMJPR16P6 and JPMJFR201L; MEXT/JSPS KAKENHI, Grant Numbers JP21H01917, JP21H05482, JP18H01952, JP20H04625, JP19KK0357, JP18H05482, and JP18J22477; Inoue Foundation for Science; Toray Science Foundation.

**Author contributions:** H.Y. and S.S. conceived the concept. R.K., H.Y., and S.S. designed the experiments. R.K., S.Y., and S.N. performed the experiments. R.K. and S.S. conducted the quantum calculations. R.K. and S.S. analyzed the data. R.K. and S.S. wrote the manuscript, and S.Y., A.O., and H.Y. proofread it.

**Competing interests:** Authors declare no competing interests.

**Figure Legends/Captions:**

**Fig. 1.** (a) Hierarchical structures of materials and physical measurement methods of mechanical forces at different scales. (b) Target force range for the nanoscale study of polymer physics.

**Fig. 2.** (a) Flapping molecular force probe (FLAP) that enables (b) quantitative ratiometric analysis based on dual fluorescence (FL). (c) Preparation of polycarbonate (PC) from bisphenol A and triphosgene chemically doped with trace amounts of **FLAP1** (0.05 wt%). (d) FL microscopy images of the notched PC specimen under 365-nm UV irradiation. Force transmission induced before crack propagation by tensile testing (strain rate of $2.0 \times 10^{-3}$ s$^{-1}$) (top). Corresponding two-dimensional ratiometric FL images (FL$_{525}$/FL$_{470}$) obtained by a hyperspectral camera (bottom). In the last image, an unstretched region (A), a backward region of the crack (B), and the crack tip front (C) are indicated.

**Fig. 3.** DFT-calculated energy profile of a polymer-chain substructure in the vicinity of FLAP (R = *N*-hexylcarbamate in Fig 2a). Constrained geometry optimization was performed in S$_0$ at the PBE0/6-31G(d) level (bottom), in which the distance between terminal carbon atoms was fixed. From the most relaxed geometry, the relative potential energy was scanned as a function of extension, *Δx*, by 0.15 Å. Excited-state geometry optimization in S$_1$ at the TD-PBE0/6-31G(d) level was also performed (top) for each geometry obtained in S$_0$, where the extension was fixed at the same length. The presence of two relaxed conformers of the FLAP skeleton was indicated in S$_1$, whose COT bending angles were 28° (bent) and 0° (planar). Note that the conformational coordinates along the abscissa in S$_1$ (top) are different from those of S$_0$ (bottom) in this figure.

**Fig. 4.** (a) Controlled chemical doping of **FLAP1** and **FLAP2** in crosslinked polyurethanes, **PU1** and **PU2**, respectively. PTHF: poly(tetrahydrofuran); HDI: hexamethylenediisocyanate; TEA: triethanolamine; Sn catalyst: dibutyltin dilaurate. The crosslinking density can be tuned by the molar ratio of TEA to the other monomer components. (b) Engineering stress–strain curves of rubbery polyurethanes, **PU1**, **PU2** and undoped **PU0** with a crosslinking density of 10% TEA. Inset images: unstretched and stretched fluorescent specimens of **PU2** under UV light. (c, d) FL spectral changes during the tensile testing of **PU1** and **PU2**. FL intensity was normalized at 474 nm. Excitation wavelength: 365 nm. (e) An absorption spectral change of **PU2** by stretching the specimen. Oscillator strengths were calculated at the unstretched and fully stretched geometries (Supplementary Tables 12 and 13) at the TD PBE0/6-31+G(d) level. (f) Real-time and reversible FL response during the loading–unloading cycle of **PU2**. Time-dependent stress (black line) and FL ratio plots (red dots).

**Fig. 5.** (a) Stress–strain curves of undoped **PU0** with different crosslinking densities. Almost the same curves were obtained for **PU1** and **PU2** at each crosslinking density (Supplementary Fig. 27). (b,c) FL ratio (filled) and percentages of the stressed FLAP probe (empty) as a function of macroscopic strain (b) and stress (c) based on tensile testing. Error bars in the stressed probe (%) indicate the standard deviation obtained from three tensile tests. (d) Estimation of the stressed probe (%) over the threshold force of approximately 100 pN based on the observed FL spectrum. *n*: number of FLAP molecules, *ε*: molar absorption coefficient at the excitation wavelength (365 nm), and *Φ*: FL quantum yield, where the subscripts of "v" and "p" mean V-shaped (bent) and planar conformers, respectively. See the details in Supplementary Fig. 37.

**Data and materials availability:** All data are available in the main text or the supplementary information.

Supplementary Information:

**Supplementary Movie 1.** Monitoring force transmission of the stretched polycarbonate film chemically doped with the dual fluorescent molecular force probe. Rapid growth of the stressed area was visualized much earlier than crack propagation, demonstrating a lower threshold for the fluorescence switch of the force probe, well below the force required for covalent bond scission. Excitation wavelength: 365 nm. The field of view of the microscope: 1 mm diameter. Stretching rate: 5 μm s$^{-1}$. Strain rate: 2 × 10$^{-4}$ s$^{-1}$.

**Materials and Methods** (Supplementary Figs. 1–68, Supplementary Tables 1–24)

# Supplementary Information for

# Bridging Pico-to-Nanonewtons with a Ratiometric Force Probe for Monitoring Nanoscale Polymer Physics Before Damage

Ryota Kotani, Soichi Yokoyama, Shunpei Nobusue, Shigehiro Yamaguchi, Atsuhiro Osuka, Hiroshi Yabu\*, Shohei Saito\*

Correspondence to: saito.shohei.4c@kuchem.kyoto-u.ac.jp

**This PDF file includes:**

Caption for Supplementary Movie 1
Supplementary Text
Supplementary Figs. 1 to 68
Supplementary Tables 1 to 24

**The Other Supplementary Data:**

Supplementary Movie 1

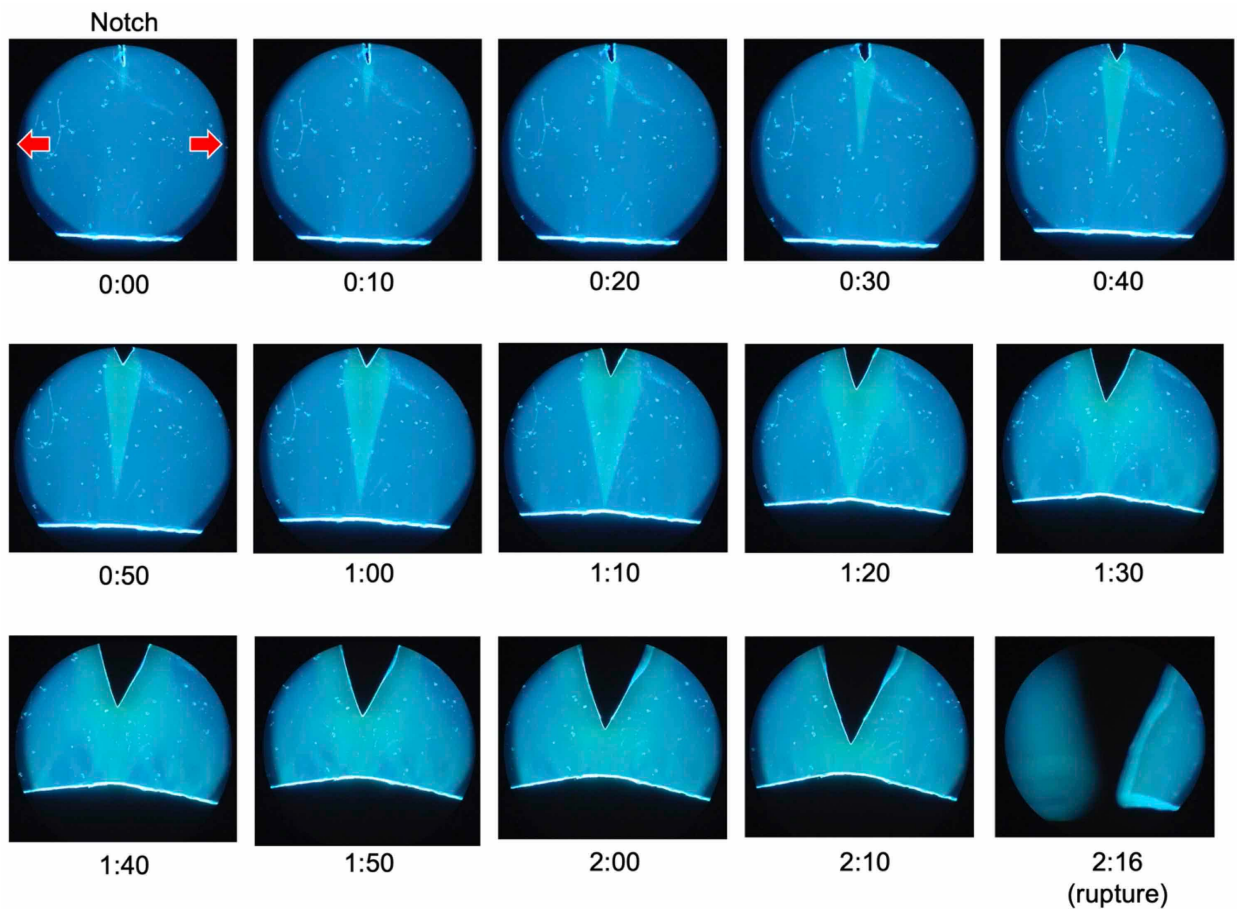

**Supplementary Movie 1.** Monitoring force transmission of the stretched polycarbonate film chemically doped with the dual fluorescent molecular force probe. Rapid growth of the stressed area was visualized much earlier than crack propagation, demonstrating a lower threshold for the fluorescence switch of the force probe, well below the force required for covalent bond scission. Excitation wavelength: 365 nm. The field of view of the microscope: 1 mm diameter. Stretching rate: 5 μm s$^{-1}$. Strain rate: $2 \times 10^{-4}$ s$^{-1}$.

## Synthesis of new compounds

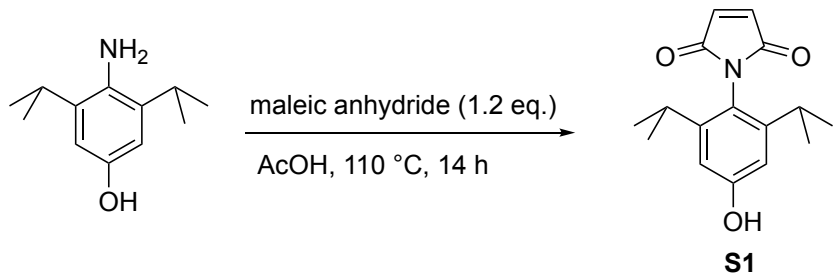

### *N*-(4-Hydroxy-2,6-diisopropylphenyl)maleimide (S1)

Under Ar atmosphere, 4-amino-3,5-diisopropylphenol[1] (2.77 g, 14.3 mmol) and maleic anhydride (1.68 g, 17.2 mmol) were added to a round-bottom flask (500 mL). After addition of AcOH (70 mL), the reaction mixture was stirred at 110 °C for 14 h. The resulting mixture diluted with EtOAc was poured into a separatory funnel, washed with saturated aqueous NaHCO$_3$ (3 times) and brine, dried over anhydrous Na$_2$SO$_4$ and evaporated. The crude product was purified by silica gel column chromatography (eluent: CH$_2$Cl$_2$/EtOAc = 20/1 by volume). Recrystallization from *n*-hexane furnished **S1** as a white solid (3.08 g, 79%).

$^1$H NMR (600 MHz, CDCl$_3$) $\delta$ (ppm) 6.87 (s, 2H), 6.69 (s, 2H), 4.80 (s, 1H), 2.56 (sept, *J* = 6.9 Hz, 2H) and 1.13 (d, *J* = 6.9 Hz, 12H); $^{13}$C NMR (151 MHz, CDCl$_3$) $\delta$ (ppm) 170.99, 157.10, 149.54, 134.43, 118.95, 111.26, 29.50 and 23.95; HR-APCI TOF-MS (*m/z*) found 273.1358, calcd for C$_{16}$H$_{19}$NO$_3$: 273.1365 [*M*]$^+$.

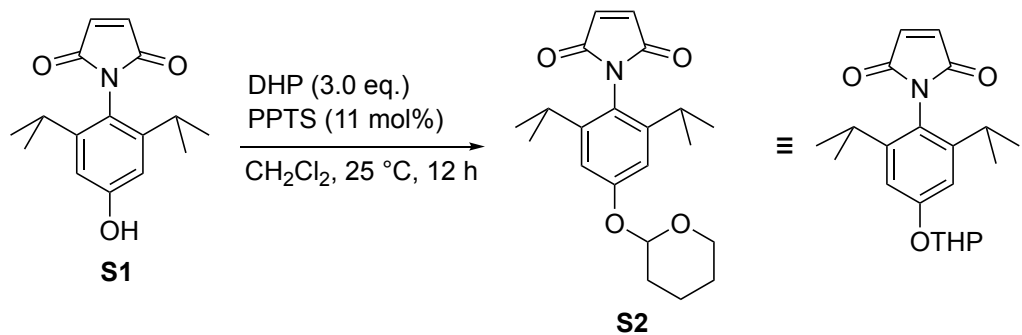

### *N*-(2,6-Diisopropyl-4-((tetrahydro-2*H*-pyran-2-yl)oxy)phenyl)maleimide (S2)

Under Ar atmosphere, **S1** (3.08 g, 11.3 mmol) and pyridinium *p*-toluenesulfonate (PPTS, 310 mg, 1.23 mmol) were added to a round-bottom flask (300 mL). After addition of CH$_2$Cl$_2$ (85 mL) and 3,4-dihydro-2*H*-pyran (DHP, 2.87 mL, 33.4 mmol) at 0 °C, the reaction mixture was stirred at 25 °C for 12 h. The reaction mixture was quenched with H$_2$O and poured into a separatory funnel, then washed with brine, dried over anhydrous Na$_2$SO$_4$ and evaporated. Recrystallization from CH$_2$Cl$_2$/*n*-hexane furnished **S2** as a white solid (3.86 g, 96%).

$^1$H NMR (600 MHz, CDCl$_3$) $\delta$ (ppm) 6.92 (s, 2H), 6.86 (s, 2H), 5.45 (m, 1H), 3.92 (m, 1H), 3.63 (m, 1H), 2.57 (sept, *J* = 6.9 Hz, 2H), 2.02 (m, 1H), 1.87 (m, 2H), 1.67–1.69 (m, 2H), 1.62–1.63 (m, 1H) and 1.13 (d, *J* = 6.9 Hz, 12H); $^{13}$C NMR (151 MHz, CDCl$_3$) $\delta$ (ppm) 170.94, 158.76, 148.89, 134.36, 119.52, 112.33, 96.34, 61.94, 30.49, 29.55, 25.36, 23.95 and 18.71; HR-APCI TOF-MS (*m/z*) found 357.1946, calcd for C$_{21}$H$_{27}$NO$_4$: 357.1946 [*M*]$^-$.

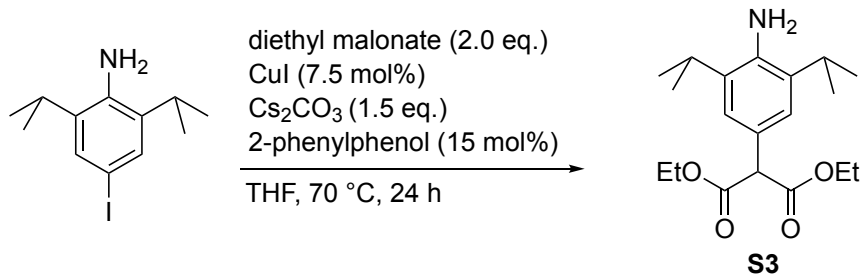

**Diethyl 2-(4-amino-3,5-diisopropylphenyl)malonate (S3)**
Under Ar atmosphere, 4-iodo-2,6-diisopropylaniline[2] (6.06 g, 20.0 mmol), CuI (285 mg, 1.50 mmol), $Cs_2CO_3$ (9.78 g, 30.0 mmol) and 2-phenylphenol (510 mg, 3.00 mmol) were added to a round-bottom flask (100 mL). After addition of THF (20 mL) and diethyl malonate (6.10 mL, 40.0 mmol), the reaction mixture was stirred at 70 °C for 24 h. The reaction mixture was quenched with $H_2O$ and diluted with EtOAc. The resulting mixture was poured into a separatory funnel, washed with brine, dried over anhydrous $Na_2SO_4$ and evaporated. The crude product was purified by silica gel column chromatography (eluent: $CH_2Cl_2$) and **S3** was obtained as a brown oil (4.00 g, 59%).
$^1$H NMR (600 MHz, $CDCl_3$) $\delta$ (ppm) 7.04 (s, 2H), 4.50 (s, 1H), 4.25–4.16 (m, 4H), 3.75 (s, 2H), 2.90 (sept, $J$ = 6.9 Hz, 2H) and 1.28–1.25 (m, 18H); $^{13}$C NMR (151 MHz, $CDCl_3$) $\delta$ (ppm) 168.96, 140.43, 132.45, 124.04, 122.55, 61.61, 58.06, 28.16, 22.50 and 14.21; HR-APCI TOF-MS ($m/z$) found 336.2183, calcd for $C_{19}H_{30}NO_4$: 336.2169 $[M+H]^+$.

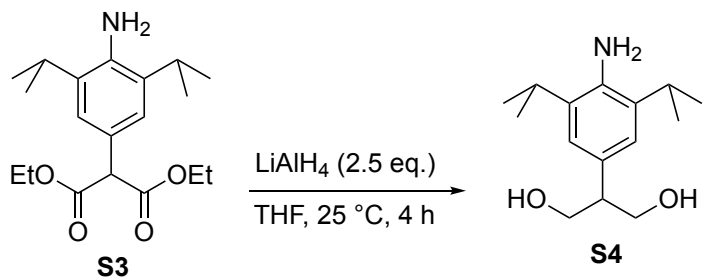

**2-(4-Amino-3,5-diisopropylphenyl)propane-1,3-diol (S4)**
Under Ar atmosphere, to the suspension of $LiAlH_4$ (396 mg, 10.4 mmol) in THF (10 mL), a THF (10 mL) solution of **S3** (1.52 g, 4.18 mmol) was added at 0 °C. The mixture was stirred at 25 °C for 4 h. The reaction mixture was quenched with saturated aqueous $Na_2SO_4$, filtered to remove aluminum salts, and washed with EtOAc (50 mL). The filtrate was purified by silica gel column chromatography (eluent: EtOAc). Recrystallization from $CH_2Cl_2$/$n$-hexane furnished **S4** as a white solid (910 mg, 87%).
$^1$H NMR (600 MHz, $CDCl_3$) $\delta$ (ppm) 6.87 (s, 2H), 3.98–3.94 (m, 2H), 3.91–3.88 (m, 2H), 3.70 (s, 2H), 3.03 (tt, $J$ = 7.8, 5.4 Hz, 1H), 2.92 (sept, $J$ = 6.6 Hz, 2H), 1.92 (t, $J$ = 5.4 Hz, 2H) and 1.26 (d, $J$ = 6.6 Hz, 12H); $^{13}$C NMR (151 MHz, $CDCl_3$) $\delta$ (ppm) 139.44, 132.97, 128.69, 122.43, 66.55, 49.75, 28.09 and 22.52; HR-APCI TOF-MS ($m/z$) found 252.1968, calcd for $C_{15}H_{26}NO_2$: 252.1958 $[M+H]^+$.

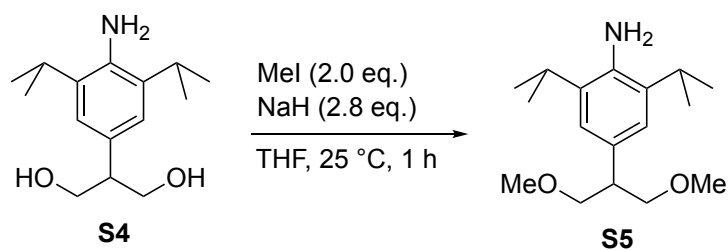

**4-(1,3-Dimethoxypropan-2-yl)-2,6-diisopropylaniline (S5)**

Under Ar atmosphere, to the suspension of NaH (60% dispersion in paraffin liquid, 401 mg, 10.0 mmol) and MeI (445 μL, 7.16 mmol) in THF (9.0 mL), a THF (4.5 mL) solution of **S4** (900 mg, 3.58 mmol) was added at 0 °C. The mixture was stirred at 25 °C for 1 h. The reaction mixture was diluted with EtOAc (20 mL) and evaporated. The crude product was purified by silica gel column chromatography (eluent: $CH_2Cl_2$:*n*-hexane:$EtO_2$ = 1:1:1 by volume) to furnish **S5** as brown oil (932 mg, 93%).

$^1$H NMR (600 MHz, CDCl$_3$) δ (ppm) 6.92 (s, 2H), 3.63 (dd, *J* = 9.2, 6.8 Hz, 2H), 3.57 (dd, *J* = 9.7, 6.4 Hz, 2H), 3.39 (s, 6H), 3.06 (tt, *J* = 9.2, 6.4 Hz, 1H), 2.91 (sept, *J* = 6.9 Hz, 2H), and 1.27 (d, *J* = 6.9 Hz, 12H); $^{13}$C NMR (151 MHz, CDCl$_3$) δ (ppm) 139.12, 132.45, 130.54, 122.43, 74.79, 58.99, 45.84, 28.13 and 22.56; HR-APCI TOF-MS (*m/z*) found 280.2271, calcd for $C_{17}H_{30}NO_2$: 280.2271 [*M*+H]$^+$.

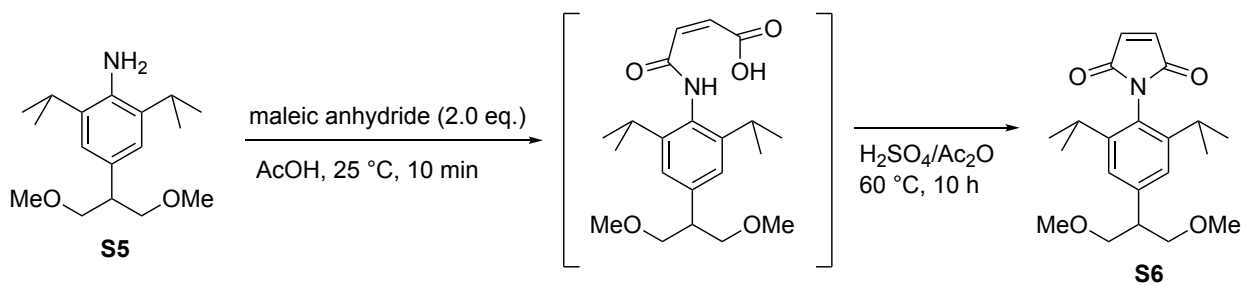

***N*-(4-(1,3-Dimethoxypropan-2-yl)-2,6-diisopropylphenyl)maleimide (S6)**

Under Ar atmosphere, **S5** (1.03 g, 3.69 mmol) and maleic anhydride (723 mg, 7.37 mmol) were added to a Schlenk flask (20 mL). After addition of AcOH (2.21 mL), the reaction mixture was stirred at 25 °C for 10 min. Then, H$_2$SO$_4$ (95%, 369 μL) and acetic anhydride (184 μL) were added to the reaction mixture, and the reaction vessel was warmed at 60 °C for 10 h. The reaction mixture was quenched with H$_2$O and diluted with EtOAc. The solution was poured into a separatory funnel, washed with brine, dried over anhydrous Na$_2$SO$_4$ and evaporated. The crude product was purified by silica gel column chromatography (eluent: CH$_2$Cl$_2$/Et$_2$O = 6/1 by volume) and **S6** was obtained as a white solid (1.06 g, 80%).

$^1$H NMR (600 MHz, CDCl$_3$) δ (ppm) 7.12 (s, 2H), 6.88 (s, 2H), 3.67–3.64 (m, 2H), 3.63–3.60 (m, 2H), 3.35 (s, 6H), 3.13 (tt, *J* =7.2, 6.0 Hz, 1H), 2.59 (sept, *J* = 6.9 Hz, 2H) and 1.15 (d, *J* = 6.9 Hz, 12H); $^{13}$C NMR (151 MHz, CDCl$_3$) δ (ppm) 170.84, 147.24, 142.95, 134.43, 124.79, 123.94, 74.04, 59.09, 46.37, 29.45, and 24.09; HR-APCI TOF-MS (*m/z*) found 359.2096, calcd for $C_{21}H_{30}NO_4$: 359.2091 [*M*]$^+$.

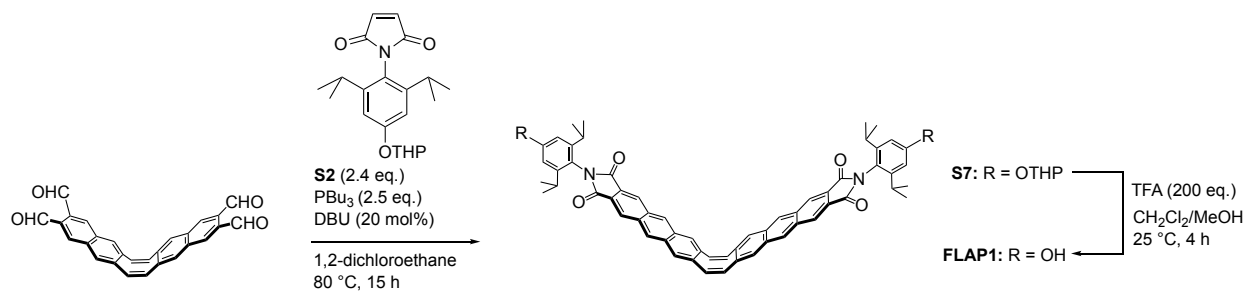

**FLAP1**

Under Ar atmosphere, **S2** (559 mg, 1.56 mmol) and 1,2-dichloroethane (12 mL) were added to a round-bottom flask (20 mL). After addition of tributylphosphine (410 μL, 1.66 mmol) at 0 °C, the reaction mixture was stirred at 25 °C for 30 min. A 1,2-dichloroethane (10 mL) solution of tetraformyl dinaphthocyclooctatetraene[3,4] (271 mg, 0.651 mmol) was added to the reaction mixture at 0 °C, followed by addition of DBU (19.5 μL, 0.130 mmol). The reaction mixture was stirred at 80 °C for 15 h, and then quenched with H$_2$O. The resulting mixture was poured into a separatory funnel, washed with brine, dried over anhydrous Na$_2$SO$_4$ and evaporated. The crude product was purified by silica gel column chromatography (eluent: CH$_2$Cl$_2$/AcOEt = 20/1 by volume), and **S7** was obtained as a yellow solid (138 mg, 20% from tetraformyl dinaphthocyclooctatetraene), which was used for the subsequent reaction. **S7** was added to a round-bottom flask (100 mL). After addition of CH$_2$Cl$_2$ (30 mL), MeOH (10 mL), and TFA (2.0 mL, 26 mmol), the reaction mixture was stirred at 25 °C for 4 h. The reaction mixture was quenched with H$_2$O and poured into a separatory funnel, then washed with brine, dried over anhydrous Na$_2$SO$_4$ and evaporated. The crude product was purified by silica gel column chromatography (eluent: CH$_2$Cl$_2$/EtOAc = 20/1 by volume). Recrystallization from CH$_2$Cl$_2$/*n*-hexane furnished **FLAP1** as a yellow solid (107 mg, 92% from **S7**).

$^1$H NMR (600 MHz, CDCl$_3$) δ (ppm) 8.55 (m, 4H + 4H), 7.94 (s, 4H), 7.24 (s, 4H), 6.75 (s, 4H), 4.90 (s, 2H), 2.70 (sept, *J* = 6.9 Hz, 4H) and 1.14 (d, *J* = 6.9 Hz, 24H); $^{13}$C NMR (151 MHz, DMSO-*d$_6$*) δ (ppm) 167.47, 158.59, 147.92, 136.12, 133.09, 131.88, 131.44, 129.93, 128.34, 126.56, 125.73, 118.40, 110.50, 28.65 and 23.56; HR-MALDI TOF-MS (*m/z*) found 895.33, calcd for C$_{60}$H$_{51}$N$_2$O$_6$: 895.37 [*M*+H]$^+$. UV/visible absorption (CH$_2$Cl$_2$): λ$_{max}$ (ε /M$^{-1}$ cm$^{-1}$) = 332 nm (9.28 × 10$^4$); fluorescence (CH$_2$Cl$_2$, λ$_{ex}$ = 340 nm): λ$_{max}$ = 523, 564, and 610 nm, *Φ*$_{FL}$ = 0.26.

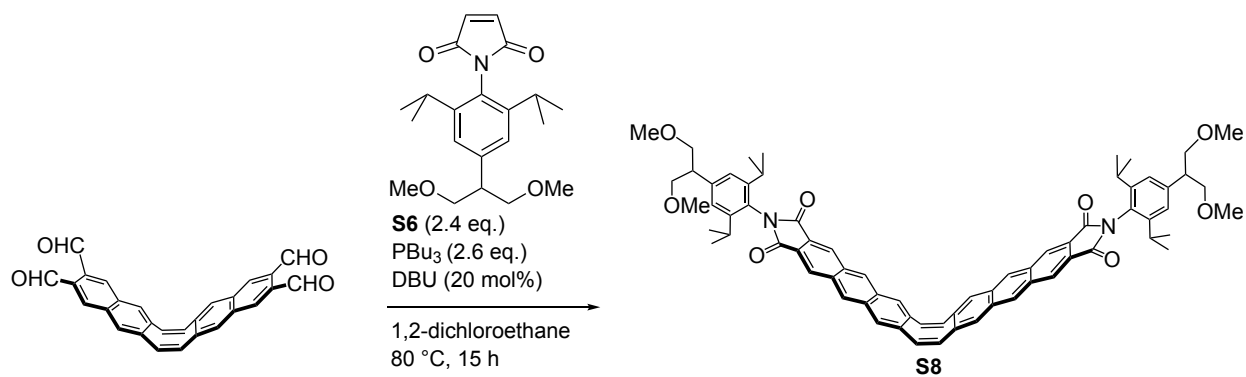

**S8**
Under Ar atmosphere, **S6** (682 mg, 1.90 mmol) and 1,2-dichloroethane (30 mL) were added to a round-bottom flask (200 mL). After addition of tributylphosphine (507 μL, 2.05 mmol) at 0 °C, the reaction mixture was stirred at 25 °C for 30 min. Then, a 1,2-dichloroethane (15 mL) solution of tetraformyl dinaphthocyclooctatetraene (329 mg, 0.79 mmol) was added to the reaction mixture at 0 °C, followed by addition of DBU (24 μL, 0.16 mmol). The reaction mixture was stirred at 80 °C for 12 h, and then quenched with $H_2O$. The resulting mixture was poured into a separatory funnel, washed with brine, dried over anhydrous $Na_2SO_4$ and evaporated. The crude product was purified by silica gel column chromatography (eluent: $CH_2Cl_2$/$Et_2O$ = 5:1 by volume) and **S8** was obtained as a yellow solid (208 mg, 25% from tetraformyl dinaphthocyclooctatetraene).
$^1$H NMR (600 MHz, CDCl$_3$) δ (ppm) 8.55 (s, 4H + 4H), 7.94 (s, 4H), 7.24 (s, 4H), 7.17 (s, 4H), 3.69–3.62 (m, 8H), 3.37 (s, 12H), 3.16 (tt, $J$ =7.2, 6.0 Hz, 2H), 2.72 (sept, $J$ = 6.8 Hz, 4H) and 1.15 (d, $J$ = 6.8 Hz, 24H); $^{13}$C NMR (151 MHz, CDCl$_3$) δ (ppm) 167.88, 146.68, 142.80, 136.63, 133.36, 132.39, 132.20, 129.84, 128.49, 126.64, 125.92, 123.88, 123.56, 74.16, 59.07, 46.37, 29.48, and 24.12; HR-MALDI TOF-MS (*m/z*) found 1089.53, calcd for $C_{70}H_{70}N_2O_8Na$: 1089.50 [*M*+Na]$^+$.

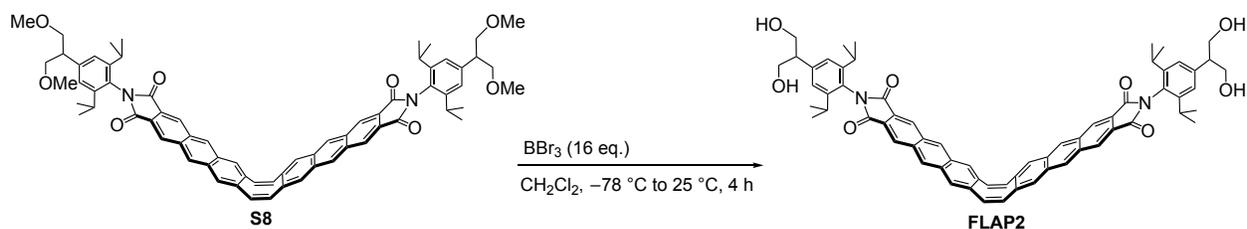

## FLAP2

Under Ar atmosphere, **S8** (9.8 mg, 9.2 μmol) and CH$_2$Cl$_2$ (0.7 mL) were added to a Schlenck flask (20 mL). After addition of a CH$_2$Cl$_2$ solution of BBr$_3$ (1.0 M, 147 μL) at –78 °C, the reaction mixture was stirred and warmed to 25 °C over 4 h. The reaction mixture was quenched with H$_2$O and poured into a separatory funnel, then washed with brine, dried over anhydrous Na$_2$SO$_4$ and evaporated. The crude product was purified by silica gel column chromatography (eluent: EtOAc) and **FLAP2** was obtained as a yellow solid (6.7 mg, 72%).

$^1$H NMR (600 MHz, CDCl$_3$) δ (ppm) 8.56 (s, 4H), 8.55 (s, 4H), 7.94 (s, 4H), 7.24 (s, 4H), 7.14 (s, 4H), 4.05–4.00 (m, 8H), 3.18–3.14 (m, 2H), 2.74 (sept, $J$ = 6.8 Hz, 4H), 2.01 (t, $J$ = 6.0 Hz, 4H) and 1.16 (d, $J$ = 6.8 Hz, 24H); $^{13}$C NMR (151 MHz, DMSO-$d_6$) δ (ppm) 167.30, 145.81, 143.99, 136.18, 133.13, 131.89, 131.47, 129.99, 128.37, 126.76, 125.69, 125.42, 123.63, 62.64, 50.82, 28.70, and 23.67; HR-MALDI TOF-MS ($m/z$) found 1033.42, calcd for C$_{66}$H$_{62}$N$_2$O$_8$Na: 1033.44 [$M$+Na]$^+$. UV/visible absorption (CH$_2$Cl$_2$): $\lambda_{max}$ ($\varepsilon$ /M$^{-1}$ cm$^{-1}$) = 331 nm (9.37 × 10$^4$); fluorescence (CH$_2$Cl$_2$, $\lambda_{ex}$ = 340 nm): $\lambda_{max}$ = 524, 565, and 610 nm, $\varPhi_{FL}$ = 0.32.

## Photophysical properties of parent FLAP molecules

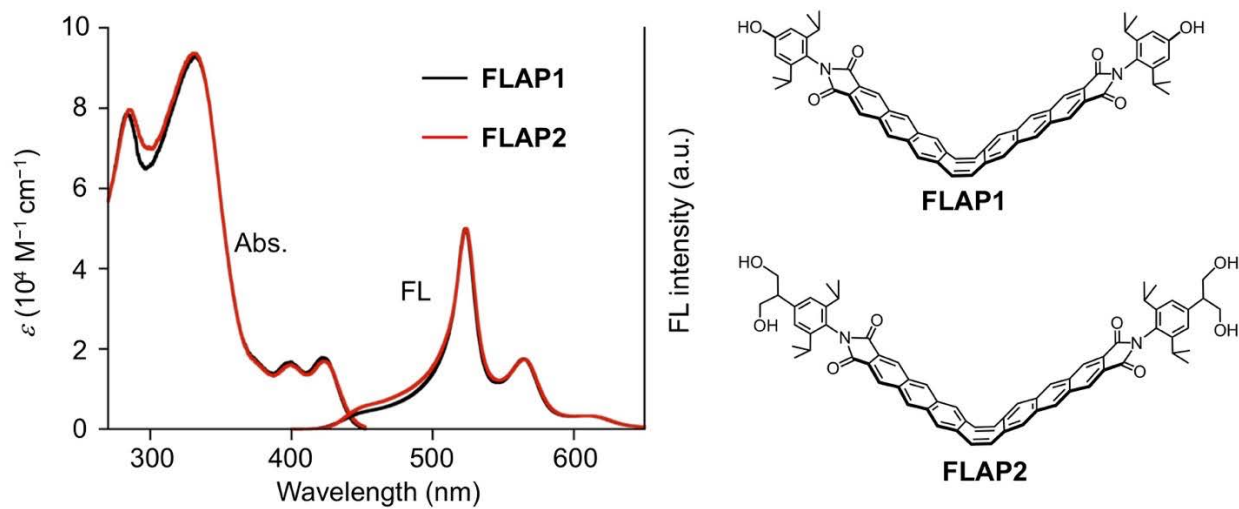

**Supplementary Fig. 1.** UV/visible absorption (Abs) and FL spectra of **FLAP1** and **FLAP2** in $CH_2Cl_2$.

**Supplementary Table 1.** Photophysical constants of **FLAP1** and **FLAP2** in $CH_2Cl_2$.
$\lambda_{ex}$ = 365 nm.

|  | $\Phi_{FL}$ | $\tau_{FL}$ (ns) [a] | $k_r$ (s$^{-1}$) [b] | $k_{nr}$ (s$^{-1}$) [c] |
|---|---|---|---|---|
| **FLAP1** | 0.26 | 11.0 | $2.4 \times 10^7$ | $6.7 \times 10^7$ |
| **FLAP2** | 0.32 | 10.6 | $3.0 \times 10^7$ | $6.4 \times 10^7$ |

[a] FL lifetime monitored at 525 nm. [b] Radiative decay constant. [c] Nonradiative decay constant. $k_r$ and $k_{nr}$ were estimated from the equations below.

$\Phi_{FL} = k_r / (k_r + k_{nr})$
$\tau_{FL} = 1 / (k_r + k_{nr})$

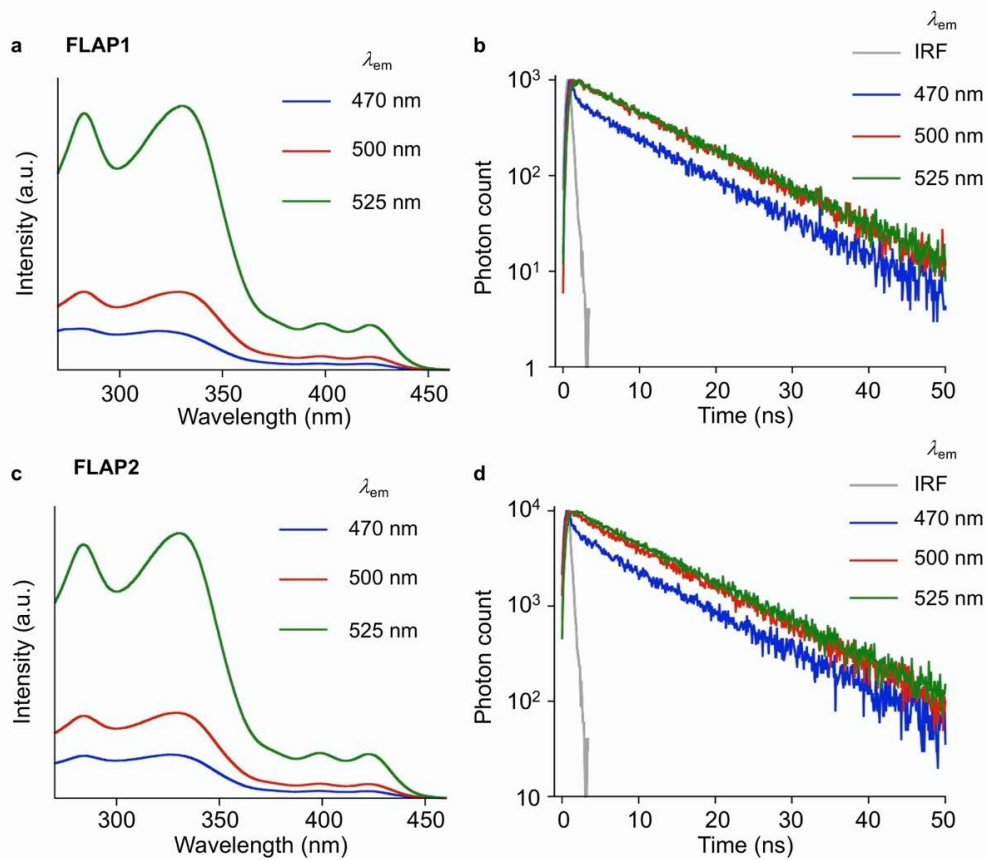

**Supplementary Fig. 2.** (a) Excitation spectra and (b) FL decay profiles of **FLAP1** in CH$_2$Cl$_2$, monitored at different emission wavelengths, and (c, d) those of **FLAP2**. Concentration: *ca.* 10$^{-6}$ M. IRF: The instrument response function.

**Supplementary Table 2.** Photophysical constants of **FLAP1** and **FLAP2** in CH$_2$Cl$_2$. $\lambda_{ex}$ = 365 nm.

|  | $\lambda_{em}$ (nm) | $\chi^2$ | $\tau_{FL}$ (ns)$^a$ | $\tau_1$ (ns) | $\tau_2$ (ns) | $A_1$ | $A_2$ |
|---|---|---|---|---|---|---|---|
| **FLAP1** | 470 | 1.06 | 9.0 | 0.2 | 9.9 | 320 | 66 |
|  | 500 | 1.16 | 10.6 | 0.3 | 10.8 | 88 | 110 |
|  | 525 | 1.03 | 11.0 |  |  |  |  |
| **FLAP2** | 470 | 1.25 | 8.9 | 0.2 | 9.8 | 270 | 66 |
|  | 500 | 1.07 | 10.4 | 0.3 | 10.7 | 94 | 104 |
|  | 525 | 1.05 | 10.6 |  |  |  |  |

$^a$ Mean FL lifetime calculated from the equation below.

$$\tau_{FL} = \frac{\tau_1{}^2 A_1 + \tau_2{}^2 A_2}{\tau_1 A_1 + \tau_2 A_2}$$

where $A_1$ and $A_2$ was determined by analysis of the fitting equation below.

$$G(t) = A_1 \exp\left(-\frac{t}{\tau_1}\right) + A_2 \exp\left(-\frac{t}{\tau_2}\right)$$

where $G(t)$ was photon count as a function of time.

**Computational studies of FLAP molecules**

Density functional theory (DFT) and time-dependent (TD) DFT calculations of the isolated molecules were performed using the Gaussian 16 program.[5] The appropriate density functional (PBE0) was selected because of reproducibility of the absorption spectrum of the FLAP series[6]. All the optimized structures gave no imaginary frequency, bearing $C_1$ symmetry. In the Supplementary Figs. 4–12, compositions (%) and oscillator strength $f$ were calculated based on the respective optimized structure. H and L stand for HOMO and LUMO, respectively. Here and elsewhere, compositions were calculated by taking the square of the dominant configuration and multiplying the result by two, in order to account for excitations involving α and β spin orbitals. In the optimization calculation in $S_1$, two energy minima at bent and planar geometries were found (Supplementary Figs. 5 and 6). Here the oscillator strength of the $S_1$ bent geometry was calculated by the PBE0 functional to be zero (Supplementary Table 4), although a non-zero oscillator strength was obtained by a different functional, CAM-B3LYP (Supplementary Fig. 11 and Supplementary Table 10). This indicates that the calculation results of the $S_1$ electronic state are sensitive to the applied calculation level.

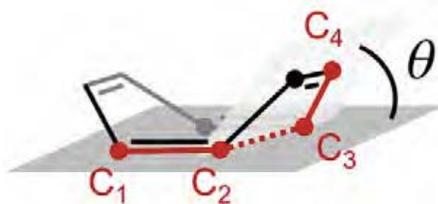

**Supplementary Fig. 3.** Definition of the COT bending angle $\theta$ used in the computational studies. The angle $\theta$ was defined as a dihedral angle between the two planes, the $C_1$–$C_2$–$C_3$ plane and the $C_2$–$C_3$–$C_4$ plane.

**Calculations performed at the PBE0/6-31+G(d) level of theory**

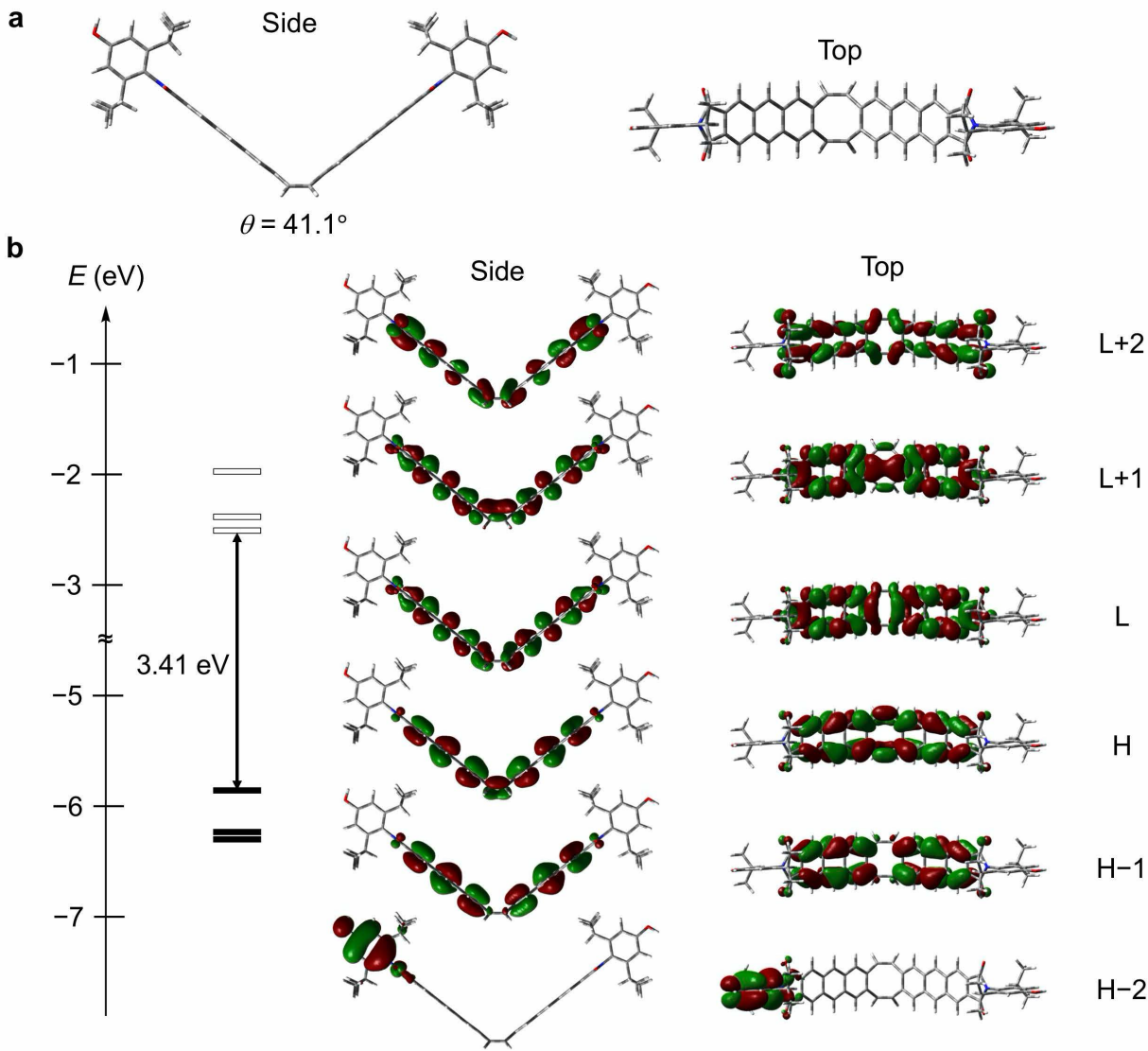

**Supplementary Fig. 4.** (a) $S_0$ optimized geometry of **FLAP1** at the PBE0/6-31+G(d) level and (b) the representative molecular orbitals and their energy levels.

**Supplementary Table 3.** Excitation energies, configurations, and oscillator strengths of the $S_0 \rightarrow S_n$ ($n \leq 3$) transitions for the $S_0$ optimized **FLAP1** at the TD PBE0/6-31+G(d) level.

| Transition | Excitation energy | Configuration(s) | Oscillator strength |
|---|---|---|---|
| $S_0 \rightarrow S_1$ | 2.89 eV (429 nm) | H → L (96%) | 0.0000 |
| $S_0 \rightarrow S_2$ | 2.99 eV (415 nm) | H−1 → L (7%)<br>H → L+1 (93%) | 0.0385 |
| $S_0 \rightarrow S_3$ | 3.19 eV (389 nm) | H−7 → L+1 (4%)<br>H−6 → L (16%)<br>H−1 → L+3 (8%)<br>H → L+2 (68%) | 0.2660 |

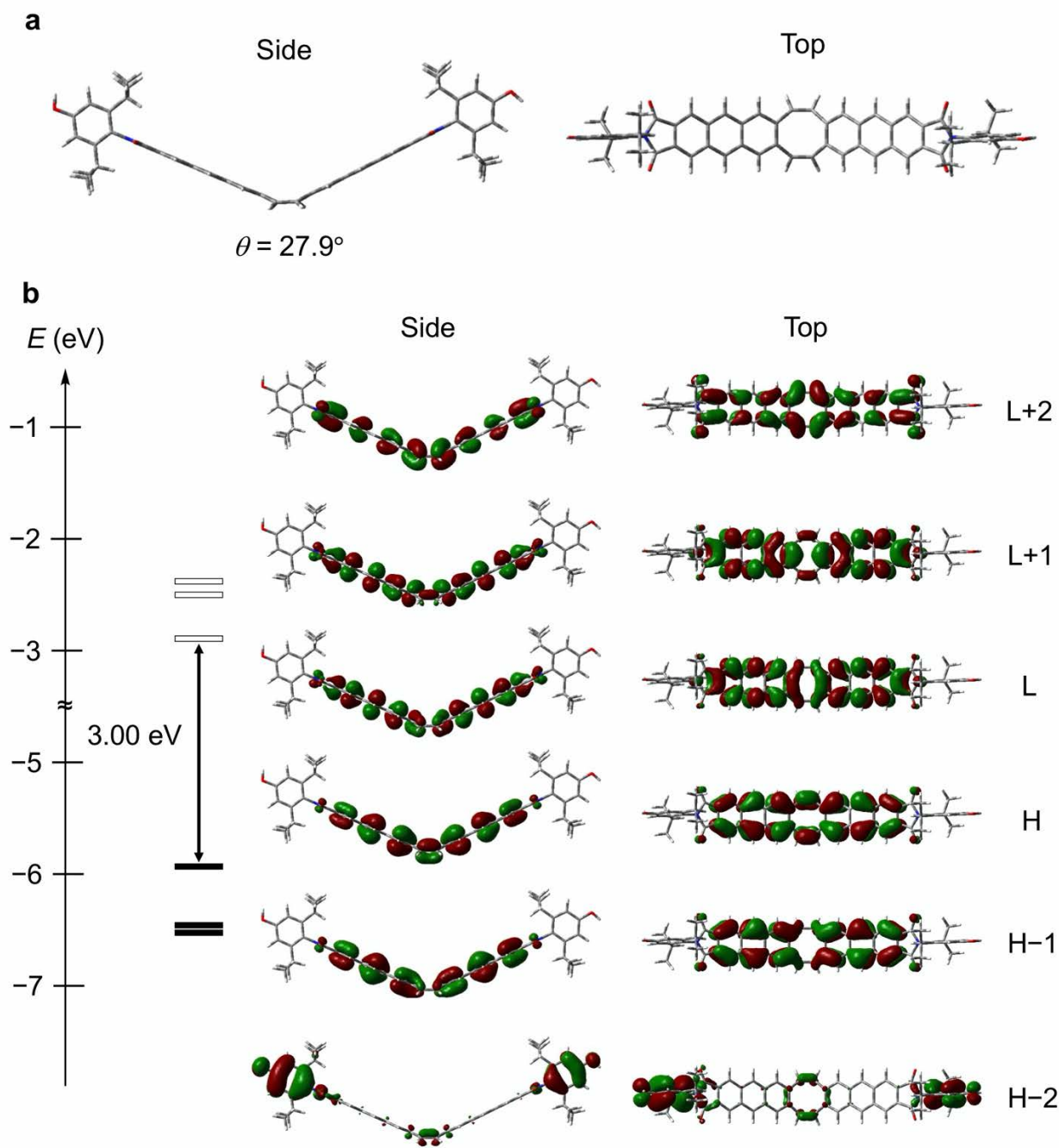

**Supplementary Fig. 5.** (a) $S_1$ optimized geometry of **FLAP1** (bent) at the TD PBE0/6-31G+(d) level and (b) the representative molecular orbitals and their energy levels.

**Supplementary Table 4.** Excitation energy, configuration, and oscillator strength of the $S_0 \rightarrow S_1$ transition for the $S_1$ optimized **FLAP1** (bent) at the TD PBE0/6-31+G(d) level.

| Transition | Excitation energy | Configuration | Oscillator strength |
|---|---|---|---|
| $S_0 \rightarrow S_1$ | 2.52 eV (492 nm) | H → L (98%) | 0.0000 |

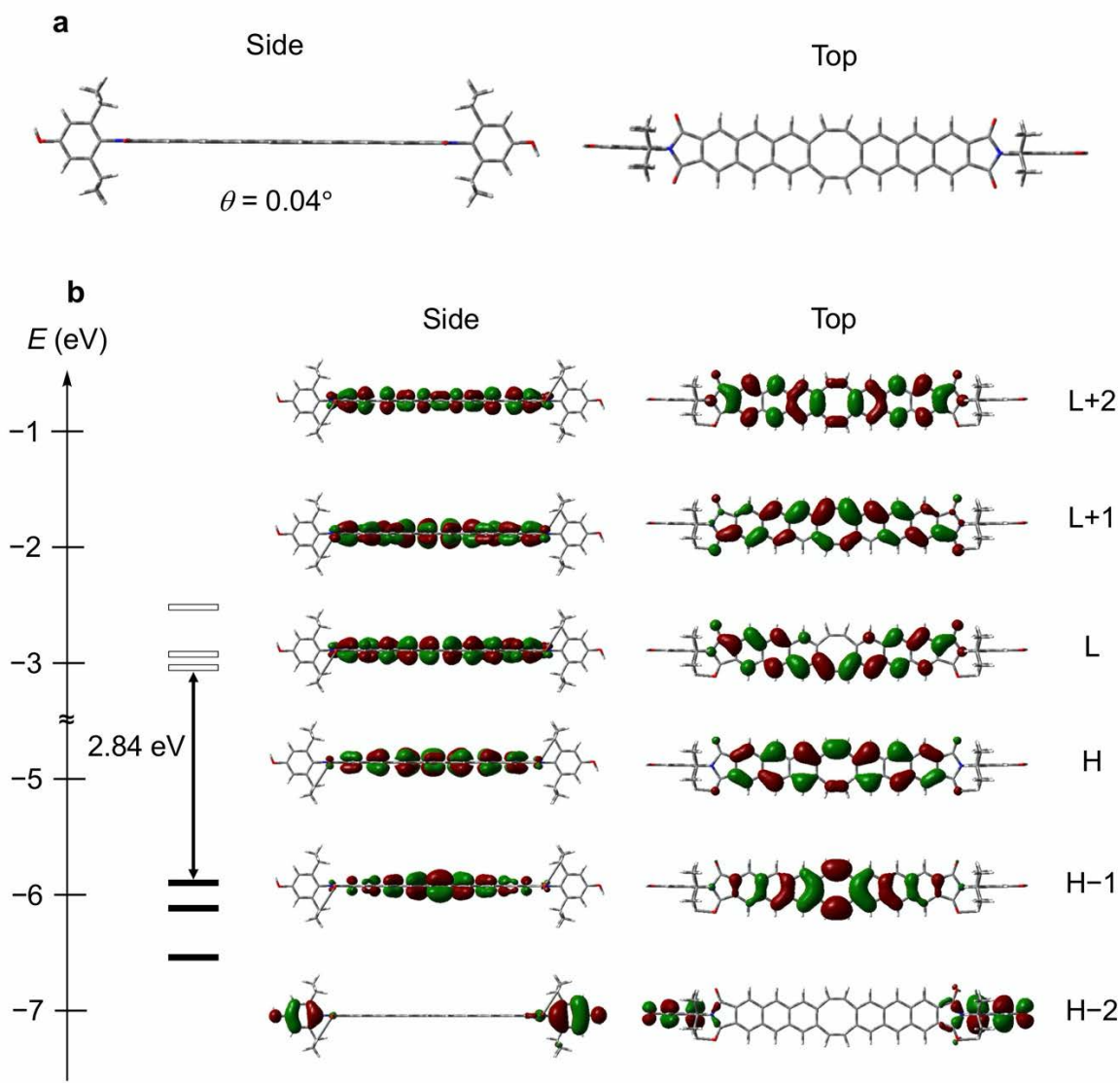

**Supplementary Fig. 6.** (a) $S_1$ optimized geometry of **FLAP1** (planar) at the TD PBE0/6-31G+(d) level and (b) the representative molecular orbitals and their energy levels.

**Supplementary Table 5.** Excitation energy, configurations, and oscillator strength of the $S_0 \rightarrow S_1$ transition for the $S_1$ optimized **FLAP1** (planar) at the TD PBE0/6-31+G(d) level.

| Transition | Excitation energy | Configurations | Oscillator strength |
|---|---|---|---|
| $S_0 \rightarrow S_1$ | 2.21 eV (561 nm) | H−1 → L (35%)<br>H → L (44%)<br>H → L+1 (19%) | 0.0731 |

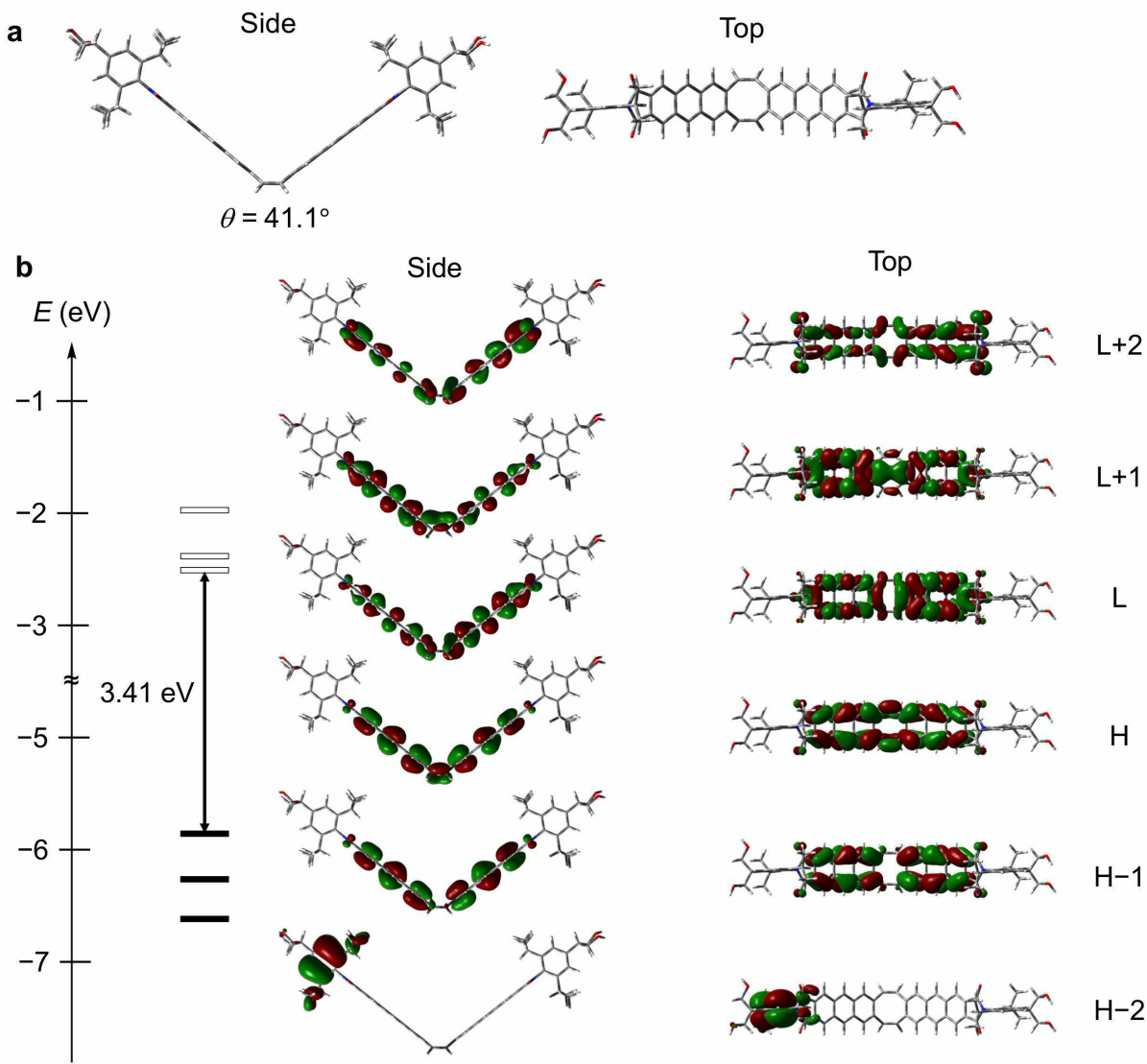

**Supplementary Fig. 7.** (a) $S_0$ optimized geometry of **FLAP2** at the PBE0/6-31+G(d) level and (b) the representative molecular orbitals and their energy levels.

**Supplementary Table 6.** Excitation energies, configurations, and oscillator strengths of the $S_0 \rightarrow S_n$ ($n \leq 3$) transitions for the $S_0$ optimized **FLAP2** at the TD PBE0/6-31+G(d) level.

| Transition | Excitation energy | Configuration(s) | Oscillator strength |
|---|---|---|---|
| $S_0 \rightarrow S_1$ | 2.89 eV (429 nm) | H → L (96%) | 0.0000 |
| $S_0 \rightarrow S_2$ | 2.99 eV (415 nm) | H−1 → L (7%)<br>H → L+1 (92%) | 0.0387 |
| $S_0 \rightarrow S_3$ | 3.19 eV (389 nm) | H−7 → L+1 (4%)<br>H−6 → L (13%)<br>H−1 → L+3 (8%)<br>H → L+2 (69%) | 0.2882 |

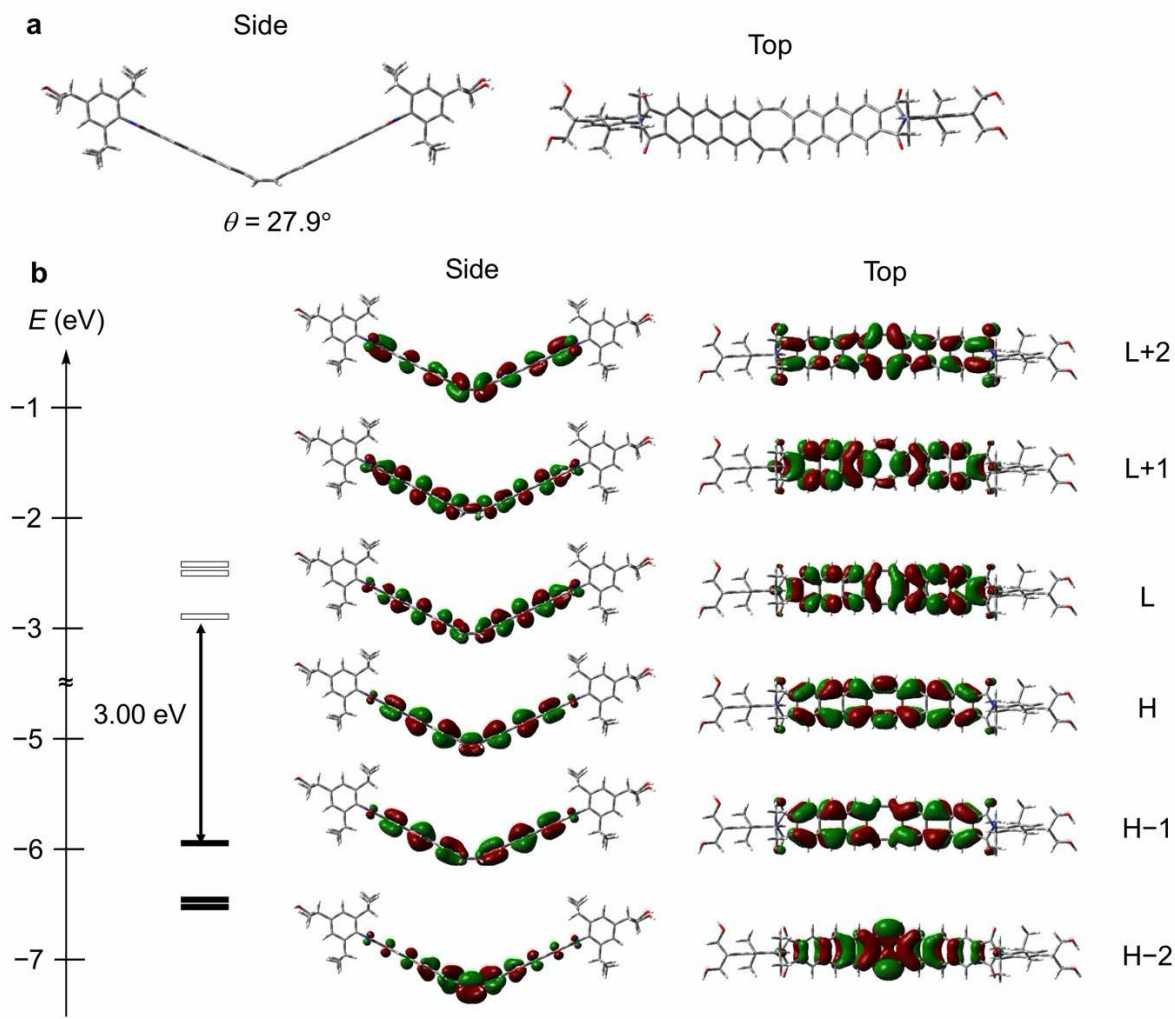

**Supplementary Fig. 8.** (a) $S_1$ optimized geometry of **FLAP2** (bent) at the TD PBE0/6-31G+(d) level and (b) the representative molecular orbitals and their energy levels.

**Supplementary Table 7.** Excitation energy, configuration, and oscillator strength of the $S_0 \rightarrow S_1$ transition for the $S_1$ optimized **FLAP2** (bent) at the TD PBE0/6-31+G(d) level.

| Transition | Excitation energy | Configuration | Oscillator strength |
|---|---|---|---|
| $S_0 \rightarrow S_1$ | 2.52 eV (492 nm) | H → L (98%) | 0.0005 |

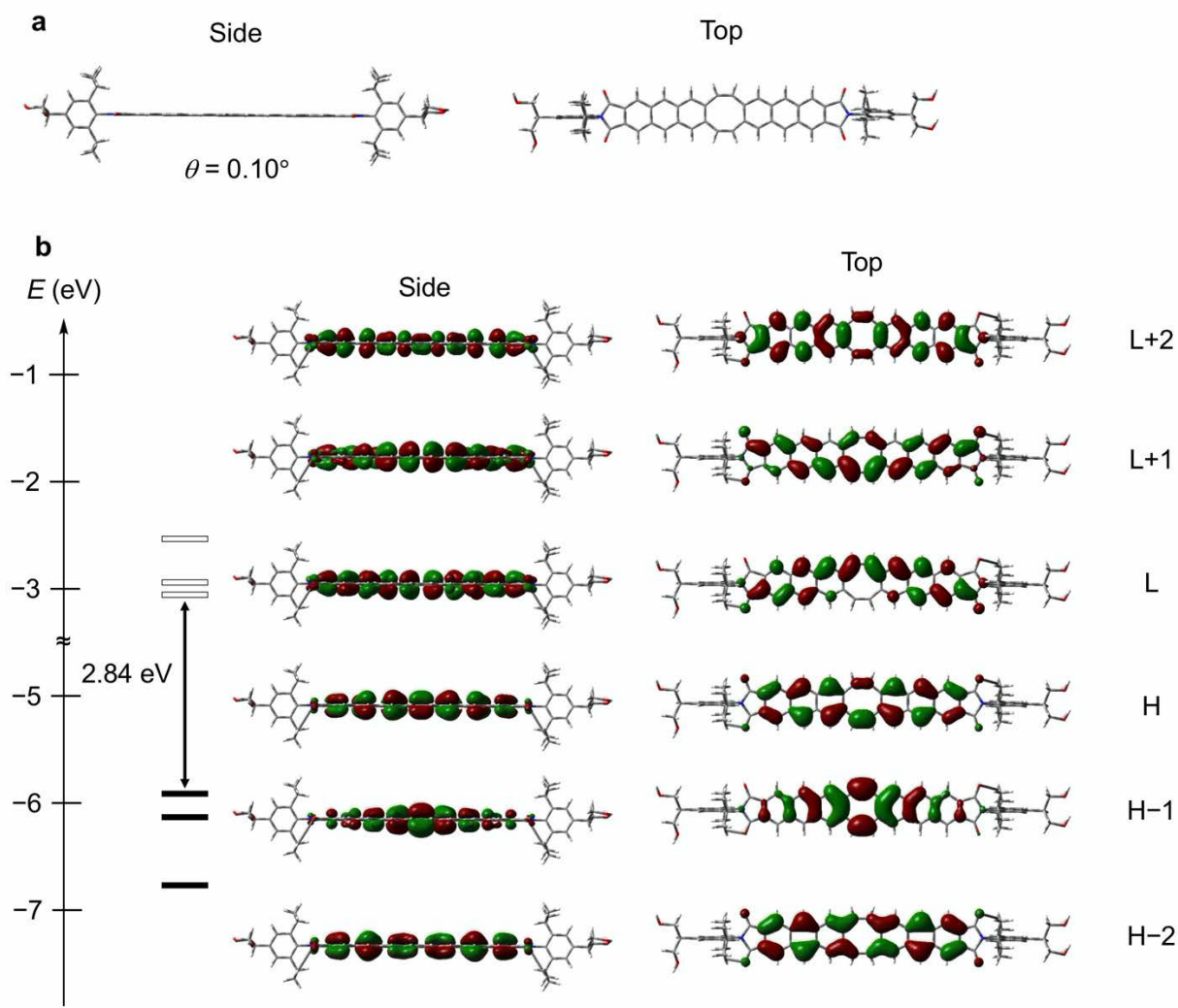

**Supplementary Fig. 9.** (a) $S_1$ optimized planar geometry of **FLAP2** (planar) at the TD PBE0/6-31G+(d) level and (b) representative molecular orbitals and their energy levels.

**Supplementary Table 8.** Excitation energy, configurations, and oscillator strength of the $S_0 \rightarrow S_1$ transition for planar $S_1$ optimized **FLAP2** at the TD PBE0/6-31+G(d) level.

| Transition | Excitation energy | Configurations | Oscillator strength |
|---|---|---|---|
| $S_0 \rightarrow S_1$ | 2.21 eV (561 nm) | H−1 → L (35%)<br>H → L (44%)<br>H → L+1 (19%) | 0.0749 |

**Calculations performed at the CAM-B3LYP/6-31+G(d) level of theory**

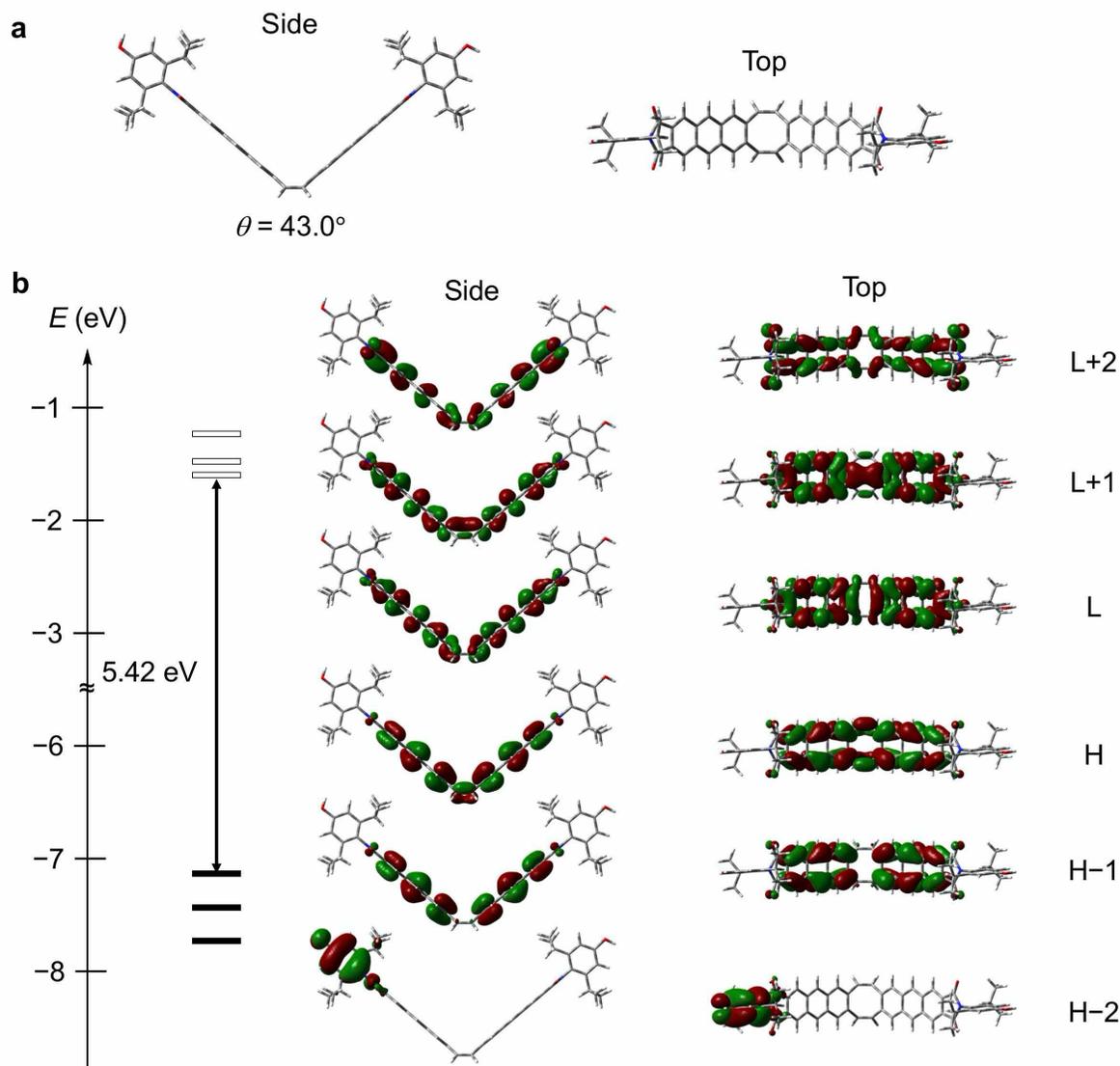

**Supplementary Fig. 10.** (a) $S_0$ optimized geometry of **FLAP1** at the CAM-B3LYP/6-31+G(d) level and (b) the representative molecular orbitals and their energy levels.

**Supplementary Table 9.** Excitation energies, configurations and oscillator strengths of the $S_0 \rightarrow S_n$ ($n \leq 3$) transitions for the $S_0$ optimized **FLAP1** at the TD CAM-B3LYP/6-31+G(d) level.

| Transition | Excitation energy | Configurations | Oscillator strength |
|---|---|---|---|
| $S_0 \rightarrow S_1$ | 3.38 eV (367 nm) | H−1 → L+1 (33%)<br>H → L (63%) | 0.0000 |
| $S_0 \rightarrow S_2$ | 3.40 eV (365 nm) | H−1 → L (40%)<br>H → L+1 (57%) | 0.0847 |
| $S_0 \rightarrow S_3$ | 3.65 eV (340 nm) | H−7 → L+1 (11%)<br>H−5 → L (15%)<br>H−1 → L+3 (22%)<br>H → L+2 (43%) | 0.2185 |

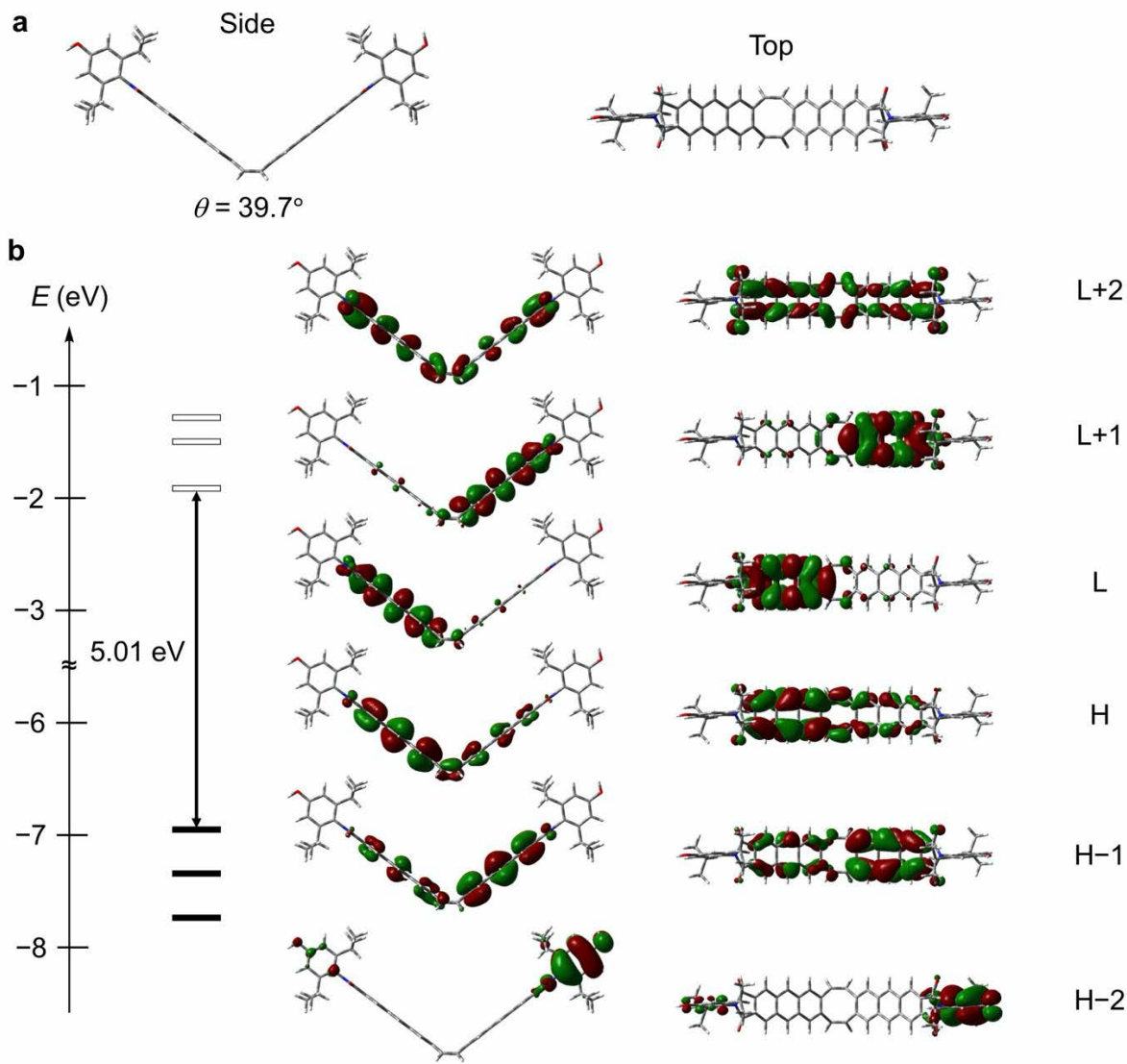

**Supplementary Fig. 11.** (a) $S_1$ optimized geometry of **FLAP1** at the TD CAM-B3LYP/6-31G+(d) level and (b) the representative molecular orbitals and their energy levels.

**Supplementary Table 10.** Excitation energy, configurations, and oscillator strength of the $S_0 \rightarrow S_1$ transition for the $S_1$ optimized **FLAP1** (bent) at the TD CAM-B3LYP/6-31+G(d) level.

| Transition | Excitation energy | Configurations | Oscillator strength |
|---|---|---|---|
| $S_0 \rightarrow S_1$ | 2.90 eV (428 nm) | H−1 → L (8%)<br>H → L (88%) | 0.0423 |

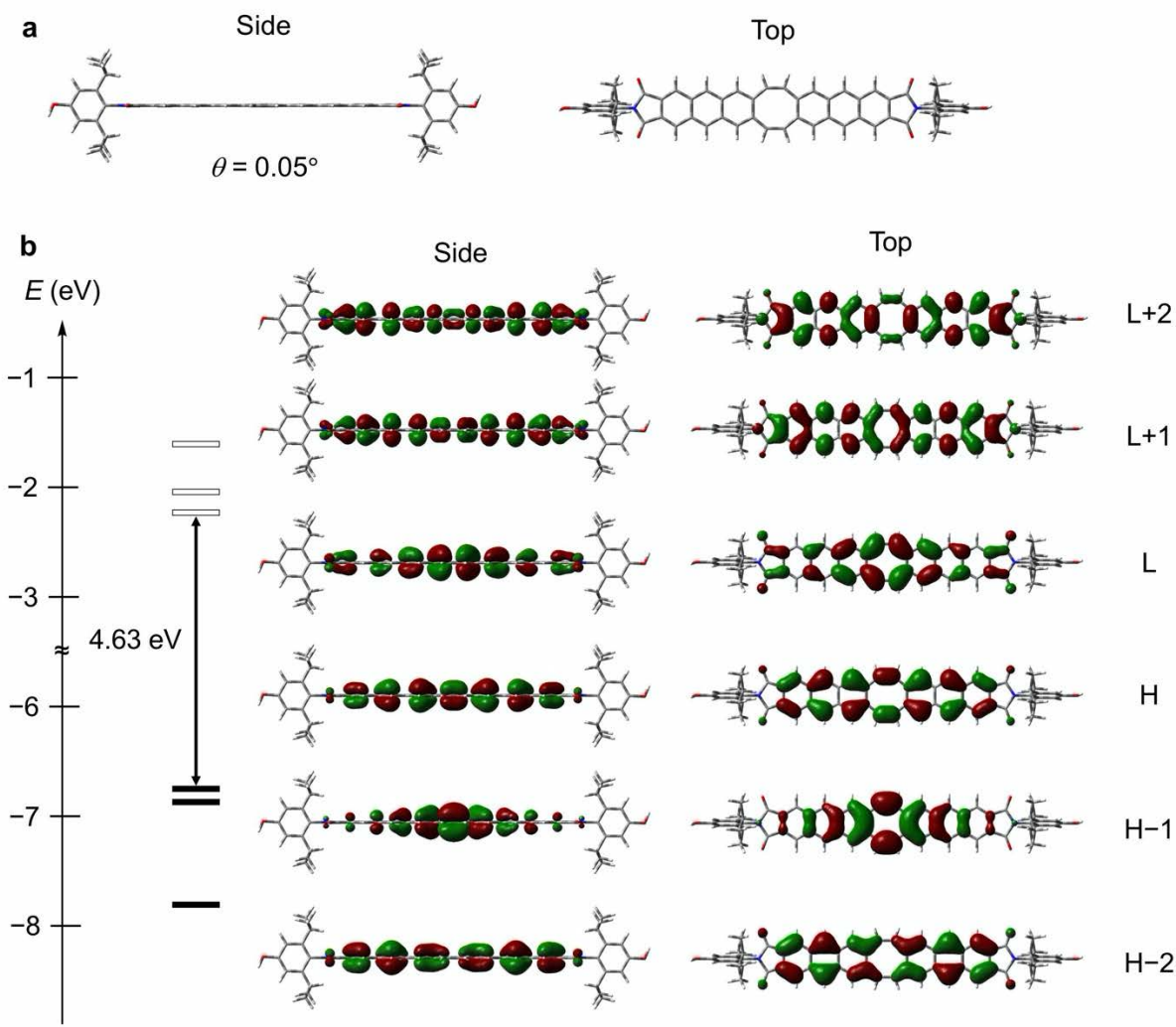

**Supplementary Fig. 12.** (a) $S_1$ optimized geometry of **FLAP1** (planar) at the TD CAM-B3LYP/6-31G+(d) level and (b) the representative molecular orbitals and their energy levels.

**Supplementary Table 11.** Excitation energy, configuration, and oscillator strength of the $S_0 \rightarrow S_1$ transition for the $S_1$ optimized **FLAP1** (planar) at the TD CAM-B3LYP/6-31+G(d) level.

| Transition | Excitation energy | Configuration | Oscillator strength |
|---|---|---|---|
| $S_0 \rightarrow S_1$ | 2.22 eV (558 nm) | H → L (94%) | 0.0000 |

## Computational studies of the polymer-chain substructure in the vicinity of FLAP

The following substructure of a FLAP-doped polyurethane chain, **FLAP1′**, was calculated in Fig. 3 of the main text. The distance between the $C^A$ and $C^B$ atoms was fixed in each optimization by using the *opt=modredundant* keyword of Gaussian16.

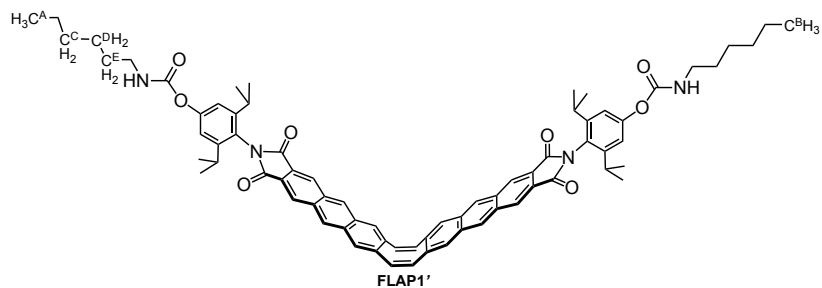

**Supplementary Fig. 13.** Substructure of a FLAP-doped polyurethane chain, **FLAP1′**, in Fig. 3.

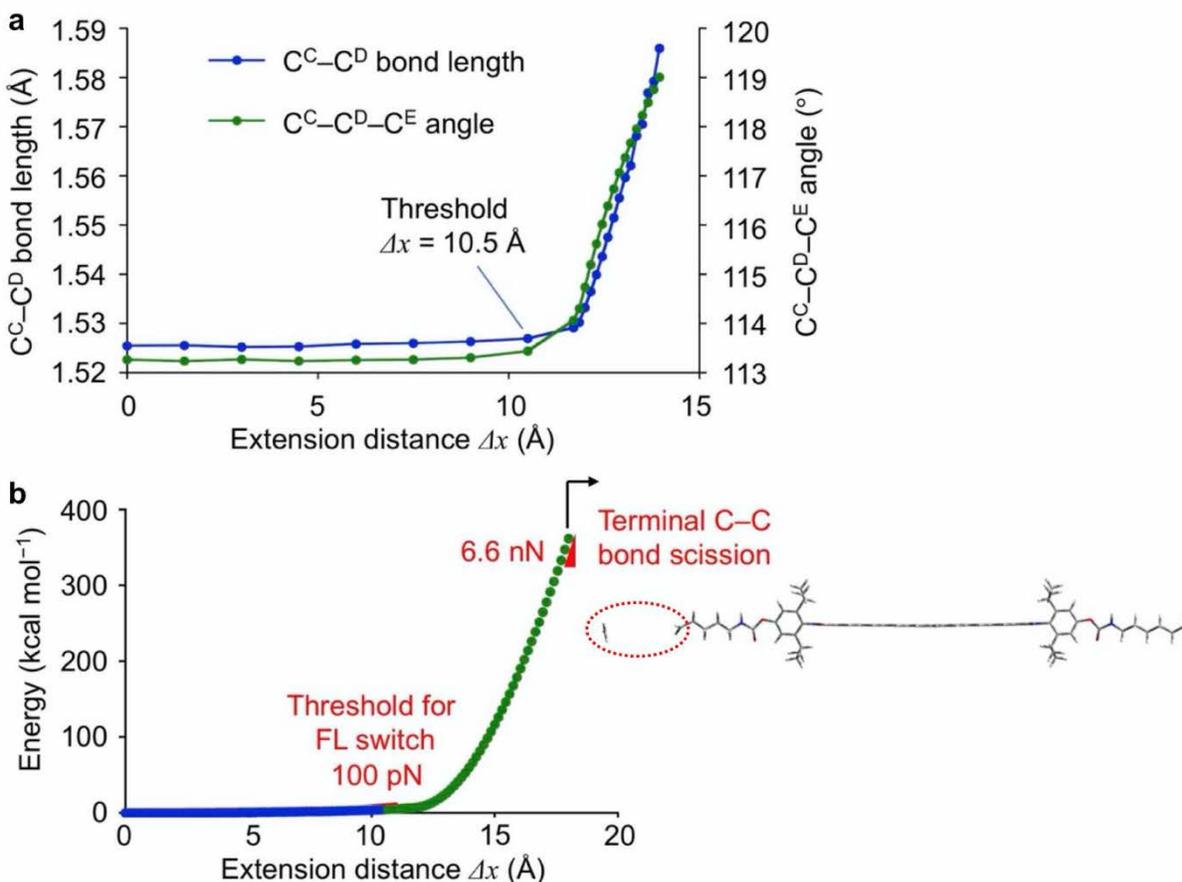

**Supplementary Fig. 14.** (a) DFT calculated changes in the $C^C$–$C^D$ bond length and the $C^C$–$C^D$–$C^E$ bond angle of **FLAP1′**. The bond length and angle are not remarkably increased at the threshold for the FL switch, indicating the FL response of the FLAP force probe before mechanical damage. (b) Calculated energy diagram of **FLAP1′** until the formal C–C bond scission at the PBE0/6-31G(d) level.

**Note:** Experimental evidence for such bond elongation has not been demonstrated in the literature.[7] One of the reviewers of this manuscript gave a comment that the C–C bonds are the stiffest elements of the molecule and they will be least strained by applied force; the C–C bond doesn't need to elongate to experience accelerated dissociation, since (a) C–C bond distance and bond dissociation energy don't generally correlate (except in introductory chemistry textbooks), nor should they based on any known law of physics and (b) acceleration of C–C bond homolysis in strained molecules proceeds not by elongation (or "weakening") of the bond but by stabilization of the TS that is longer than the reactant along the pulling axis.

## Calculated electronic transitions for the absorption

To interpret the absorption spectral change of the polyurethane film upon mechanical testing, TD-DFT calculations of **FLAP1′** were performed for the unstretched ($\Delta x = 0$ Å) and fully stretched ($\Delta x = 13.0$ Å) geometries. Details of the calculation results in Fig. 3E are shown below.

**Supplementary Table 12.** Excitation energies (> 360 nm), configurations, and oscillator strengths of the $S_0 \to S_n$ transitions for the unstretched ($\Delta x = 0$ Å) geometry at the TD PBE0/6-31+G(d) level.

| Transition | Excitation energy | Configuration(s) | Oscillator strength |
| --- | --- | --- | --- |
| $S_0 \to S_1$ | 2.87 eV (431 nm) | H → L (97%) | 0.0000 |
| $S_0 \to S_2$ | 2.99 eV (415 nm) | H−1 → L (7%)<br>H → L+1 (92%) | 0.0377 |
| $S_0 \to S_3$ | 3.16 eV (391 nm) | H−6 → L (16%)<br>H−2 → L (2%)<br>H−1 → L+3 (7%)<br>H → L+2 (69%) | 0.2858 |
| $S_0 \to S_4$ | 3.20 eV (389 nm) | H−1 → L (91%)<br>H → L+1 (7%) | 0.0149 |
| $S_0 \to S_5$ | 3.30 eV (376 nm) | H−6 → L+1 (22%)<br>H−2 → L+1 (4%)<br>H−1 → L+2 (29%)<br>H → L+3 (33%) | 0.0299 |
| $S_0 \to S_6$ | 3.31 eV (375 nm) | H−1 → L+1 (96%) | 0.0000 |

**Supplementary Table 13.** Excitation energies (> 360 nm), configurations, and oscillator strengths of the $S_0 \to S_n$ transitions for the fully stretched ($\Delta x = 13.0$ Å) geometry at the TD PBE0/6-31+G(d) level.

| Transition | Excitation energy | Configurations | Oscillator strength |
| --- | --- | --- | --- |
| $S_0 \to S_1$ | 2.55 eV (486 nm) | H → L (98%) | 0.0000 |
| $S_0 \to S_2$ | 2.58 eV (481 nm) | H−1 → L (38%)<br>H → L+1 (59%) | 0.1196 |
| $S_0 \to S_3$ | 2.89 eV (429 nm) | H−1 → L+1 (95%)<br>H−1 → L+4 (3%) | 0.0000 |
| $S_0 \to S_4$ | 2.92 eV (425 nm) | H → L+2 (96%) | 0.0228 |
| $S_0 \to S_5$ | 3.09 eV (401 nm) | H−2 → L+1 (37%)<br>H−1 → L+2 (49%)<br>H → L+3 (9%) | 0.0000 |
| $S_0 \to S_6$ | 3.14 eV (395 nm) | H−1 → L (59%)<br>H → L+1 (38%) | 5.4348 |
| $S_0 \to S_7$ | 3.19 eV (389 nm) | H−2 → L (95%)<br>H → L+2 (3%) | 0.0185 |
| $S_0 \to S_8$ | 3.39 eV (366 nm) | H−4 → L+2 (10%)<br>H−3 → L (87%) | 0.0000 |
| $S_0 \to S_9$ | 3.39 eV (366 nm) | H−4 → L (87%)<br>H−3 → L+2 (10%) | 0.0236 |

## Synthesis and characterization of the linear polycarbonates (PCs)

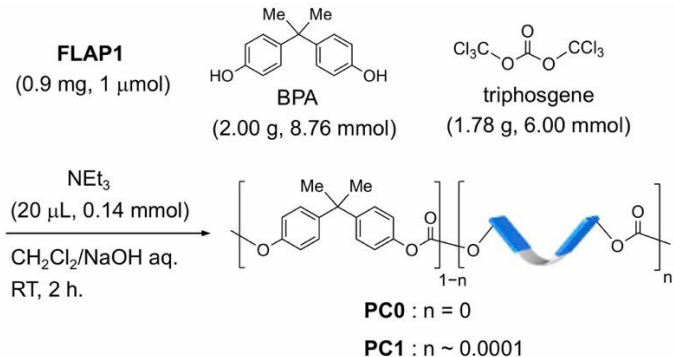

**Supplementary Fig. 15.** Synthesis of the linear polycarbonates (PCs). The amount of **FLAP1** is given for the **PC1** synthesis, while **PC0** does not contain **FLAP1**.

Bisphenol A (BPA) was recrystallized in advance from AcOH, and washed with $H_2O$. To the mixture of bisphenol A (BPA) (2.00 g, 8.76 mmol) and NaOH (2.73 g, 68.2 mmol) in $H_2O$ (20 mL), **FLAP1** (0.9 mg, 1 μmol) and triphosgene (2.60 g, 8.76 mmol) in $CH_2Cl_2$ (20 mL) were added. After addition of triethylamine (TEA) (20 μL, 0.14 mmol), the reaction mixture was stirred vigorously at 25 °C for 2 h. Then, the aqueous layer was removed by decantation and the remaining organic layer was poured into MeOH (100 mL). The precipitation was washed with MeOH (100 mL) followed by $H_2O$ (100 mL), and dried under vacuum to give a lump of **PC1** ($\approx$2.3 g). To obtain **PC0**, the same procedure was performed without **FLAP1**. The resulting linear PCs were soluble in common organic solvents, and therefore shape of the PC samples can be easily designed. By punching a PC sheet, a dumbbell-shaped specimen, JIS K 6251 (No. 3) meeting the requirements of ISO 37, was obtained.

**PC0:** $^1$H NMR (600 MHz, $CDCl_3$) $\delta$ (ppm) 7.25–7.23 (m, 4H), 7.17–7.15 (m, 4H), and 1.67 (s, 6H).

**PC1:** $^1$H NMR (600 MHz, $CDCl_3$) $\delta$ (ppm) 7.25–7.24 (m, 4H), 7.17–7.15 (m, 4H), and 1.68 (s, 6H). Note that peaks arising from the FLAP dopant were too small to detect.

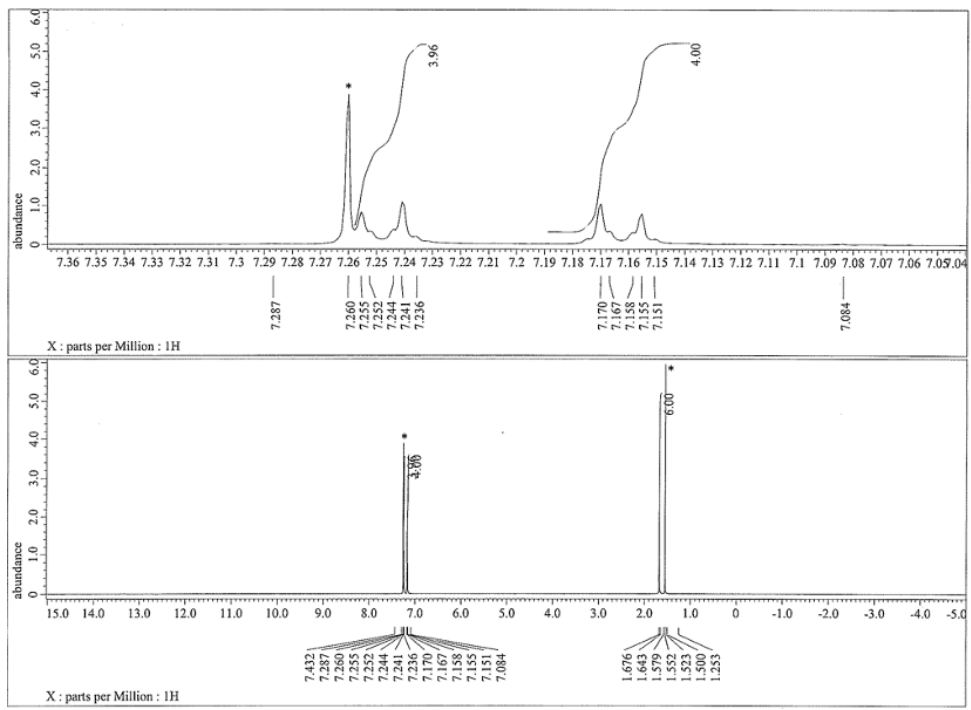

**Supplementary Fig. 16.** $^1$H NMR spectrum of **PC0** in CDCl$_3$ at 25 °C. Peaks marked with * indicate residual solvents.

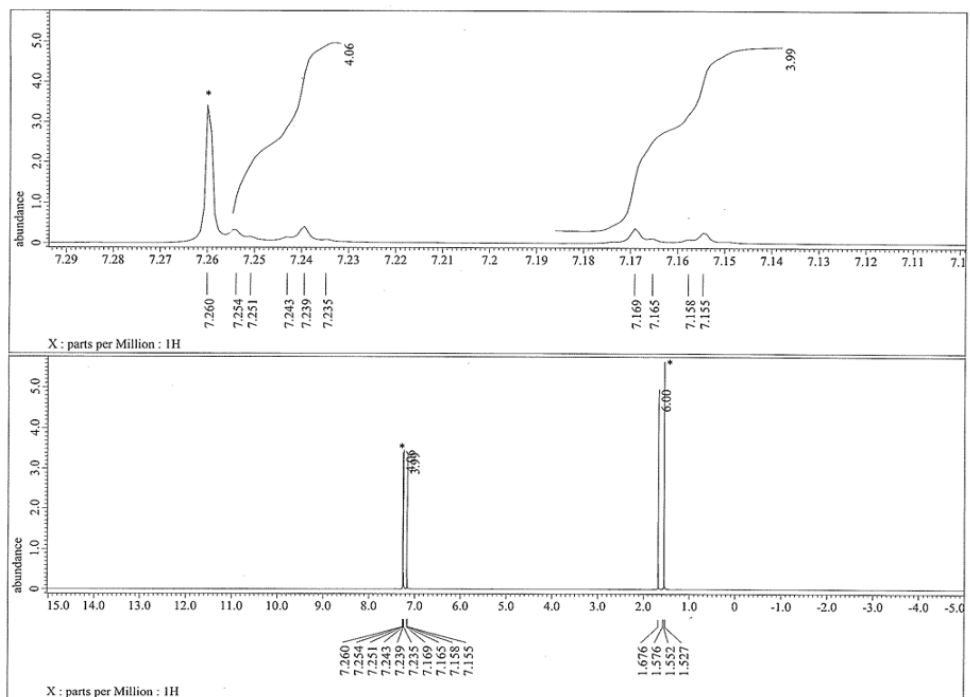

**Supplementary Fig. 17.** $^1$H NMR spectrum of **PC1** in CDCl$_3$ at 25 °C. Peaks marked with * indicate residual solvents.

## DSC analysis of the linear polycarbonates (PCs)

Differential scanning calorimetry (DSC) measurement of the linear polycarbonates (PCs) was conducted under $N_2$ atmosphere at a flow speed of 30 mL min$^{-1}$. Programmed heating and cooling cycles are shown in Fig S18. Fig S19 shows DSC profiles of the 2$^{nd}$ cooling (**E→F**) and heating (**G→H**) cycles. Each sample was placed into an aluminum pan and covered with an aluminum cover, which was well pressed with a designated pressing tool. As a refence sample, a vacant pan was also pressed with the cover.

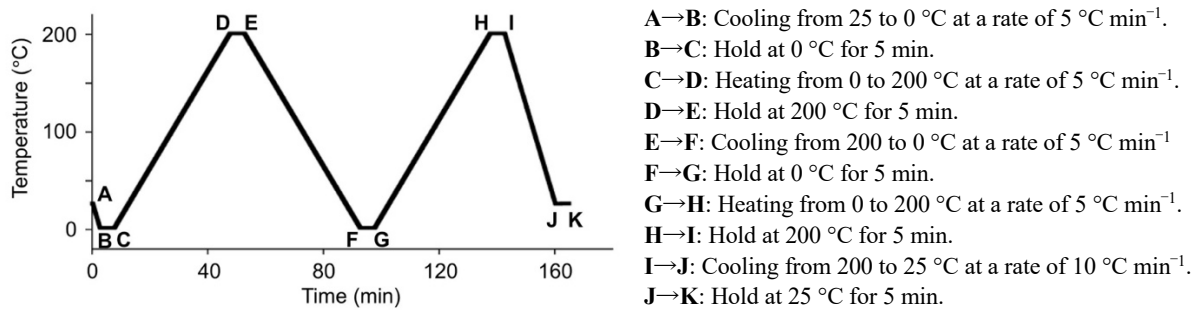

**A→B**: Cooling from 25 to 0 °C at a rate of 5 °C min$^{-1}$.
**B→C**: Hold at 0 °C for 5 min.
**C→D**: Heating from 0 to 200 °C at a rate of 5 °C min$^{-1}$.
**D→E**: Hold at 200 °C for 5 min.
**E→F**: Cooling from 200 to 0 °C at a rate of 5 °C min$^{-1}$
**F→G**: Hold at 0 °C for 5 min.
**G→H**: Heating from 0 to 200 °C at a rate of 5 °C min$^{-1}$.
**H→I**: Hold at 200 °C for 5 min.
**I→J**: Cooling from 200 to 25 °C at a rate of 10 °C min$^{-1}$.
**J→K**: Hold at 25 °C for 5 min.

**Supplementary Fig. 18.** Programmed heating and cooling cycles in the DSC measurement of the PCs.

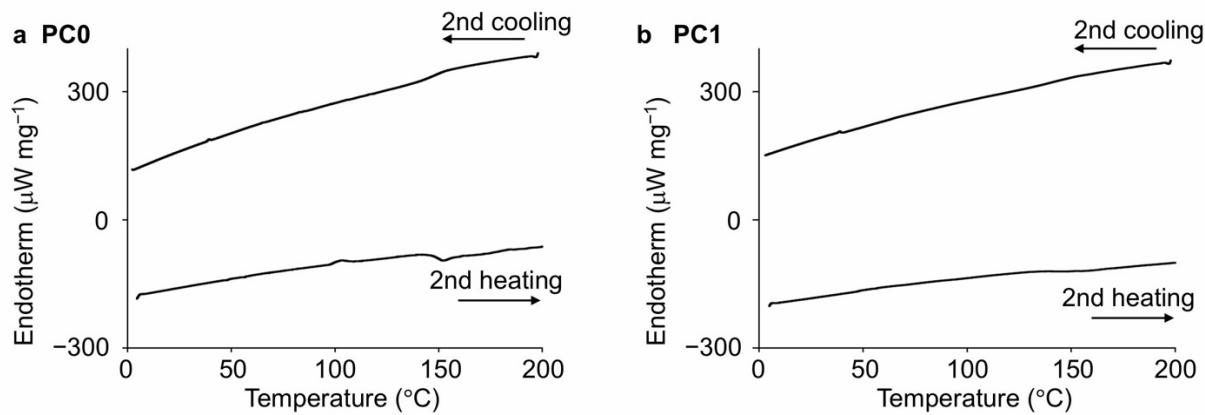

**Supplementary Fig. 19.** DSC traces of **PC0** and **PC1**. Glass transition temperatures ($T_g$) were estimated to be 151 °C for **PC0** and 153 °C for **PC1**.

## Mechanical properties of the linear polycarbonates (PCs)

Uniaxial tensile tests were carried out on dumbbell-shaped specimens. True strain ($\varepsilon_{true}$) is calculated by $\varepsilon_{true} = \ln(L/L_0)$, where $L$ means crosshead displacement and the initial length $L_0 = 30$ mm. True stress ($\sigma_{true}$) is calculated by $\sigma_{true} = F/A$, where $F$ is the recorded force and $A$ is the cross-sectional area of the deformed specimen, approximated by $A = A_0(L_0/L)$. Toughness was calculated as the integrated area under the true stress–strain curve. Young's modulus was defined as the slope of stress–strain curves at 0–1% true strain. Stretching rate was fixed at 5 mm min$^{-1}$, corresponding to the strain rate of $2.8 \times 10^{-3}$ s$^{-1}$.

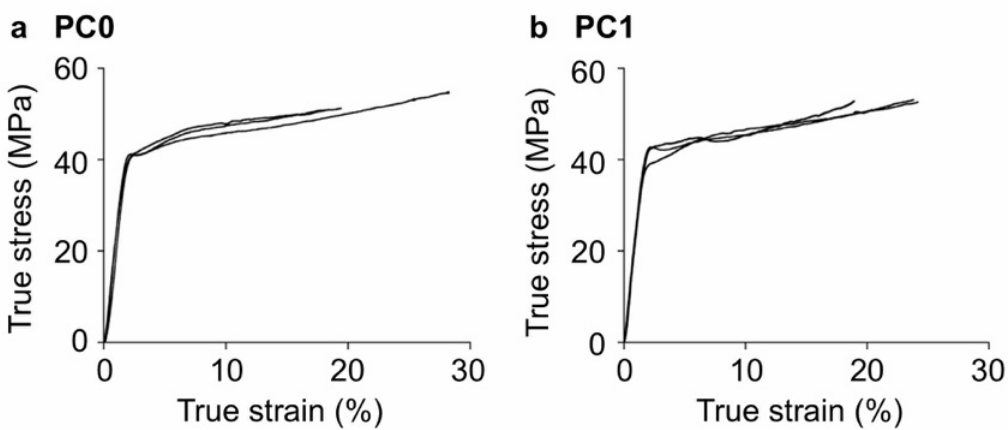

**Supplementary Fig. 20.** Stress-strain curves of **PC0** and **PC1**.

**Supplementary Table 14.** Mechanical properties for the PCs.

|  | True rupture strain (%) | True rupture stress (MPa) | Toughness (MJ m$^{-3}$) | Young's modulus (GPa) |
|---|---|---|---|---|
| **PC0** | 22.0 ± 5.4 | 52.3 ± 2.1 | 10.3 ± 2.3 | 2.20 ± 0.28 |
| **PC1** | 22.3 ± 2.9 | 52.9 ± 0.3 | 10.0 ± 1.4 | 2.49 ± 0.10 |

## Photophysical properties of the unstretched polycarbonate (PC)

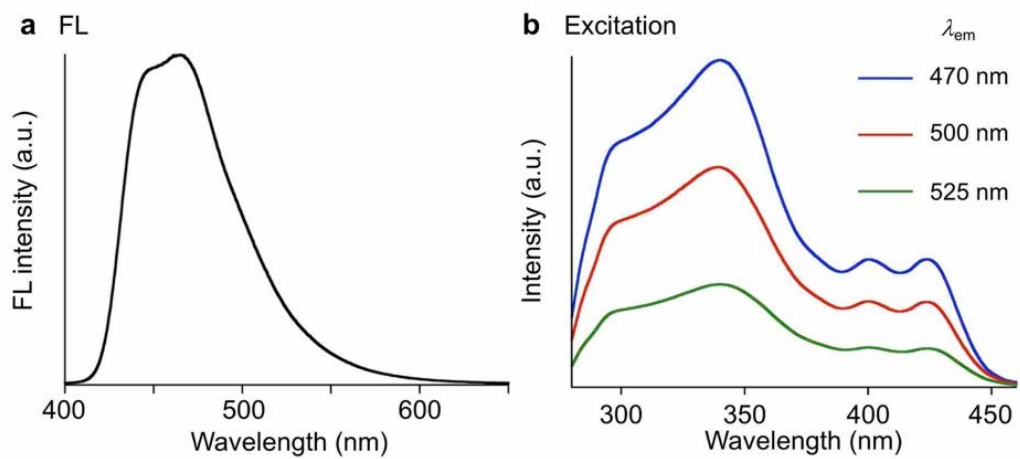

**Supplementary Fig. 21.** (a) FL spectrum ($\lambda_{ex}$ = 365 nm) and (b) excitation spectra of the **PC1** film. No excimer emission was observed.

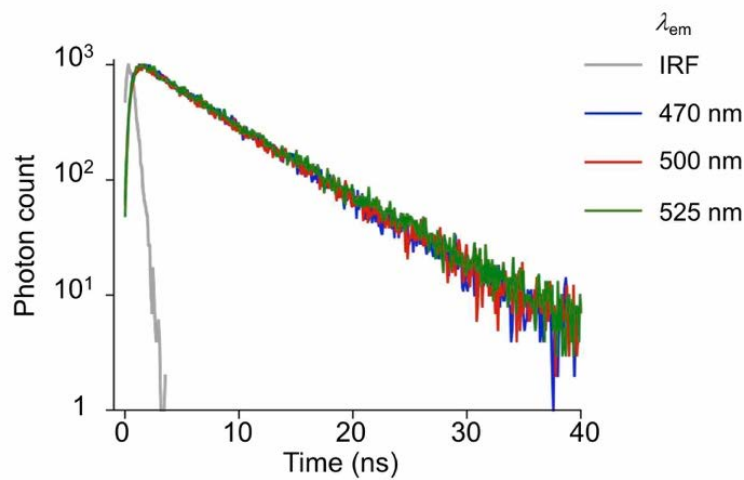

**Supplementary Fig. 22.** FL decay profiles of **PC1** ($\lambda_{ex}$ = 365 nm), in which the FL lifetime is almost constant ($\tau_{FL}$ = 6.8–7.0 ns) regardless of the monitored FL wavelengths.

**Supplementary Table 15.** Photophysical constants of **PC1**. $\lambda_{ex}$ = 365 nm.

| $\Phi_{FL}$ | $\tau_{FL}$ (ns)[a] | $k_r$ (s$^{-1}$) | $k_{nr}$ (s$^{-1}$) |
|---|---|---|---|
| 0.30 | 6.8 | $4.4 \times 10^7$ | $1.0 \times 10^8$ |

[a] $\lambda_{em}$ = 470 nm.

## Synthesis of the crosslinked polyurethanes (PUs)

Poly(tetrahydrofuran) (PTHF, $M_n \sim 650$, Aldrich) was dried under vacuum (~ 2 torr) for 2 h at 70 °C. After cooling to 25 °C, dimethylformamide (DMF, Super dehydrated grade, Wako Chemicals), a trace of FLAP dopant (**FLAP1** or **FLAP2**), hexamethylene diisocyanate (HDI, TCI chemicals) and triethanolamine (TEA, Wako Chemicals) were added, and the reaction mixture was stirred at 25 °C for 10 min. Then, dibutyltin dilaurate (DBTDL, Aldrich) (30 μL, 50 μmol) in THF (0.5 mL) was added to the reaction and stirred at 25 °C for 2 min. The reaction mixture was poured into a custom-made mold of PTFE (polytetrafluoroethylene), and polymerized under $N_2$ atmosphere at 25 °C for 48 h. Then, dumbbell-shaped polyurethane (PU) specimens were washed with $H_2O$ and dried under vacuum for 8 h at 70 °C to obtain transparent PU films, **PU1** or **PU2**, with thickness of 1.5 mm and width of 6 mm. **PU0** was obtained by the same procedure without the FLAP dopant.

The average distance between the FLAP molecules doped in PU was estimated to be 20 nm as follows; the weight ratio of FLAP to the polymer was 0.2 mg g$^{-1}$ equal to 0.2 μmol m$^{-3}$ (polymer density: 1.0 g cm$^{-3}$). Therefore, the average volume occupied by a single FLAP molecule was calculated to be $8.0 \times 10^{-24}$ m$^3$. Taking the cube root, the average distance between the FLAP molecules was determined to be $2.0 \times 10^{-8}$ m$^3$ (20 nm).

**Supplementary Table 16.** Components and solvent in the synthesis of **PU0**.

|  | PU0 (8.7%TEA) | PU0 (10%TEA) | PU0 (13%TEA) |
|---|---|---|---|
| PTHF | 5.22 g (8.03 mmol) | 4.80 g (7.38 mmol) | 4.99 g (7.68 mmol) |
| HDI | 1.73 mL (10.8 mmol) | 1.65 mL (10.3 mmol) | 1.94 mL (12.1 mmol) |
| TEA | 268 mg (1.80 mmol) | 292 mg (1.96 mmol) | 439 mg (2.94 mmol) |
| DMF | 14.9 mL | 13.9 mL | 14.5 mL |

**Supplementary Table 17.** Components and solvent in the synthesis of **PU1**.

|  | PU1 (8.7%TEA) | PU1 (10%TEA) | PU1 (13%TEA) |
|---|---|---|---|
| **FLAP1** | 1.5 mg (1.7 μmol) | 1.3 mg (1.5 μmol) | 1.6 mg (1.8 μmol) |
| PTHF | 5.14 g (7.91 mmol) | 4.67 g (7.18 mmol) | 5.03 g (7.74 mmol) |
| HDI | 1.69 mL (10.5 mmol) | 1.61 mL (10.0 mmol) | 1.96 mL (12.2 mmol) |
| TEA | 262 mg (1.75 mmol) | 285 mg (1.91 mmol) | 443 mg (2.97 mmol) |
| DMF | 14.7 mL | 13.5 mL | 14.5 mL |
| FLAP/polymer[a] | 0.21 mg g$^{-1}$ | 0.20 mg g$^{-1}$ | 0.21 mg g$^{-1}$ |

[a] Weight ratio of FLAP to the polymer.

**Supplementary Table 18.** Components and solvent in the synthesis of **PU2**.

|  | PU2 (8.7%TEA) | PU2 (10%TEA) | PU2 (13%TEA) |
|---|---|---|---|
| **FLAP2** | 1.5 mg (1.5 μmol) | 1.5 mg (1.5 μmol) | 1.5 mg (1.5 μmol) |
| PTHF | 5.21 g (8.01 mmol) | 5.49 g (8.45 mmol) | 4.92 g (7.57 mmol) |
| HDI | 1.71 mL (10.6 mmol) | 1.89 mL (11.8 mmol) | 1.91 mL (11.9 mmol) |
| TEA | 266 mg (1.78 mmol) | 335 mg (2.24 mmol) | 430 mg (2.88 mmol) |
| DMF | 14.9 mL | 15.9 mL | 14.2 mL |
| FLAP/polymer[a] | 0.21 mg g$^{-1}$ | 0.19 mg g$^{-1}$ | 0.20 mg g$^{-1}$ |

[a] Weight ratio of FLAP to the polymer.

## DSC analysis of the crosslinked polyurethanes (PUs)

Differential scanning calorimetry (DSC) measurement of the crosslinked polyurethane (PU) specimen was conducted under $N_2$ atmosphere at a flow speed of 30 mL min$^{-1}$. Programmed heating and cooling cycles are shown in Fig S23. Fig S25 shows DSC profiles of the 2nd cooling (**E→F**) and heating (**G→H**) cycles. Each PU sample was placed into an aluminum pan, covered with an aluminum cover, and then pressed with a designated tool. As a refence sample, a vacant pan was also pressed with the cover.

Spontaneous crystallization behavior was more pronounced in the PU samples with a smaller crosslinking density, as suggested in the values of $\Delta H_{r \to c}$ and $\Delta H_{c \to r}$. Cross-Nicol images in Fig. S24 also suggest that the semicrystalline sample is transformed into the rubbery one upon heating.

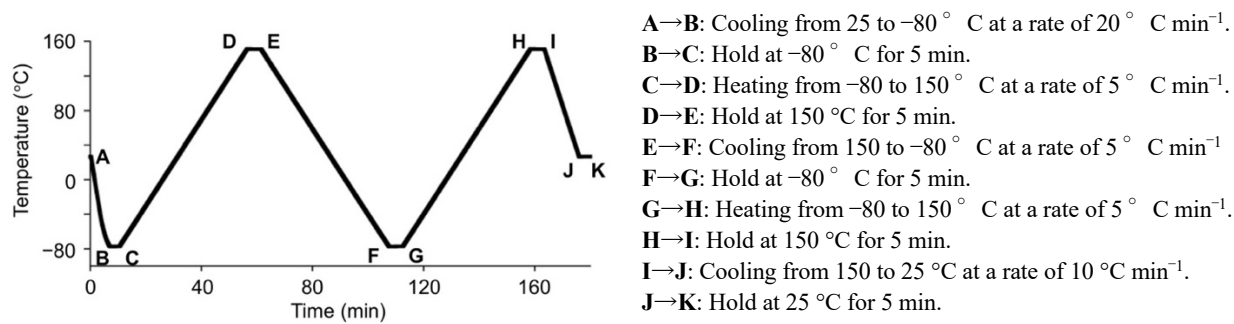

**A→B**: Cooling from 25 to −80 °C at a rate of 20 °C min$^{-1}$.
**B→C**: Hold at −80 °C for 5 min.
**C→D**: Heating from −80 to 150 °C at a rate of 5 °C min$^{-1}$.
**D→E**: Hold at 150 °C for 5 min.
**E→F**: Cooling from 150 to −80 °C at a rate of 5 °C min$^{-1}$
**F→G**: Hold at −80 °C for 5 min.
**G→H**: Heating from −80 to 150 °C at a rate of 5 °C min$^{-1}$.
**H→I**: Hold at 150 °C for 5 min.
**I→J**: Cooling from 150 to 25 °C at a rate of 10 °C min$^{-1}$.
**J→K**: Hold at 25 °C for 5 min.

**Supplementary Fig. 23.** Programmed heating and cooling cycles in the DSC measurement of the PUs.

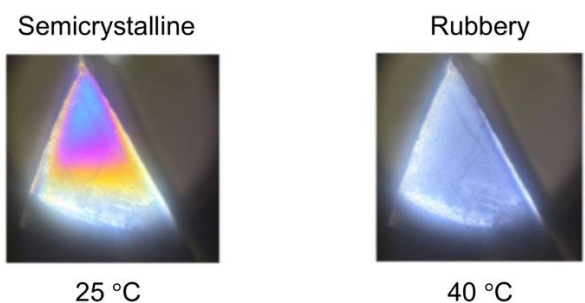

**Supplementary Fig. 24.** Cross-Nicol images of **PU0** (8.7%TEA) at 20 and 40 °C.

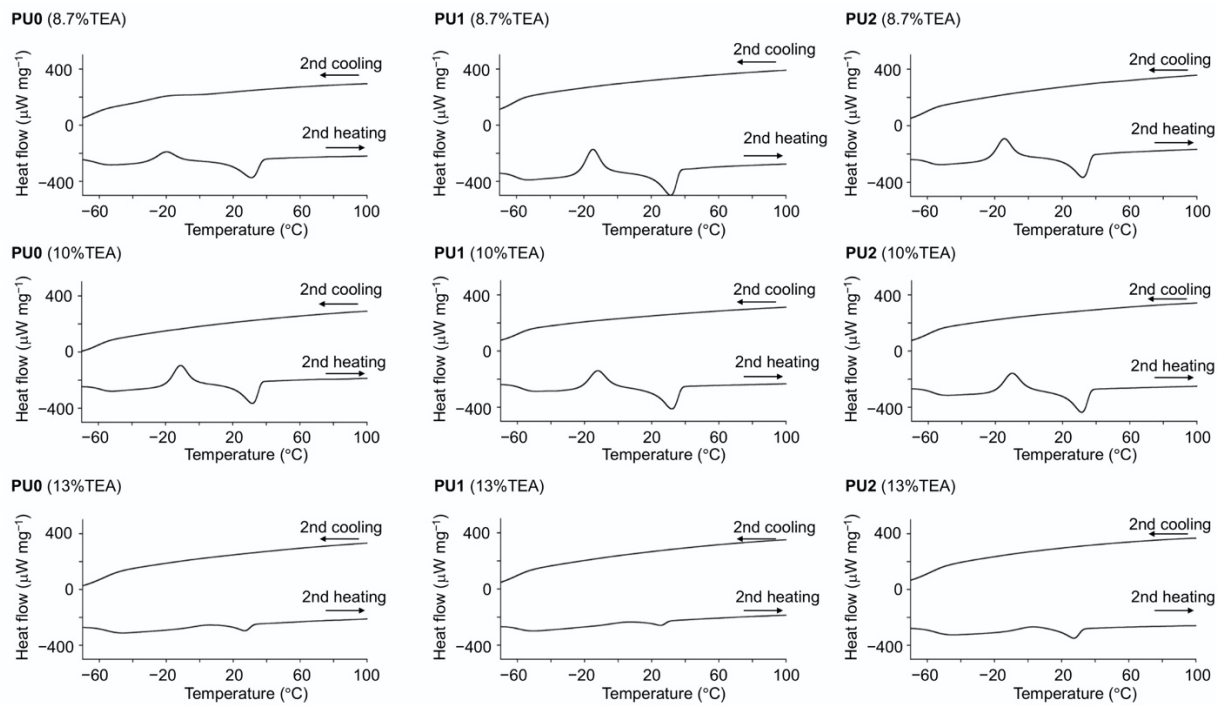

**Supplementary Fig. 25.** DSC traces of the crosslinked PUs.

**Supplementary Table 19.** DSC parameters of the PU films with each crosslinking density (%TEA).

| | weight (mg) | $T_g$ [a] (°C) | $T_{r \to c}$ [b] (°C) | $\Delta H_{r \to c}$ [c] (J g$^{-1}$) | $T_{c \to r}$ [d] (°C) | $\Delta H_{c \to r}$ [e] (J g$^{-1}$) |
|---|---|---|---|---|---|---|
| **PU0 (8.7%TEA)** | 7.50 | −60 | −19 | −16 | 31 | 21 |
| **PU1 (8.7%TEA)** | 6.42 | −60 | −14 | −24 | 31 | 24 |
| **PU2 (8.7%TEA)** | 6.25 | −58 | −14 | −24 | 32 | 22 |
| **PU0 (10%TEA)** | 6.84 | −58 | −11 | −23 | 32 | 20 |
| **PU1 (10%TEA)** | 9.99 | −56 | −12 | −24 | 32 | 22 |
| **PU2 (10%TEA)** | 7.33 | −57 | −10 | −26 | 31 | 19 |
| **PU0 (13%TEA)** | 6.67 | −56 | 5 | −3.7 | 27 | 5.7 |
| **PU1 (13%TEA)** | 6.01 | −59 | 8 | −2.8 | 25 | 4.0 |
| **PU2 (13%TEA)** | 5.88 | −56 | 3 | −3.9 | 27 | 6.5 |

[a] Glass transition temperature determined by the 2nd heating profile.
[b] Rubbery-to-semicrystalline transition temperature determined by the 2nd heating profile.
[c] Enthalpy change in the rubbery-to-semicrystalline transition determined by the DSC signal area.
[d] Semicrystalline-to-rubbery transition temperature determined by the 2nd heating profile.
[e] Enthalpy change in the semicrystalline-to-rubbery transition determined by the DSC signal area.

## Rheological analysis of the crosslinked polyurethanes (PUs)

Dynamic viscoelasticity was measured for the rubbery PUs. After heating to 100 °C, the storage modulus $G'$ and loss modulus $G''$ were measured up to 0 °C at a cooling rate of 5 °C min$^{-1}$ (frequency: 1 Hz, a 10-mm parallel plate). As the temperature dropped, both $G'$ and $G''$ increased gradually. The increments were small, and the $G'$ value stayed in the range of $10^5$–$10^6$ Pa, indicating much softer properties of the crosslinked PUs than the linear PCs prepared above.

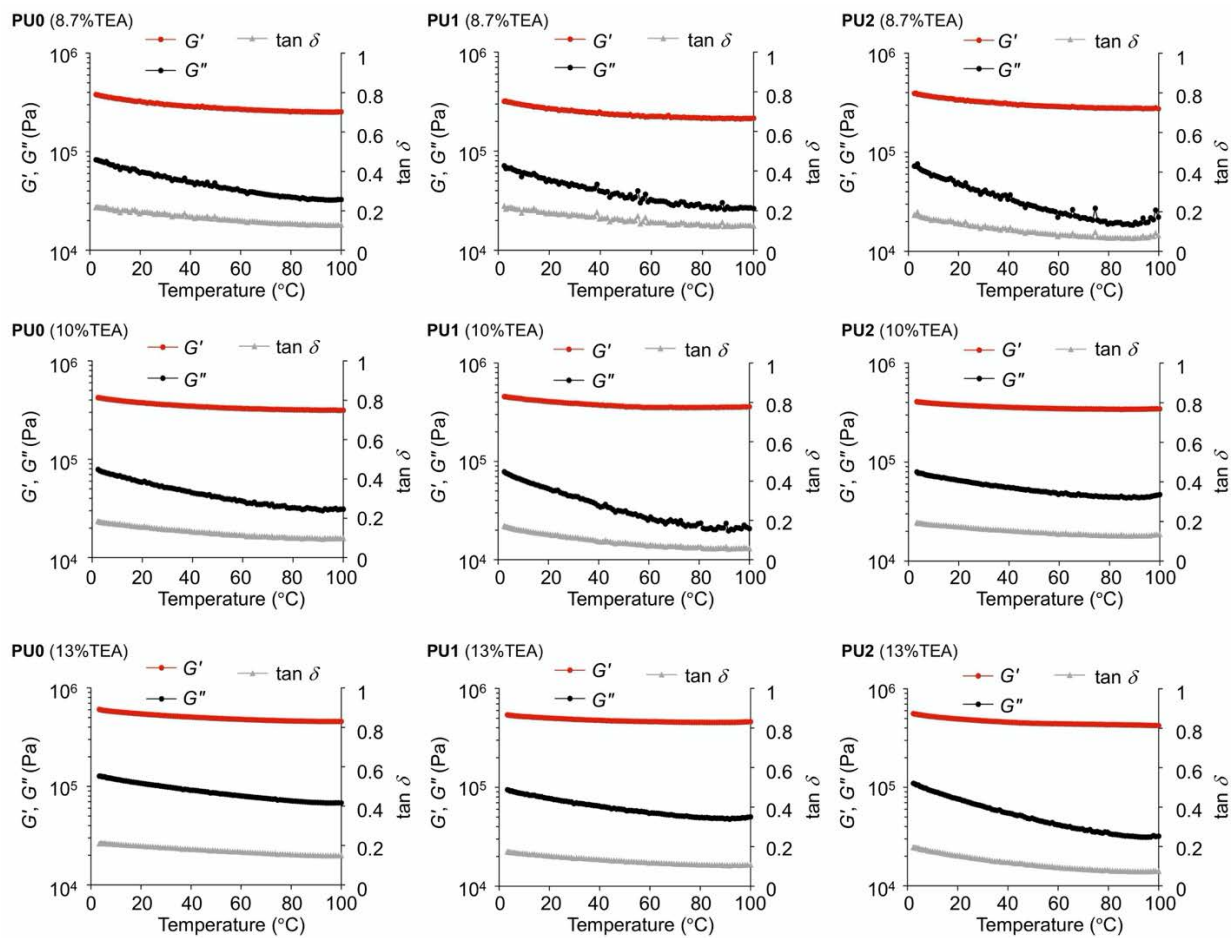

**Supplementary Fig. 26.** Dynamic viscoelasticity of the crosslinked PUs.

**Supplementary Table 20.** Dynamic viscoelasticity of the PUs at 25 °C.

|  | $G'$ (Pa) [a] | $G''$ (Pa) [b] | tan $\delta$ [c] |
|---|---|---|---|
| **PU0 (8.7%TEA)** | $3.1 \times 10^5$ | $5.9 \times 10^4$ | 0.19 |
| **PU1 (8.7%TEA)** | $2.6 \times 10^5$ | $4.8 \times 10^4$ | 0.18 |
| **PU2 (8.7%TEA)** | $3.3 \times 10^5$ | $4.2 \times 10^4$ | 0.13 |
| **PU0 (10%TEA)** | $3.7 \times 10^5$ | $5.5 \times 10^4$ | 0.15 |
| **PU1 (10%TEA)** | $3.9 \times 10^5$ | $4.8 \times 10^4$ | 0.12 |
| **PU2 (10%TEA)** | $3.7 \times 10^5$ | $6.2 \times 10^4$ | 0.17 |
| **PU0 (13%TEA)** | $5.3 \times 10^5$ | $1.0 \times 10^5$ | 0.19 |
| **PU1 (13%TEA)** | $4.9 \times 10^5$ | $7.3 \times 10^4$ | 0.15 |
| **PU2 (13%TEA)** | $4.8 \times 10^5$ | $7.0 \times 10^4$ | 0.15 |

[a] Storage modulus at 25 °C. [b] Loss modulus at 25 °C. [c] Loss tangent obtained from $G''/G'$ at 25 °C.

With higher crosslinking density (%TEA), the $G'$ values increased significantly. The FLAP dopants in **PU1** and **PU2** did not affect the viscoelasticity of the bulk PU sample.

## Mechanical properties of the rubbery polyurethanes (PUs)

Uniaxial tensile tests were carried out on dumbbell-shaped specimens. Rubbery samples were prepared by heating at 60 °C for 1 min with a blow dryer. Nominal stress, $\sigma_N$, is defined as $f/S_0$, where applied force $f$ is divided by the initial cross-sectional area $S_0$ (typically 1.5-mm thickness × 6-mm width).

$$\sigma_N = \frac{f}{S_0}$$

Nominal strain, $\varepsilon_N$, is defined as $\Delta L/L_0$, where sample gauge length ($\Delta L$) is divided by the initial length ($L_0$ = 10 mm).

$$\varepsilon_N = \frac{\Delta L}{L_0}$$

Toughness was calculated as the integrated area under the stress–strain curve.
Young's modulus is defined as the slope of stress–strain curves at 0–10% strain. Tensile velocity was fixed at 100 mm min$^{-1}$, corresponding to strain rate of 0.17 s$^{-1}$.

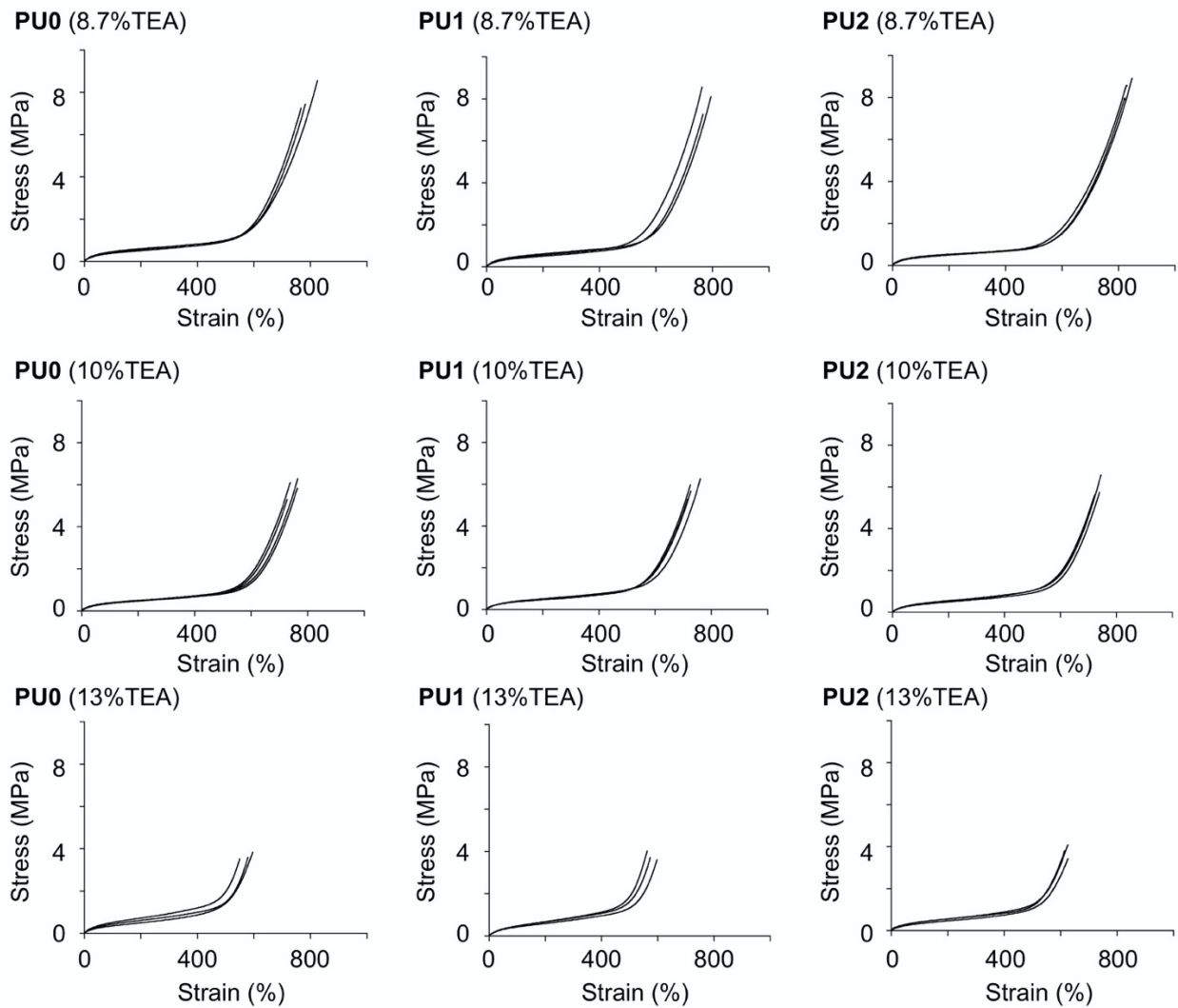

**Supplementary Fig. 27.** Stress-strain curves of the rubbery PUs.

**Supplementary Table 21.** Mechanical properties of the rubbery PUs in the uniaxial tensile testing.

|  | Rupture strain (%) | Rupture stress (MPa) | Toughness (MJ m$^{-3}$) | Young's modulus (MPa) |
|---|---|---|---|---|
| **PU0** (8.7%TEA) | 801 ± 44 | 7.2 ± 1.9 | 12.0 ± 3.3 | 1.2 ± 0.1 |
| **PU1** (8.7%TEA) | 791 ± 30 | 7.8 ± 0.7 | 12.3 ± 1.2 | 1.2 ± 0.2 |
| **PU2** (8.7%TEA) | 824 ± 8 | 8.2 ± 0.3 | 12.6 ± 0.5 | 1.1 ± 0.1 |
| **PU0** (10%TEA) | 749 ± 19 | 5.9 ± 0.4 | 8.7 ± 0.7 | 1.1 ± 0.1 |
| **PU1** (10%TEA) | 732 ± 20 | 5.8 ± 0.4 | 8.6 ± 0.7 | 1.0 ± 0.1 |
| **PU2** (10%TEA) | 733 ± 11 | 6.0 ± 0.4 | 9.2 ± 1.0 | 1.1 ± 0.1 |
| **PU0** (13%TEA) | 575 ± 24 | 3.6 ± 0.2 | 5.4 ± 0.2 | 1.1 ± 0.1 |
| **PU1** (13%TEA) | 577 ± 18 | 3.6 ± 0.4 | 5.6 ± 0.1 | 1.1 ± 0.1 |
| **PU2** (13%TEA) | 616 ± 12 | 4.0 ± 0.5 | 5.8 ± 0.6 | 1.1 ± 0.2 |

Values of average ± standard deviation of 3–4 specimen were shown.

## Photophysical properties of the unstretched polyurethanes (PUs)

FL spectra in Supplementary Figs. 28–33 were recorded on a JASCO Spectrofluorometer FP-8500. The synthesized polyurethane (PU) films become semicrystalline when stored at room temperature (25 °C) for more than 2 days. Therefore, the rubbery samples were prepared by heating at 60 °C for 1 min with a blow dryer. While mechanical properties were largely dependent on the crystallinity (Supplementary Fig. 39), FL spectra, FL quantum yields, and FL lifetimes of the unstretched PU films showed small differences (Supplementary Table 22, Supplementary Figs. 28–33). No excimer formation was observed in these conditions.

**Supplementary Table 22.** Photophysical constants of semicrystalline and rubbery PU films with each crosslinking density (%TEA). $\lambda_{ex}$ = 365 nm.

|  | $\Phi_{FL}$ | $\tau_{FL}$ (ns)[a] | $k_r$ (s$^{-1}$) | $k_{nr}$ (s$^{-1}$) |
|---|---|---|---|---|
| semicrystalline **PU1** (8.7%TEA) | 0.30 | 6.8 | 4.4 × 10$^7$ | 1.0 × 10$^8$ |
| rubbery **PU1** (8.7%TEA) | 0.26 | 6.9 | 3.8 × 10$^7$ | 1.1 × 10$^8$ |
| semicrystalline **PU1** (10%TEA) | 0.30 | 6.7 | 4.4 × 10$^7$ | 1.0 × 10$^8$ |
| rubbery **PU1** (10%TEA) | 0.28 | 6.9 | 4.0 × 10$^7$ | 1.0 × 10$^8$ |
| semicrystalline **PU1** (13%TEA) | 0.33 | 6.9 | 4.8 × 10$^7$ | 1.0 × 10$^8$ |
| rubbery **PU1** (13%TEA) | 0.31 | 7.1 | 4.3 × 10$^7$ | 1.0 × 10$^8$ |
| semicrystalline **PU2** (8.7%TEA) | 0.35 | 6.8 | 5.1 × 10$^7$ | 1.0 × 10$^8$ |
| rubbery **PU2** (8.7%TEA) | 0.37 | 7.0 | 3.8 × 10$^7$ | 0.9 × 10$^8$ |
| semicrystalline **PU2** (10%TEA) | 0.35 | 6.7 | 5.2 × 10$^7$ | 1.0 × 10$^8$ |
| rubbery **PU2** (10%TEA) | 0.35 | 7.1 | 5.0 × 10$^7$ | 0.9 × 10$^8$ |
| semicrystalline **PU2** (13%TEA) | 0.33 | 7.0 | 4.6 × 10$^7$ | 1.0 × 10$^8$ |
| rubbery **PU2** (13%TEA) | 0.33 | 6.9 | 4.8 × 10$^7$ | 1.0 × 10$^8$ |

[a] $\lambda_{em}$ = 470 nm.

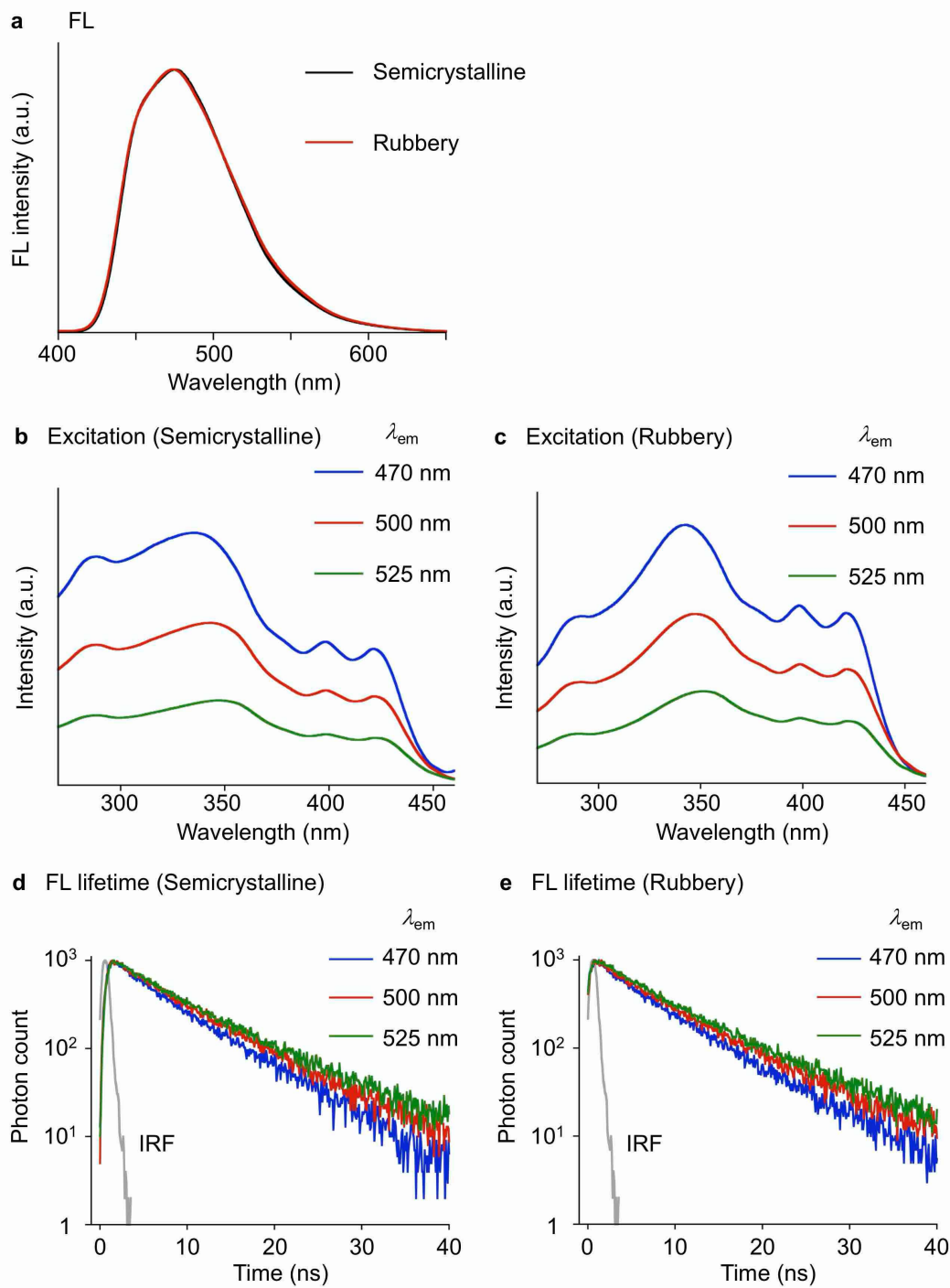

**Supplementary Fig. 28.** Photophysical properties of semicrystalline and rubbery **PU1** films (8.7%TEA). (a) FL spectra ($\lambda_{ex}$ = 365 nm), (b, c) excitation spectra, and (d, e) FL decay profiles.

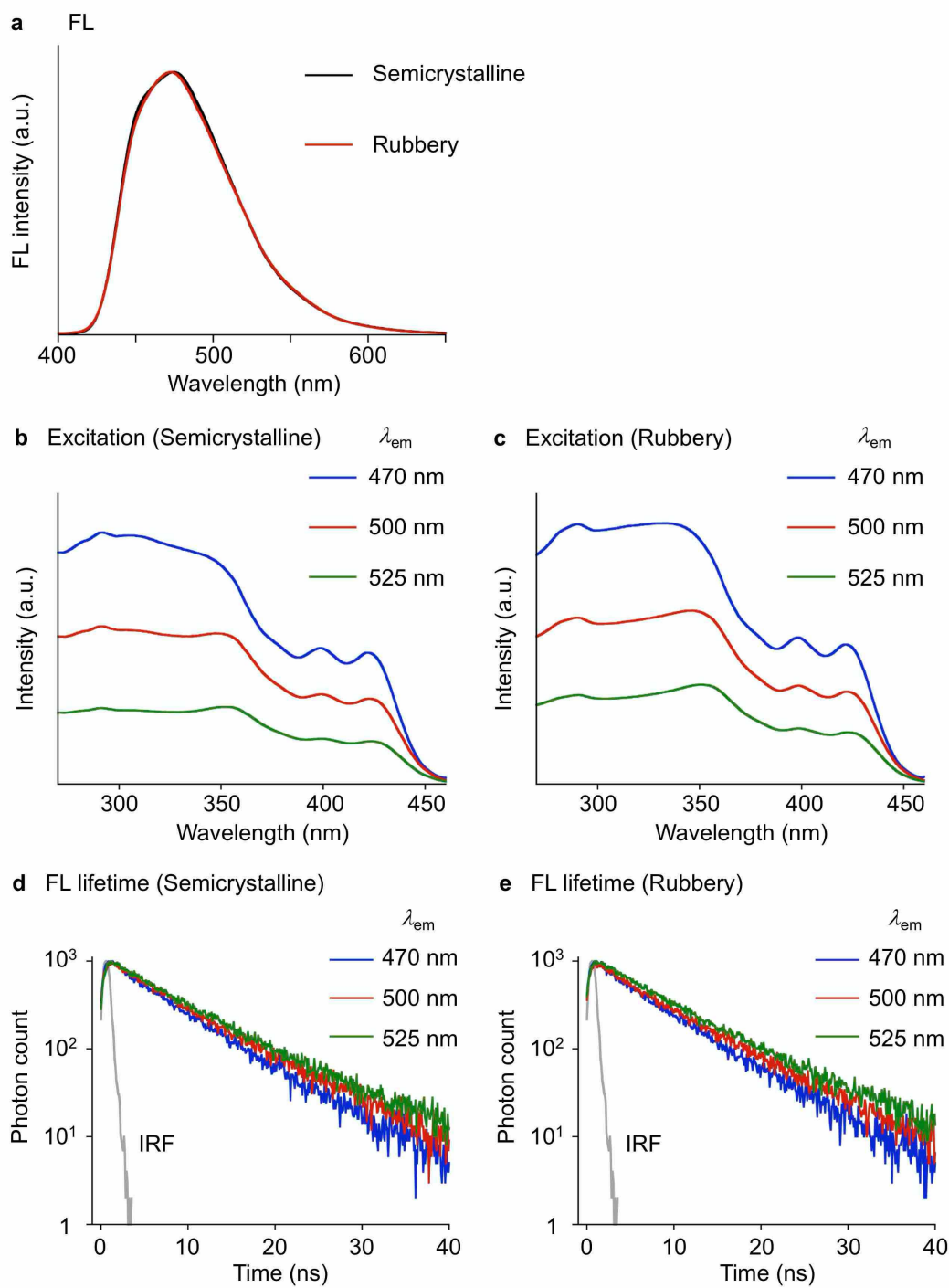

**Supplementary Fig. 29.** Photophysical properties of semicrystalline and rubbery **PU1** films (10%TEA). (a) FL spectra ($\lambda_{ex}$ = 365 nm), (b, c) excitation spectra, and (d, e) FL decay profiles.

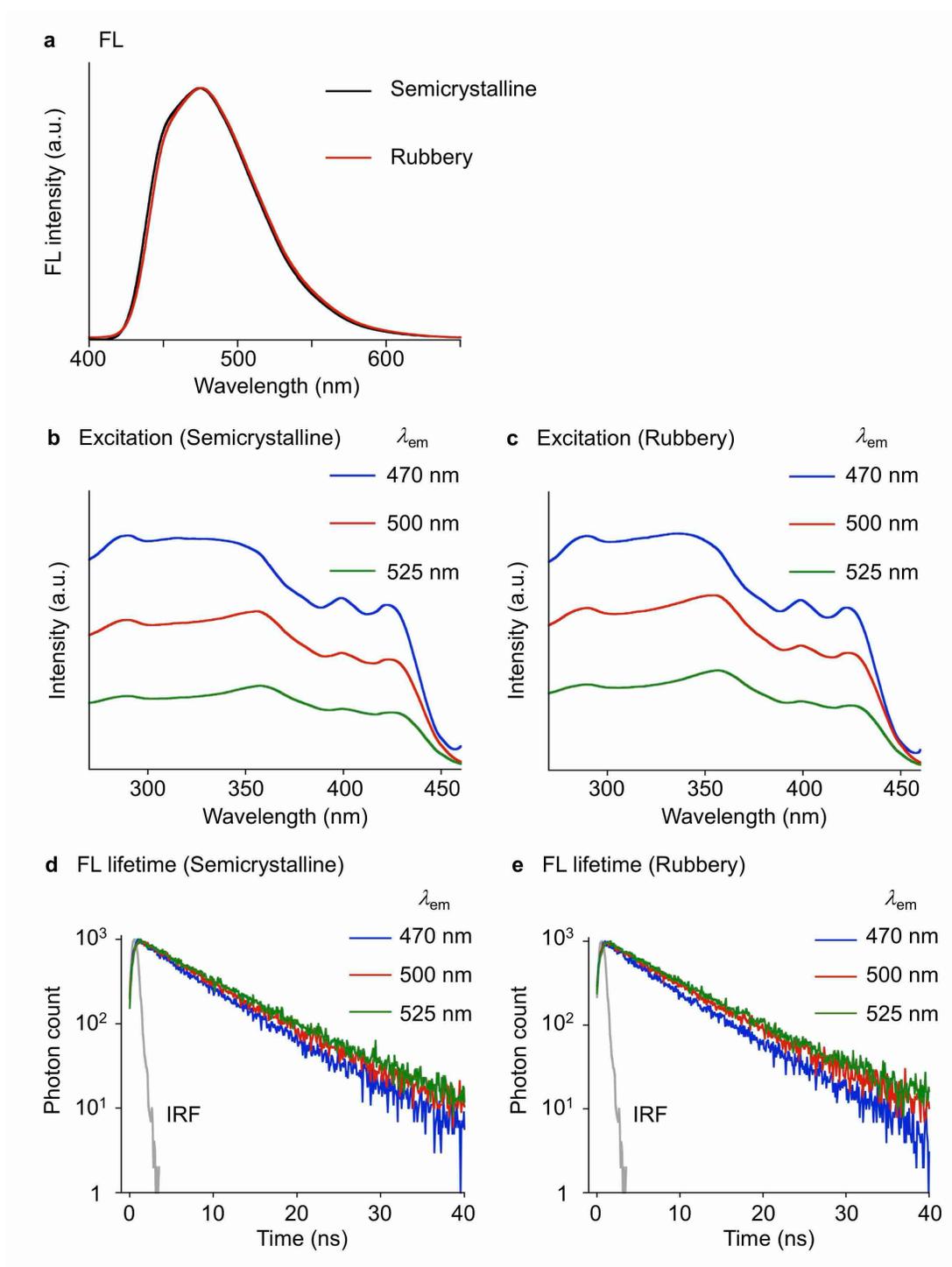

**Supplementary Fig. 30.** Photophysical properties of semicrystalline and rubbery **PU1** films (13%TEA). (a) FL spectra ($\lambda_{ex}$ = 365 nm), (b, c) excitation spectra, and (d, e) FL decay profiles.

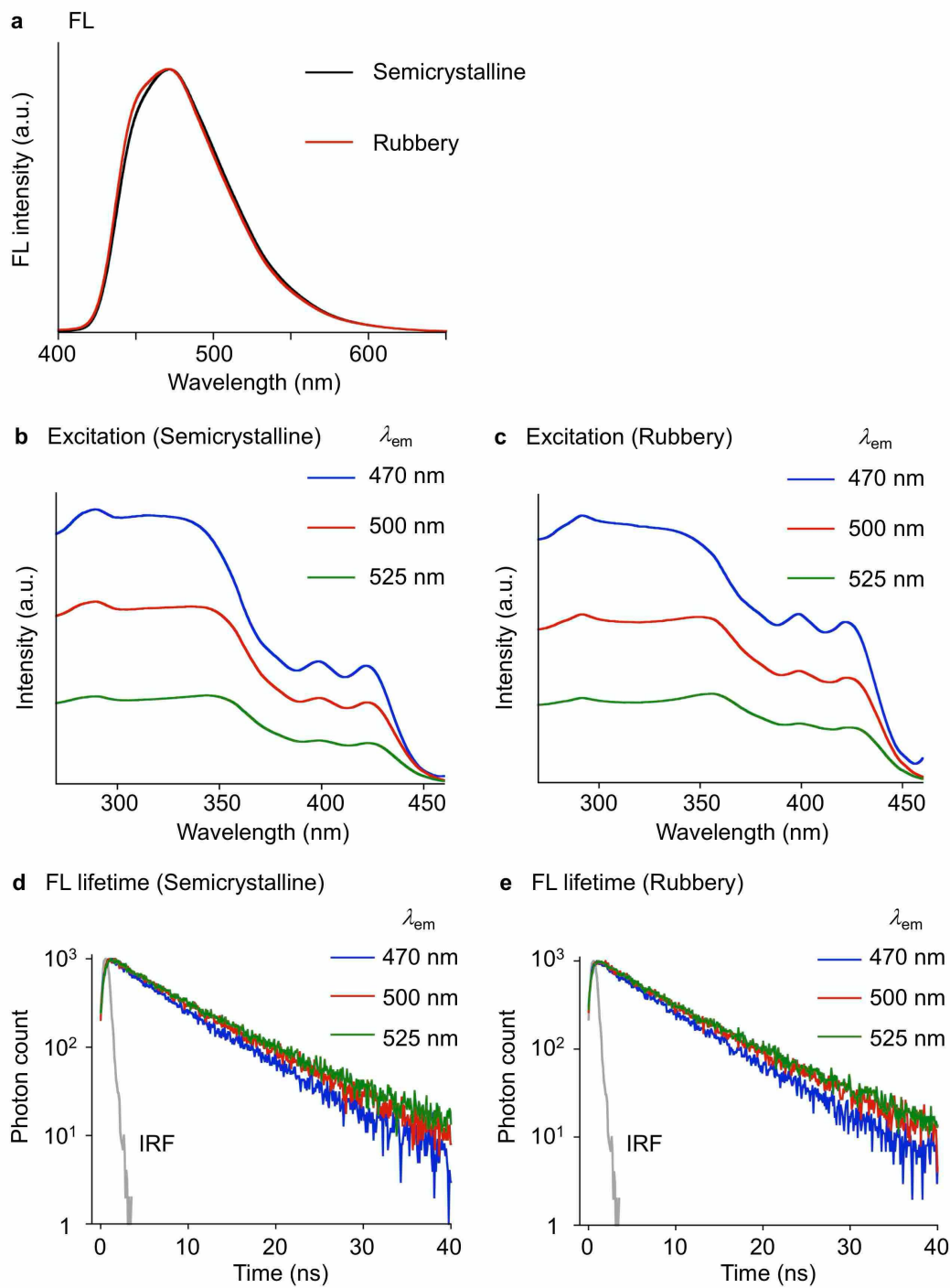

**Supplementary Fig. 31.** Photophysical properties of semicrystalline and rubbery **PU2** films (8.7%TEA). (a) FL spectra ($\lambda_{ex}$ = 365 nm), (b, c) excitation spectra, and (d, e) FL decay profiles.

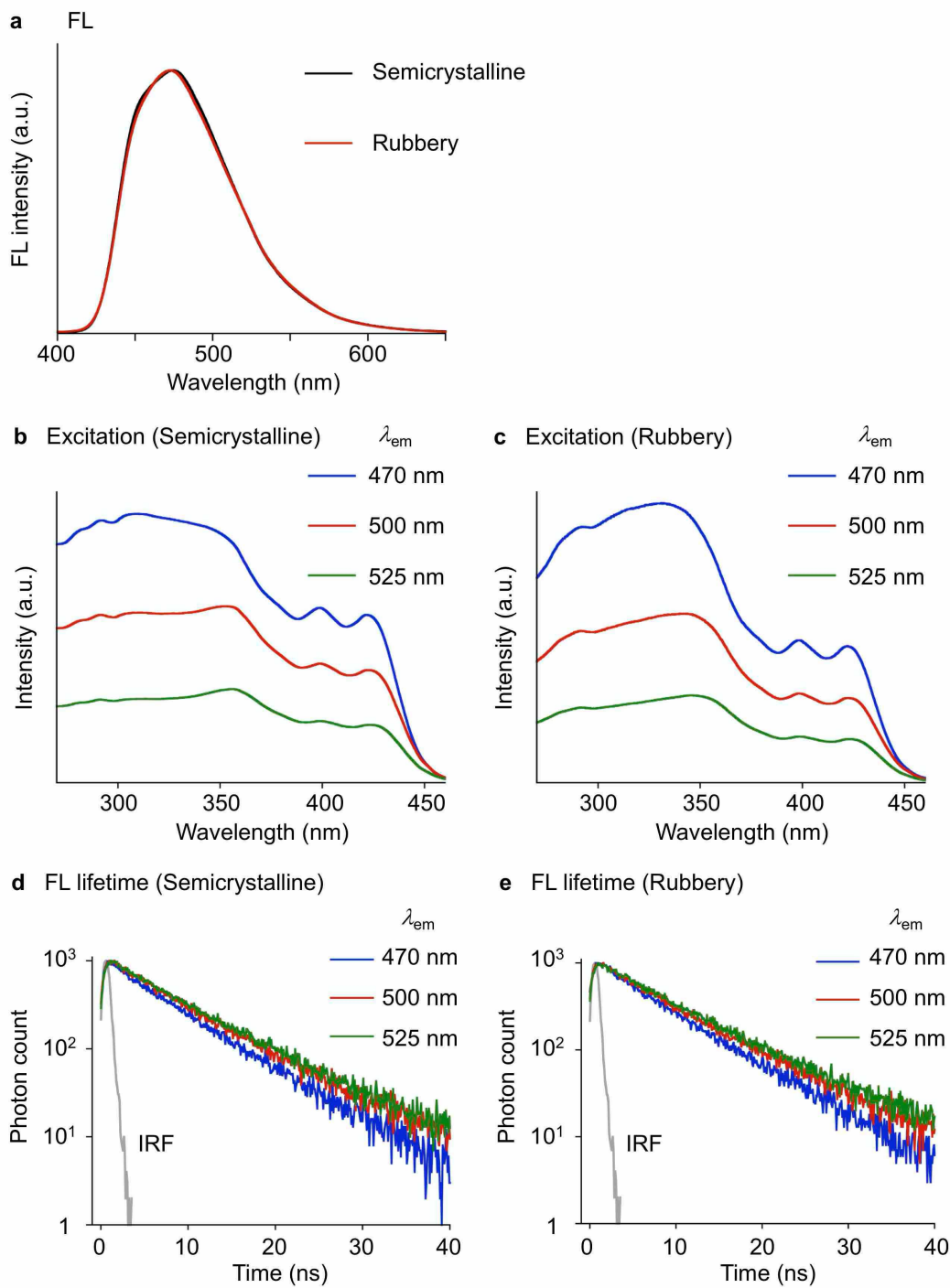

**Supplementary Fig. 32.** Photophysical properties of semicrystalline and rubbery **PU2** films (10%TEA). (a) FL spectra ($\lambda_{ex}$ = 365 nm), (b, c) excitation spectra, and (d, e) FL decay profiles.

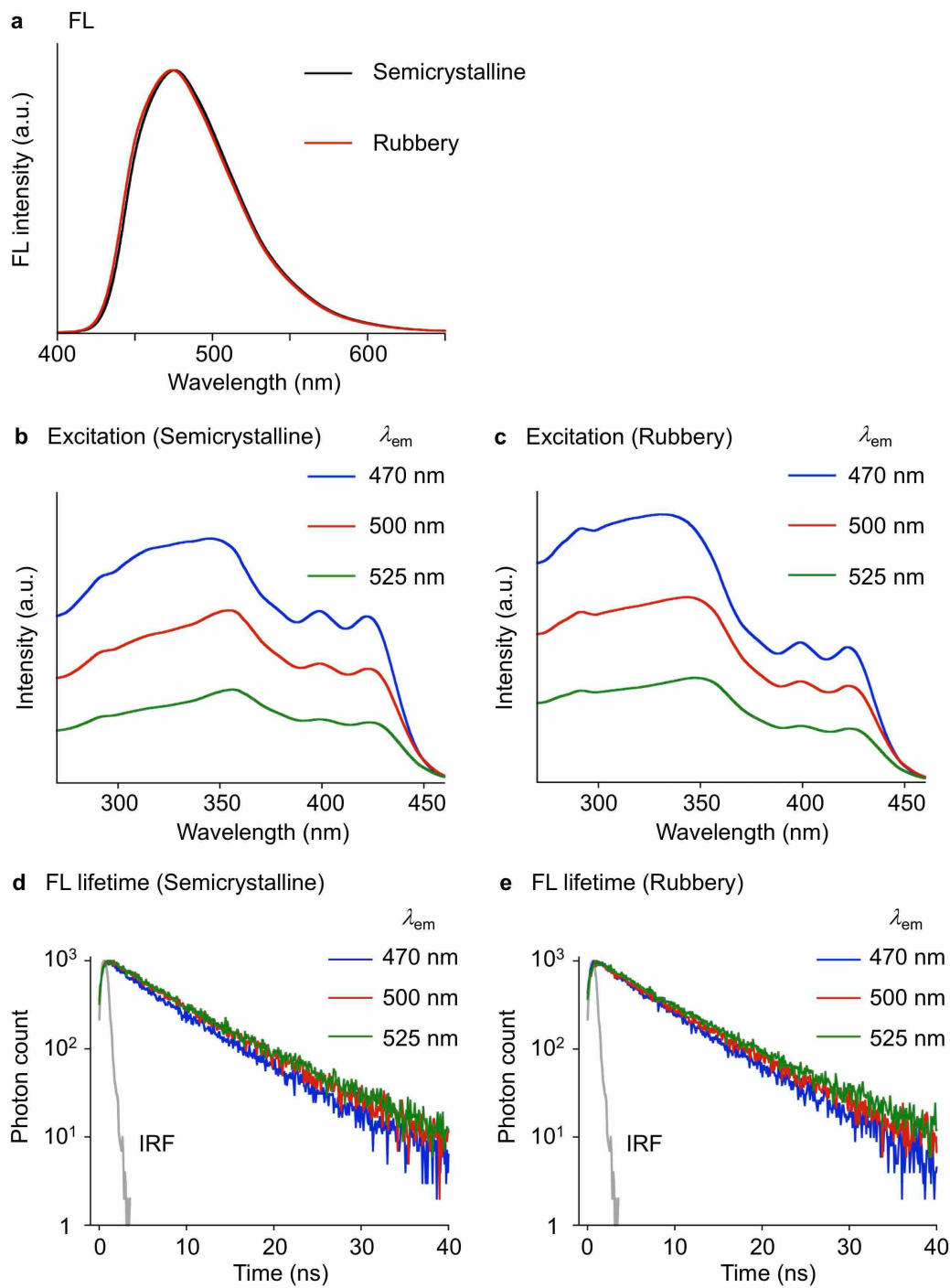

**Supplementary Fig. 33.** Photophysical properties of semicrystalline and rubbery **PU2** films (13%TEA). (a) FL spectra ($\lambda_{ex}$ = 365 nm), (b, c) excitation spectra, and (d, e) FL decay profiles.

**Real-time monitoring of the FL and absorption spectra of the stretched PUs**

During the tensile testing, time-dependent FL and absorption spectra were monitored using a multi-channel photodetector (Otsuka Electronics, MCPD-6800) equipped with an optical fiber. The FL spectra were recorded with 0.8-s exposure time and 2 accumulations (at 1.6-s intervals), in which a 365-nm LED was used for excitation. Absorption spectra were recorded on the same photodetecting system with a 1.6-s exposure time and 2 accumulations (at 3.2-s intervals). Before the tensile testing, reference absorption spectrum $I_0(\lambda)$ was measured with a 100-W tungsten light source fixed at a constant distance from the optical fiber head. Then, a PU specimen was set (and stretched) so that the specimen crossed the optical path to acquire absorption spectrum $I(\lambda)$. Absorbance $A$ at each wavelength was obtained from $I_0(\lambda)$ and $I(\lambda)$ as below.

$$A = -\log\left(\frac{I}{I_0}\right)$$

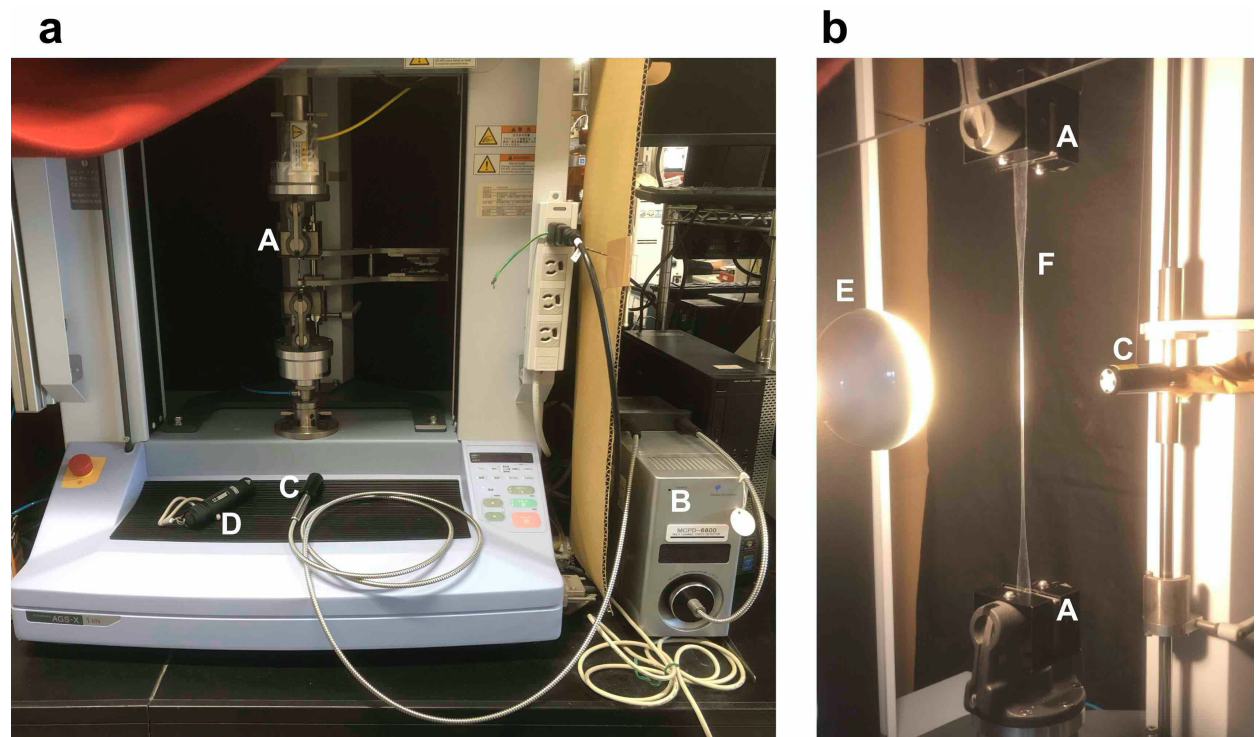

**Supplementary Fig. 34.** Experimental setup for the real-time collection of (a) FL spectra and (b) absorption spectra of the stretched PU films.

**A**: Upper and lower grips of a tensile testing machine to hold a specimen.
**B**: Multi-channel photodetector.
**C**: Optical fiber head connected to **B**.
**D**: 365-nm LED (NICHIA, JAXMAN U1).
**E**: 100-W tungsten light (METRO).
**F:** PU specimen.

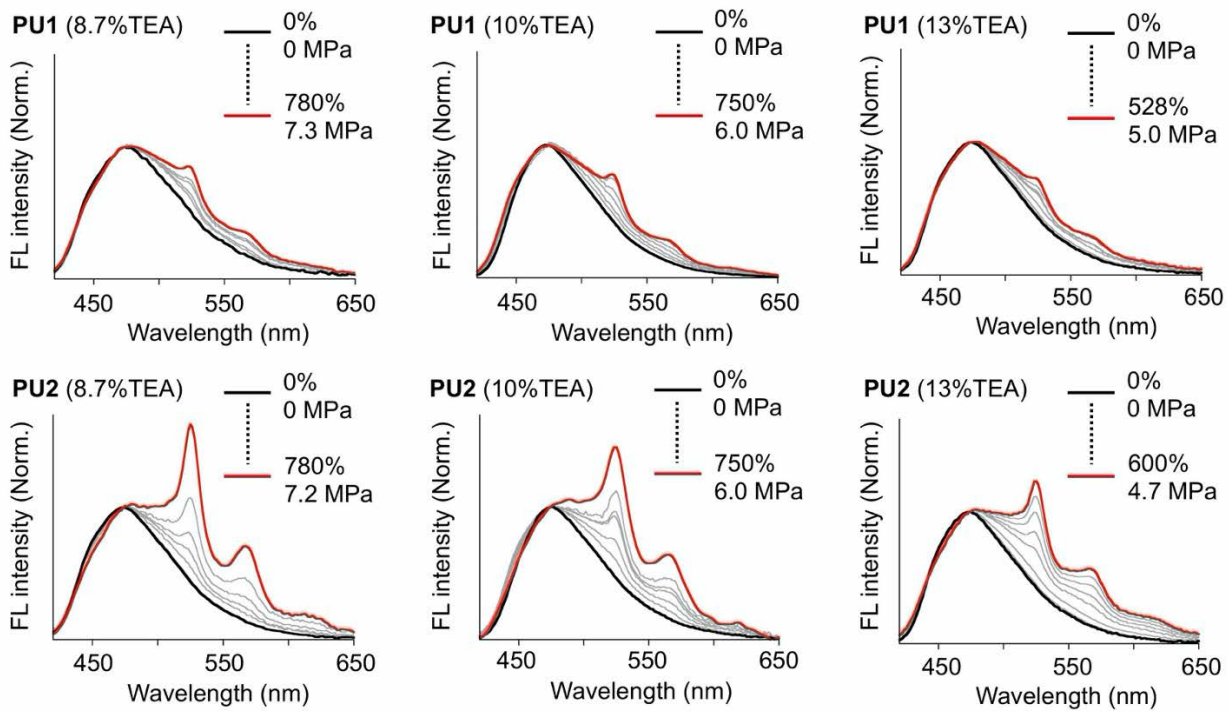

**Supplementary Fig. 35.** FL spectral changes of the rubbery PUs. Normalized at 474 nm. $\lambda_{ex}$ = 365 nm.

**Loading–unloading cycle testing**

To confirm the rapid reversibility of the stress-induced FL response, the loading and unloading cycle testing of **PU2** (10%TEA) was performed with ratiometric FL analysis (Fig. 4F in the main text). Supplementary Fig. 36 shows mechanical profiles in the cycle test.

Start → **A**: Loading up to 300% strain (purple line).
**A** → **B**: Unloading up to stress of 0.1 MPa (blue line).
**B** → **C**: Loading up to 600% strain (green line).
**C** → **D**: Unloading up to stress of 0.1 MPa (orange line).
**D** → **E**: Loading up to rupture (red line).

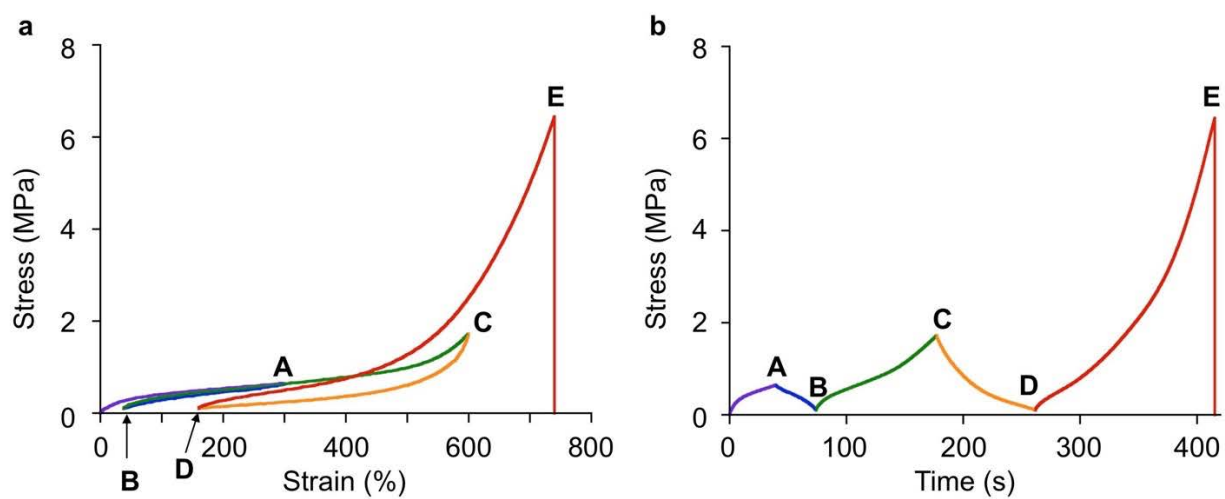

**Supplementary Fig. 36.** Stress profile as a function of (a) strain and (b) time for the cycle testing.

## Estimation of the stressed FLAP probe (%)

Percentage of the stressed FLAP probe over the FL switching threshold was estimated from the ratiometric FL analysis of the stretched PU films (Fig. 5 in the main text).

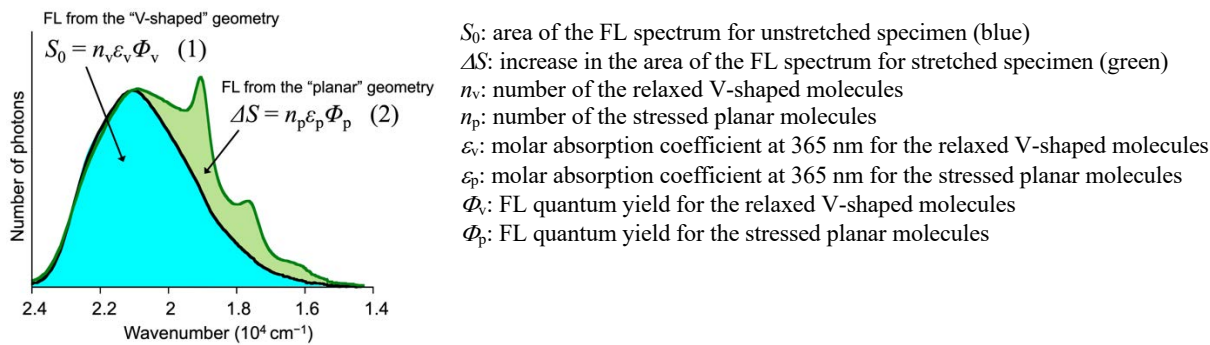

FL from the "V-shaped" geometry
$S_0 = n_v \varepsilon_v \Phi_v$   (1)

FL from the "planar" geometry
$\Delta S = n_p \varepsilon_p \Phi_p$   (2)

$S_0$: area of the FL spectrum for unstretched specimen (blue)
$\Delta S$: increase in the area of the FL spectrum for stretched specimen (green)
$n_v$: number of the relaxed V-shaped molecules
$n_p$: number of the stressed planar molecules
$\varepsilon_v$: molar absorption coefficient at 365 nm for the relaxed V-shaped molecules
$\varepsilon_p$: molar absorption coefficient at 365 nm for the stressed planar molecules
$\Phi_v$: FL quantum yield for the relaxed V-shaped molecules
$\Phi_p$: FL quantum yield for the stressed planar molecules

**Supplementary Fig. 37.** Optical parameters for the estimation.

For the estimation, it was assumed that the FL emission was observed mainly from two types of independent species bearing the bent and planar geometries. Considering (1) and (2) in Supplementary Fig. 37, the proportion of the stressed planar FLAP molecules can be expressed as follows,

$$\frac{n_p}{n_p+n_v} = \frac{1}{1+\frac{n_v}{n_p}} = \frac{1}{1+\frac{\varepsilon_p \Phi_p}{\varepsilon_v \Phi_v} \cdot \frac{S_0}{\Delta S}} \quad (3)$$

Here, molar absorption coefficients were roughly estimated as $\varepsilon_p$ = 140000 M$^{-1}$ cm$^{-1}$ for the planar geometry and $\varepsilon_v$ = 70000 M$^{-1}$ cm$^{-1}$ for the bent geometry, considering TD-DFT calculation results at the TD PBE0/6-31+G(d) level of theory. The FL quantum yields of the independent planar and bent species were assumed to be comparable ($\Phi_v \approx \Phi_p$), because $\Phi_v$ was determined to be about 0.3 from the FL analysis of unstretched PU polymers with the FLAP dopant (Supplementary Table 22), and $\Phi_p$ was referenced from the FL quantum yield (about 0.3) of the parent FLAP compounds in solution (Supplementary Table 1).

## Control experiments: Physical doping of FLAP into the polyurethane (PU)

**PU3** was synthesized in the same procedure as **PU1** (10%TEA), in which **FLAP1** was replaced by **FLAP3**[8]. **FLAP3** is not covalently connected to the polymer chain network of **PU3**, but simply dispersed inside **PU3**. Without extension, mechanical and photophysical properties of the **PU3** film are almost the same as those of **PU1** and **PU2**. On the other hand, no characteristic FL response was observed under the tensile testing, clearly indicating that the conformational planarization of the FLAP dopant cannot be induced without covalent connection between the FLAP molecule and the polymer chains.

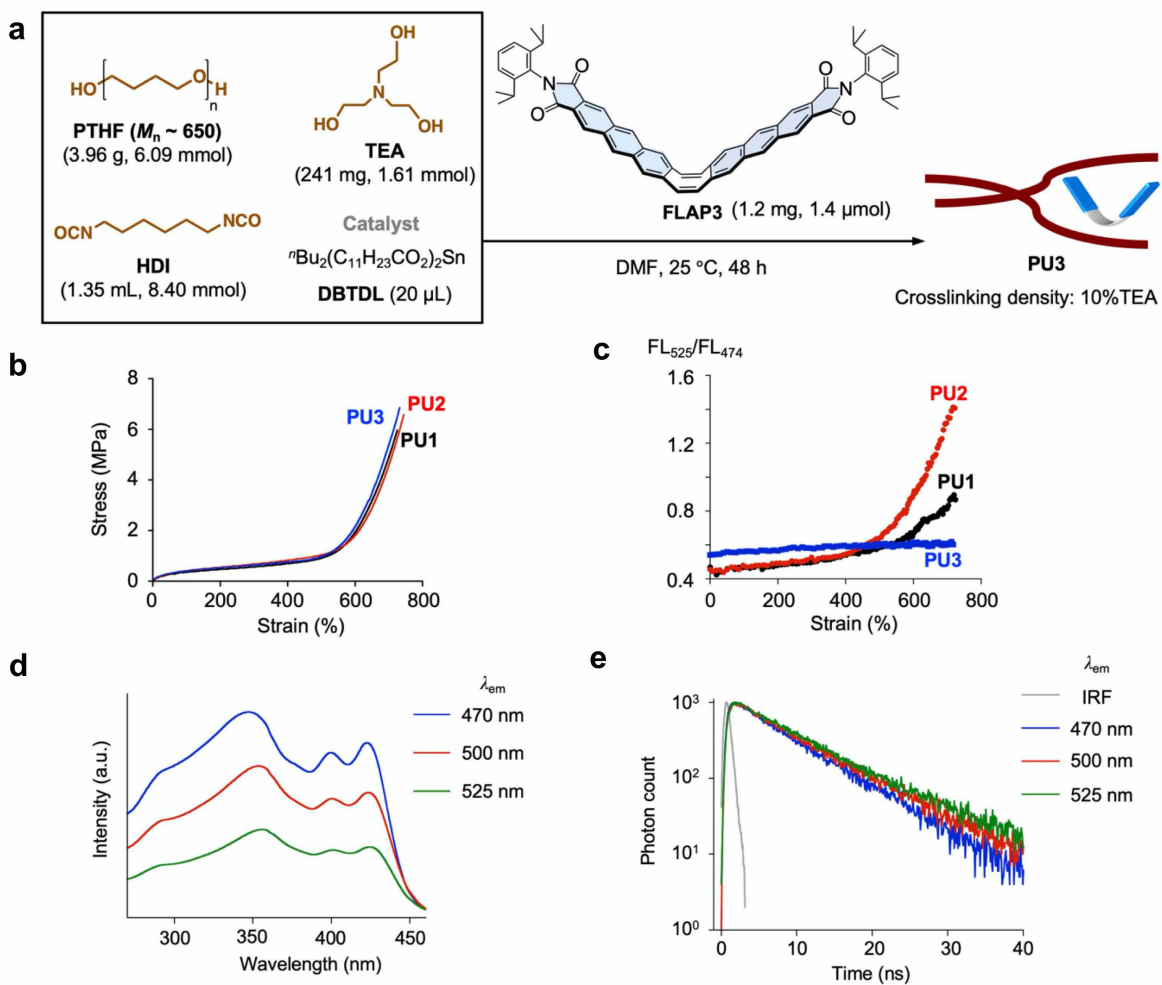

**Supplementary Fig. 38.** (a) Synthesis of **PU3** (10%TEA). (b) Stress-strain curves of **PU1**–**PU3** (10%TEA). Strain rate: 0.17 s$^{-1}$. (c) FL ratiometric analysis of the stretched **PU1**–**PU3** films. (d) Excitation spectra of the unstretched **PU3** film. (e) FL lifetime profiles of the unstretched **PU3** film.

**Supplementary Table 23.** Photophysical constants of the rubbery **PU3** films. $\lambda_{ex}$ = 365 nm.

| $\Phi_{FL}$ | $\tau_{FL}$ (ns)$^a$ | $k_r$ (s$^{-1}$) | $k_{nr}$ (s$^{-1}$) |
|---|---|---|---|
| 0.30 | 7.1 | 4.2 × 10$^7$ | 1.0 × 10$^8$ |

$^a\lambda_{em}$ = 470 nm. These values were comparable those of **PU1** and **PU2** in Supplementary Table 22.

## Semicrystalline polyurethanes (PUs)

Mechanical and photophysical properties of the semicrystalline PU samples, stored at room temperature (25 °C) for more than 2 days, were evaluated in the same protocols as the rubbery PU samples. Young's modulus is more than 10 times higher than that of the rubbery PU (Supplementary Table 24). Accordingly, yield point at the early stage of the strain became more pronounced for the semicrystalline samples (Supplementary Fig. 39), followed by typical necking behavior (Supplementary Fig. 40). Since the necking occurs in a specific position, acquired FL spectra depend on the UV-irradiated area. Rupture strain significantly decreased in the sample with the higher crosslinking density (Supplementary Table 24).

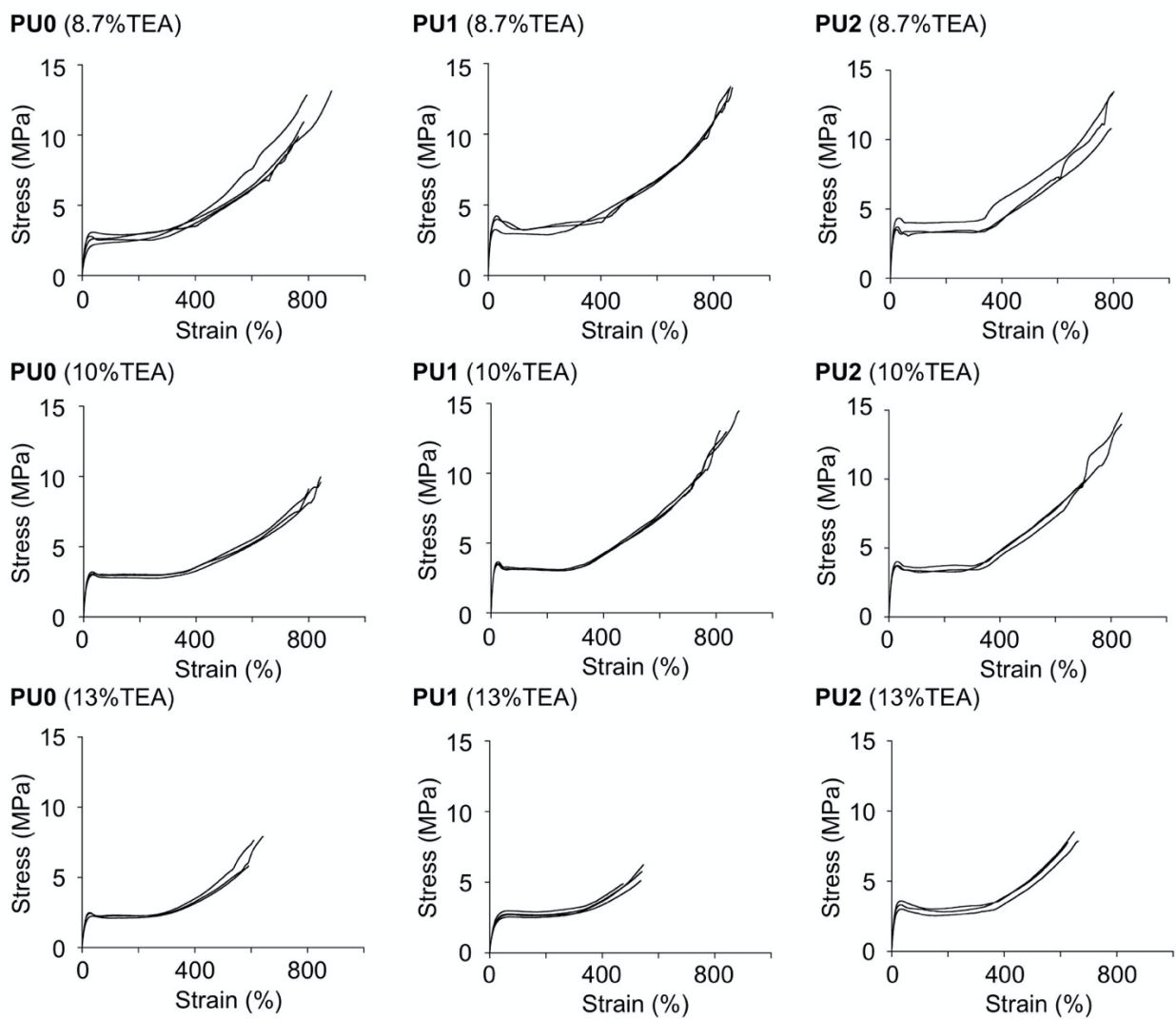

**Supplementary Fig. 39.** Stress-strain curves of the semicrystalline PUs.

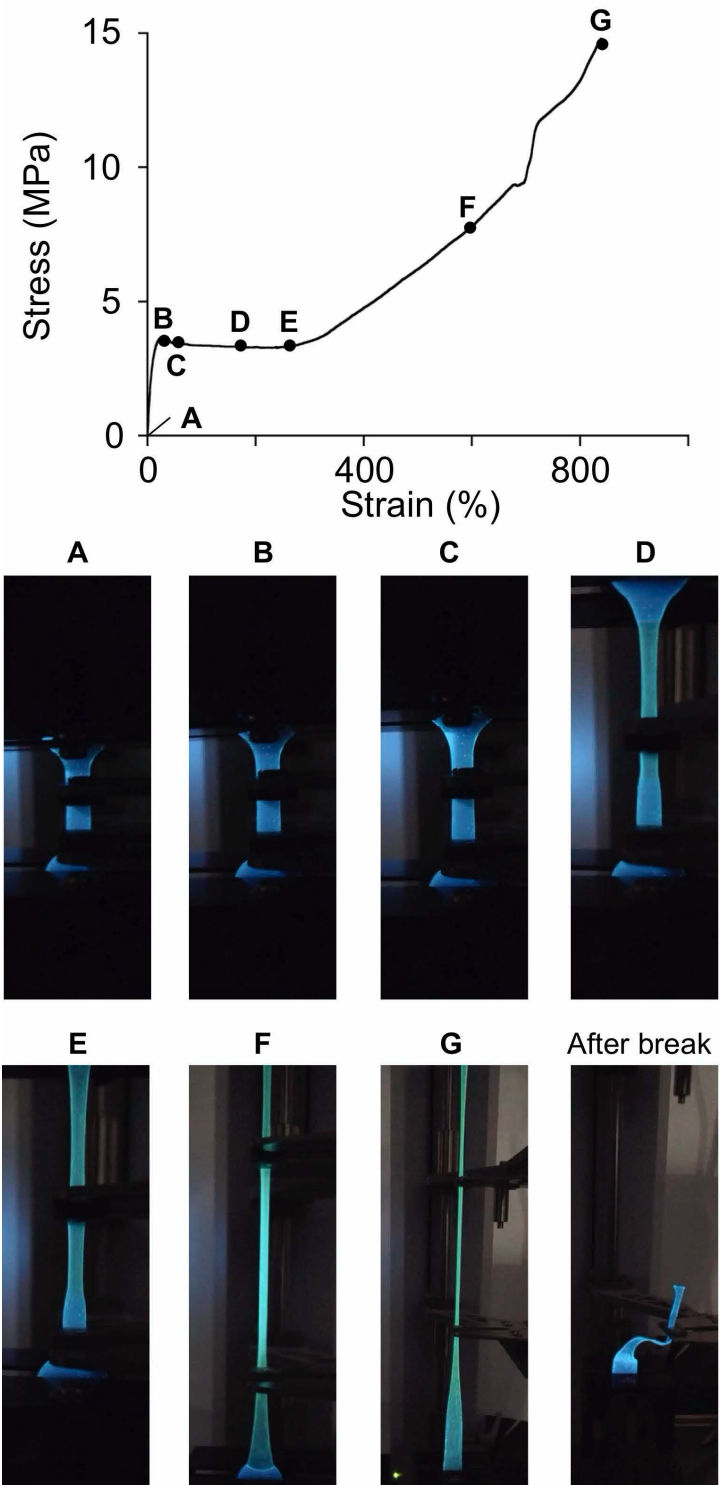

**Supplementary Fig. 40.** A typical example of the stress-strain curve and the corresponding photographs of the stretched fluorescent specimen, in which the semicrystalline **PU2** (10%TEA) was applied to the tensile testing.

**Supplementary Table 24.** Mechanical properties of the semicrystalline PUs in the uniaxial tensile testing.

|  | Rupture strain (%) | Rupture stress (MPa) | Toughness (MJ m$^{-3}$) | Young's modulus (MPa) |
| --- | --- | --- | --- | --- |
| **PU0 (8.7%TEA)** | 807 ± 52 | 11.7 ± 1.6 | 40.1 ± 5.3 | 18 ± 4 |
| **PU1 (8.7%TEA)** | 862 ± 5 | 13.3 ± 0.1 | 48.9 ± 0.9 | 30 ± 2 |
| **PU2 (8.7%TEA)** | 797 ± 5 | 12.5 ± 1.5 | 46.0 ± 5.0 | 30 ± 3 |
| **PU0 (10%TEA)** | 829 ± 25 | 9.6 ± 0.4 | 36.5 ± 1.4 | 22 ± 1 |
| **PU1 (10%TEA)** | 844 ± 34 | 13.5 ± 0.8 | 47.7 ± 4.0 | 28 ± 1 |
| **PU2 (10%TEA)** | 797 ± 72 | 12.9 ± 2.6 | 46.6 ± 8.5 | 28 ± 2 |
| **PU0 (13%TEA)** | 614 ± 26 | 7.1 ± 1.1 | 19.9 ± 2.4 | 17 ± 2 |
| **PU1 (13%TEA)** | 525 ± 34 | 5.5 ± 0.6 | 16.7 ± 1.1 | 15 ± 1 |
| **PU2 (13%TEA)** | 645 ± 18 | 8.1 ± 0.4 | 25.7 ± 1.8 | 24 ± 3 |

Average ± standard deviation of 3–4 specimens was shown.

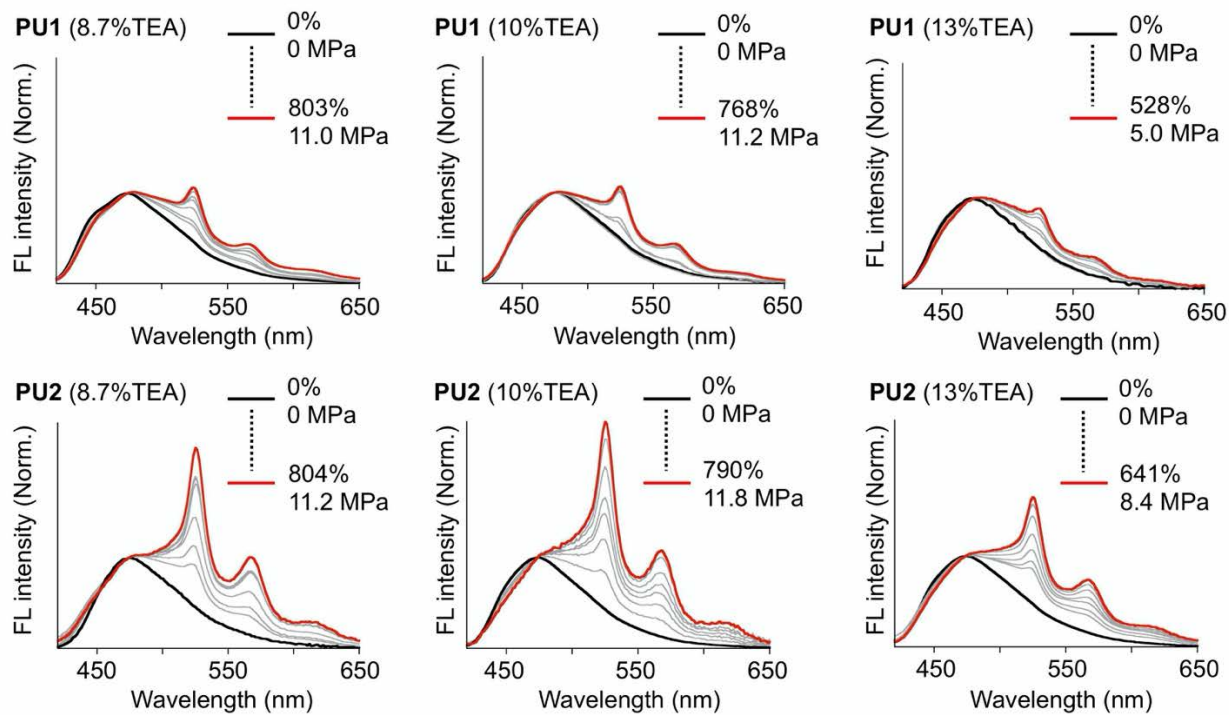

**Supplementary Fig. 41.** FL spectral changes of the semicrystalline PUs. Normalized at 474 nm. $\lambda_{ex}$ = 365 nm.

**NMR spectra of new compounds**

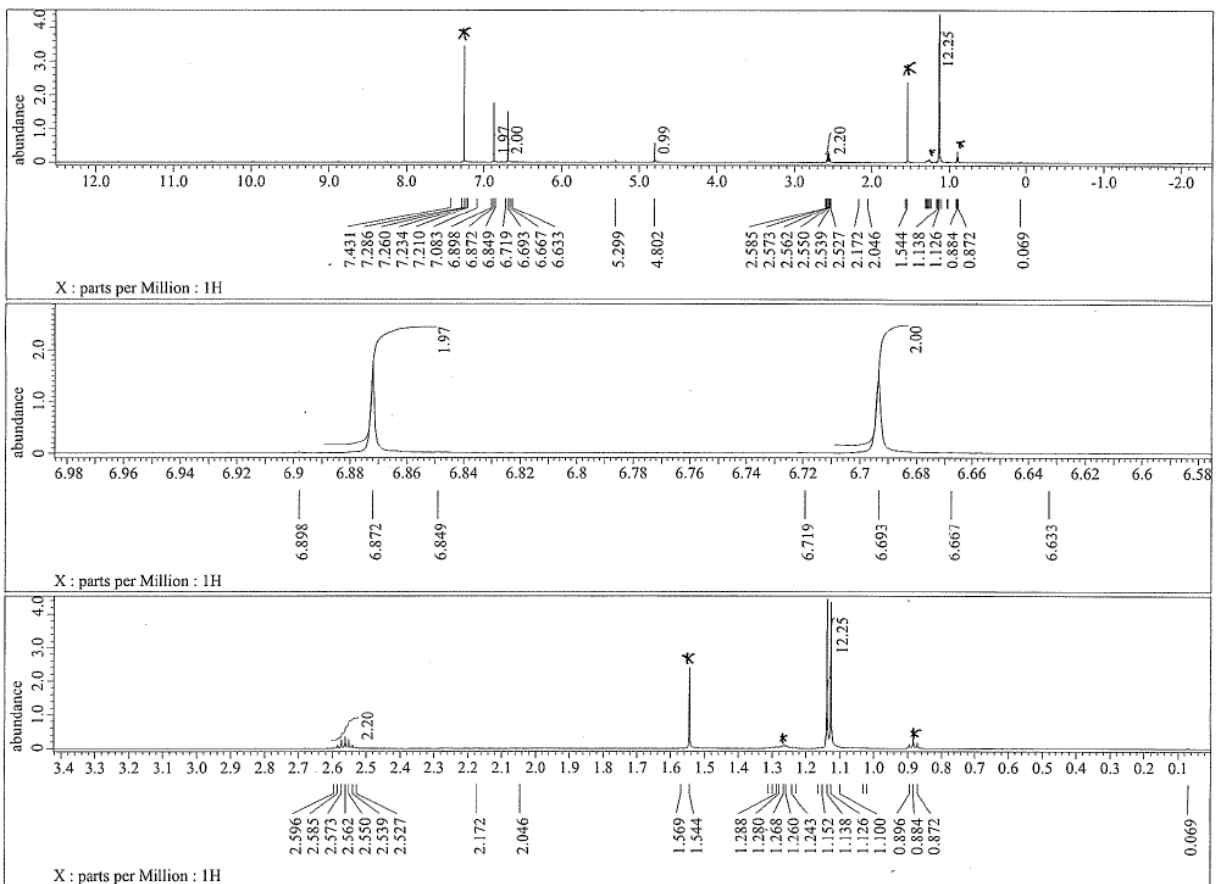

**Supplementary Fig. 42.** $^1$H NMR spectra of **S1** in CDCl$_3$ at 25 °C. The peaks marked with * indicate residual solvents.

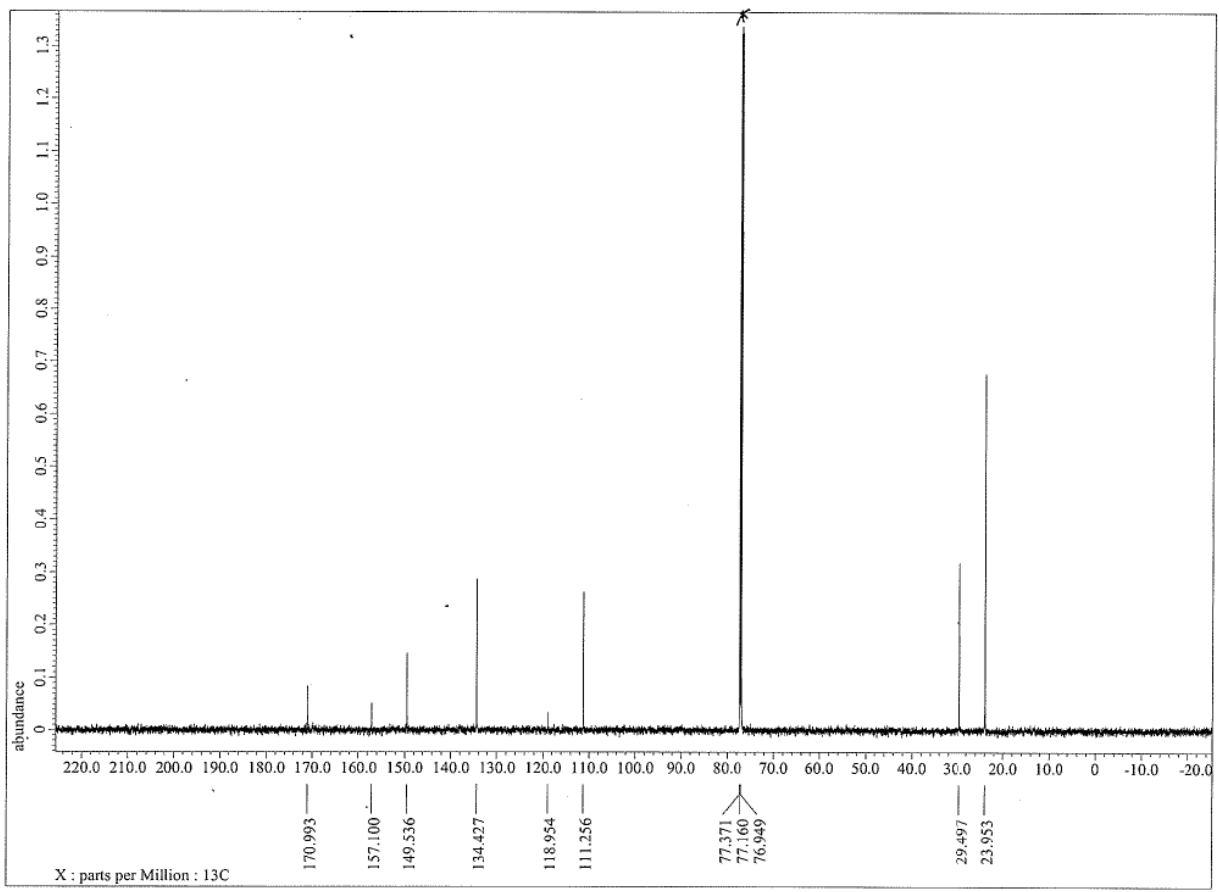

**Supplementary Fig. 43.** $^{13}$C NMR spectrum of **S1** in CDCl$_3$ at 25 °C. The peak marked with * indicates residual solvents.

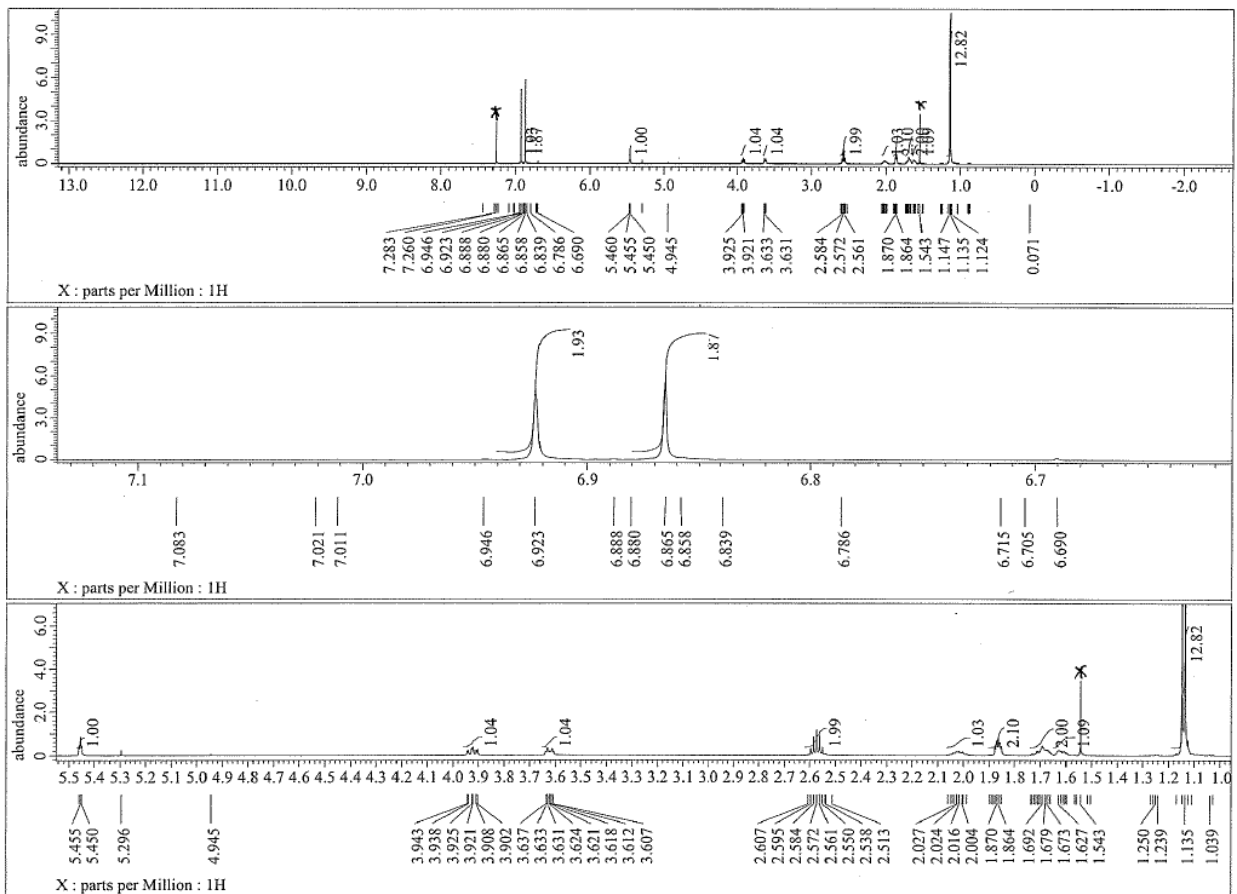

**Supplementary Fig. 44.** $^1$H NMR spectra of **S2** in CDCl$_3$ at 25 °C. The peaks marked with * indicate residual solvents.

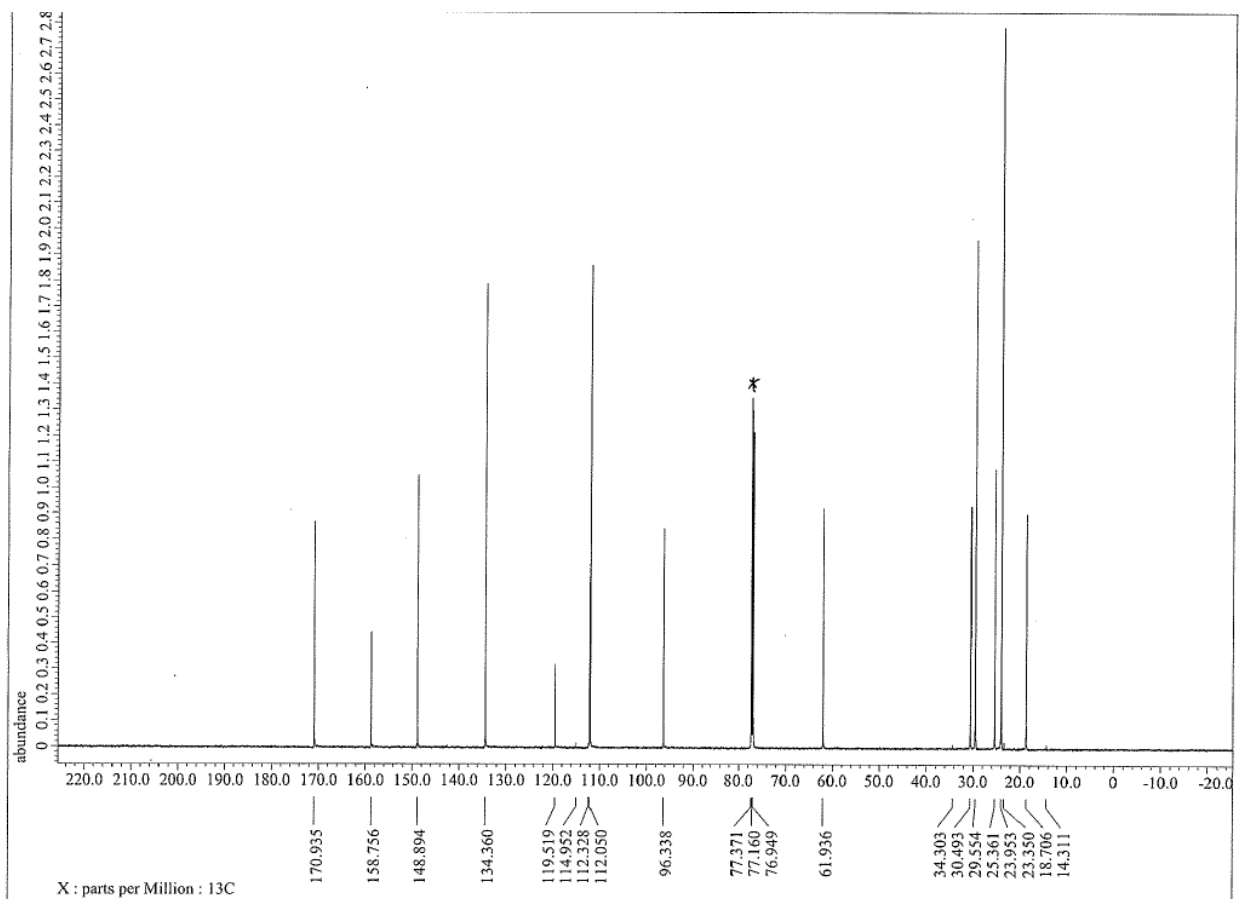

**Supplementary Fig. 45.** $^{13}$C NMR spectrum of **S2** in CDCl$_3$ at 25 °C. The peak marked with * indicates residual solvents.

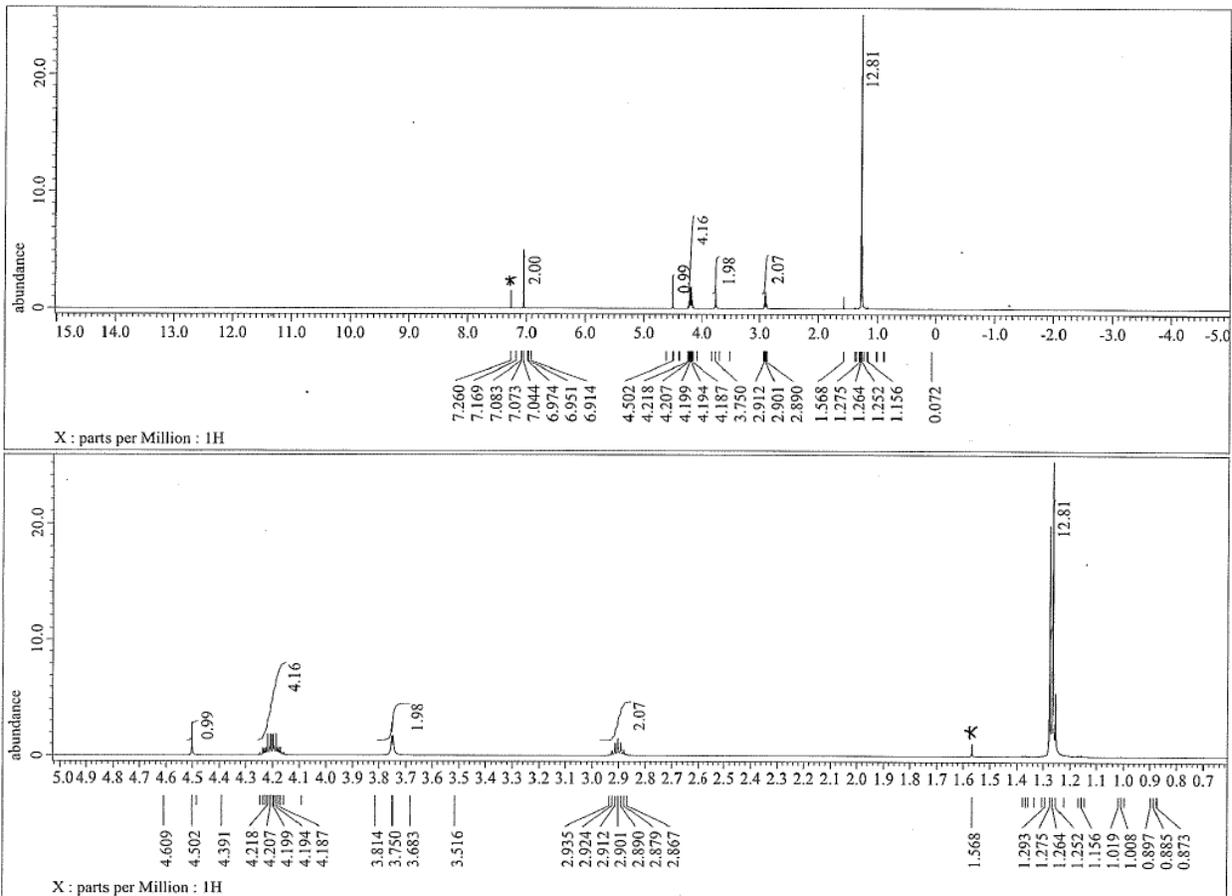

**Supplementary Fig. 46.** $^1$H NMR spectra of **S3** in CDCl$_3$ at 25 °C. The peaks marked with * indicate residual solvents.

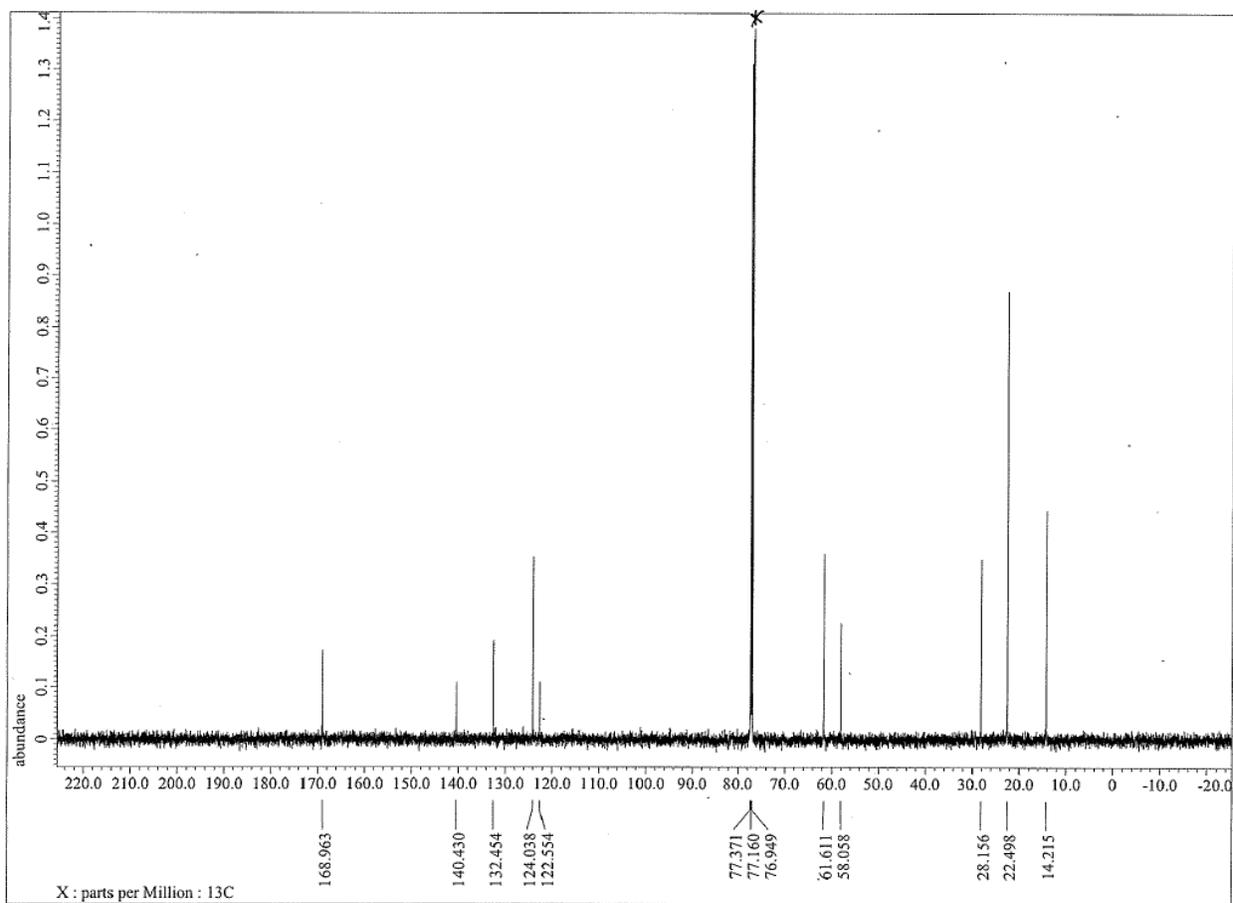

**Supplementary Fig. 47.** $^{13}$C NMR spectrum of **S3** in CDCl$_3$ at 25 °C. The peak marked with * indicates residual solvents.

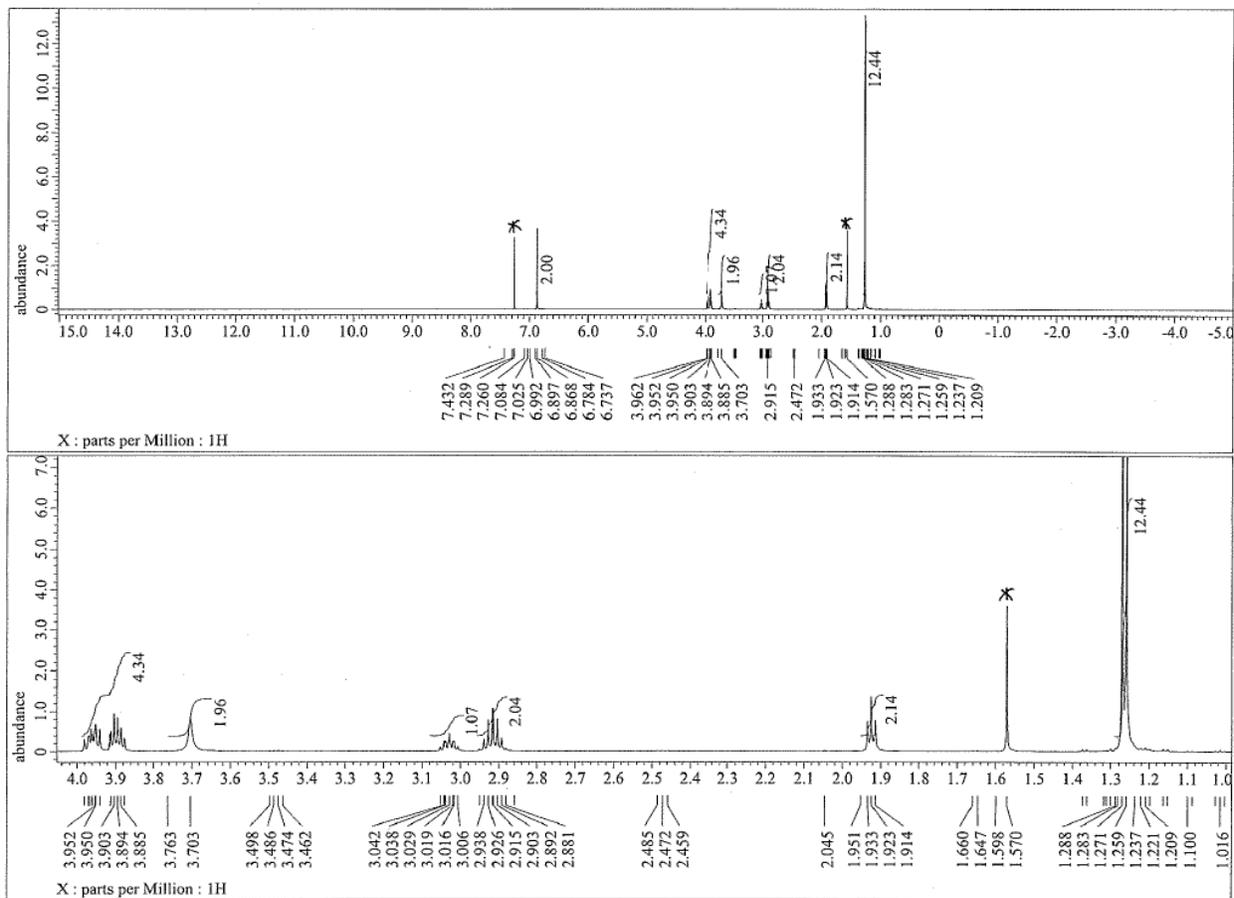

**Supplementary Fig. 48.** $^1$H NMR spectra of **S4** in CDCl$_3$ at 25 °C. The peaks marked with * indicate residual solvents.

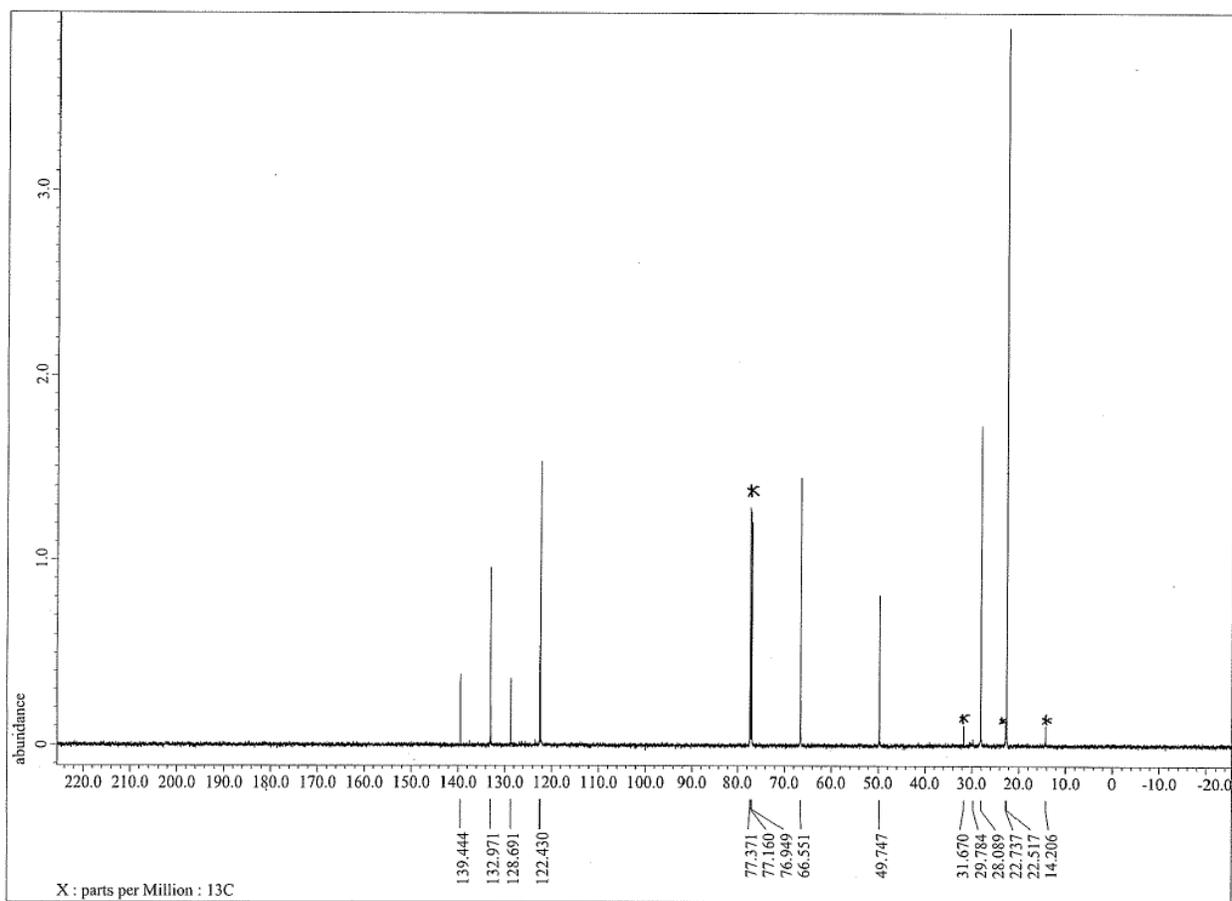

**Supplementary Fig. 49.** $^{13}$C NMR spectrum of **S4** in CDCl$_3$ at 25 °C. The peaks marked with * indicate residual solvents.

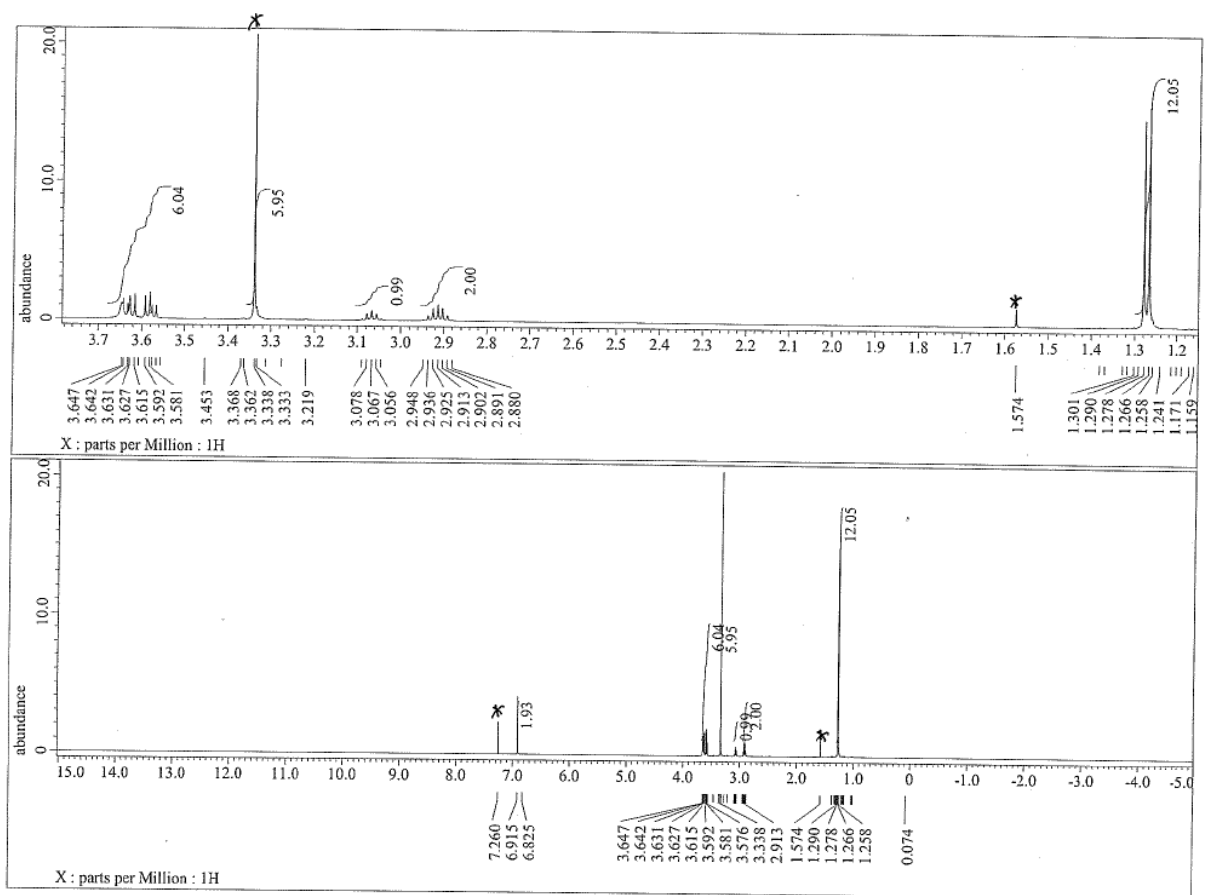

**Supplementary Fig. 50.** $^1$H NMR spectra of **S5** in CDCl$_3$ at 25 °C. The peaks marked with * indicate residual solvents.

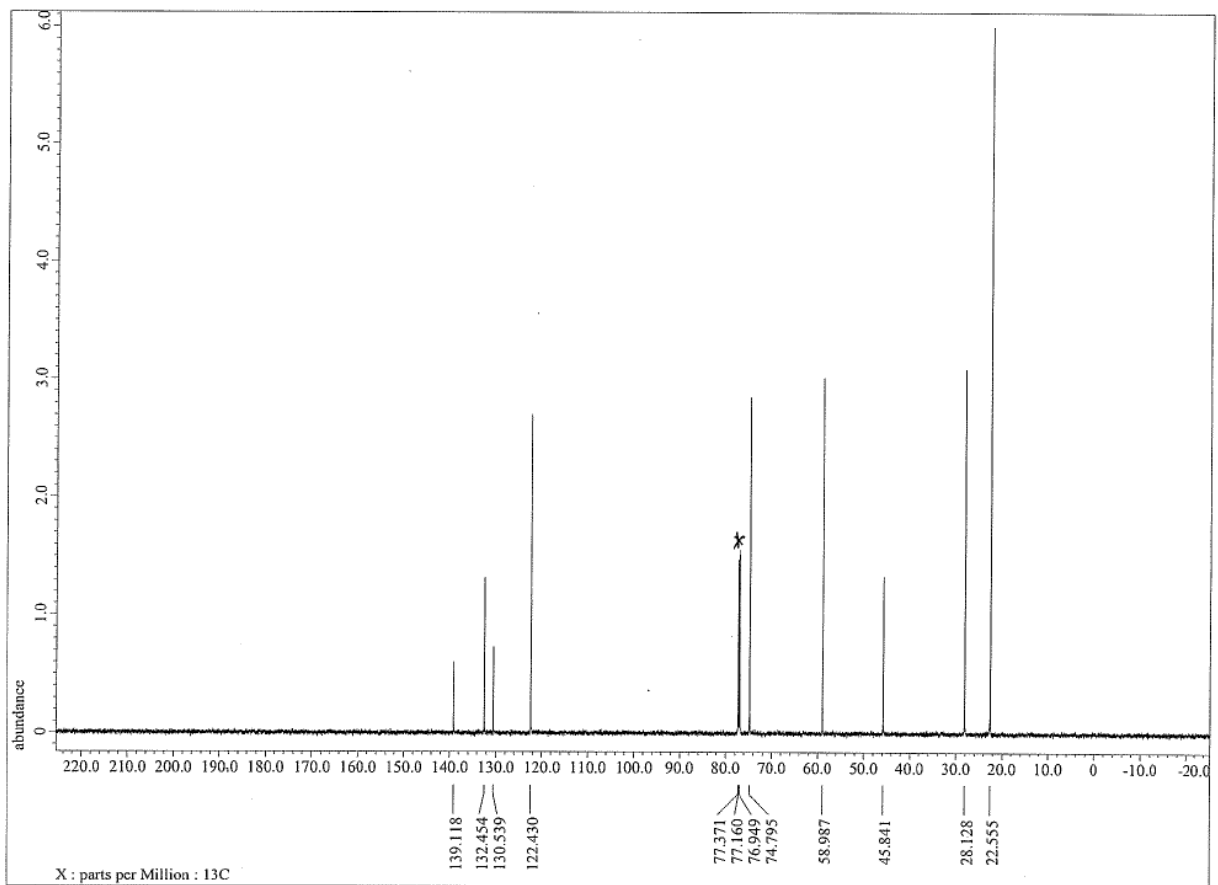

**Supplementary Fig. 51.** $^{13}$C NMR spectrum of **S5** in CDCl$_3$ at 25 °C. The peak marked with * indicates residual solvents.

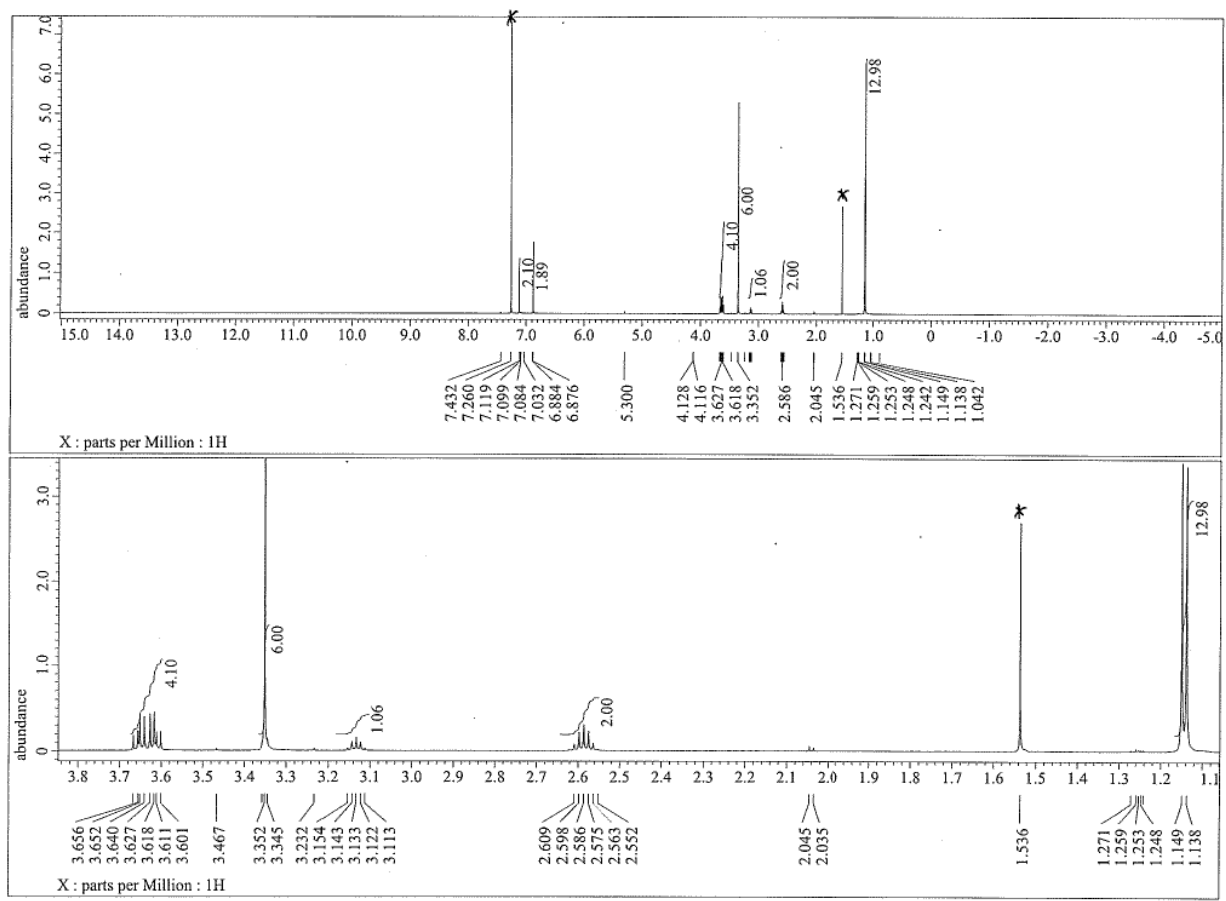

**Supplementary Fig. 52.** $^1$H NMR spectra of **S6** in CDCl$_3$ at 25 °C. The peaks marked with * indicate residual solvents.

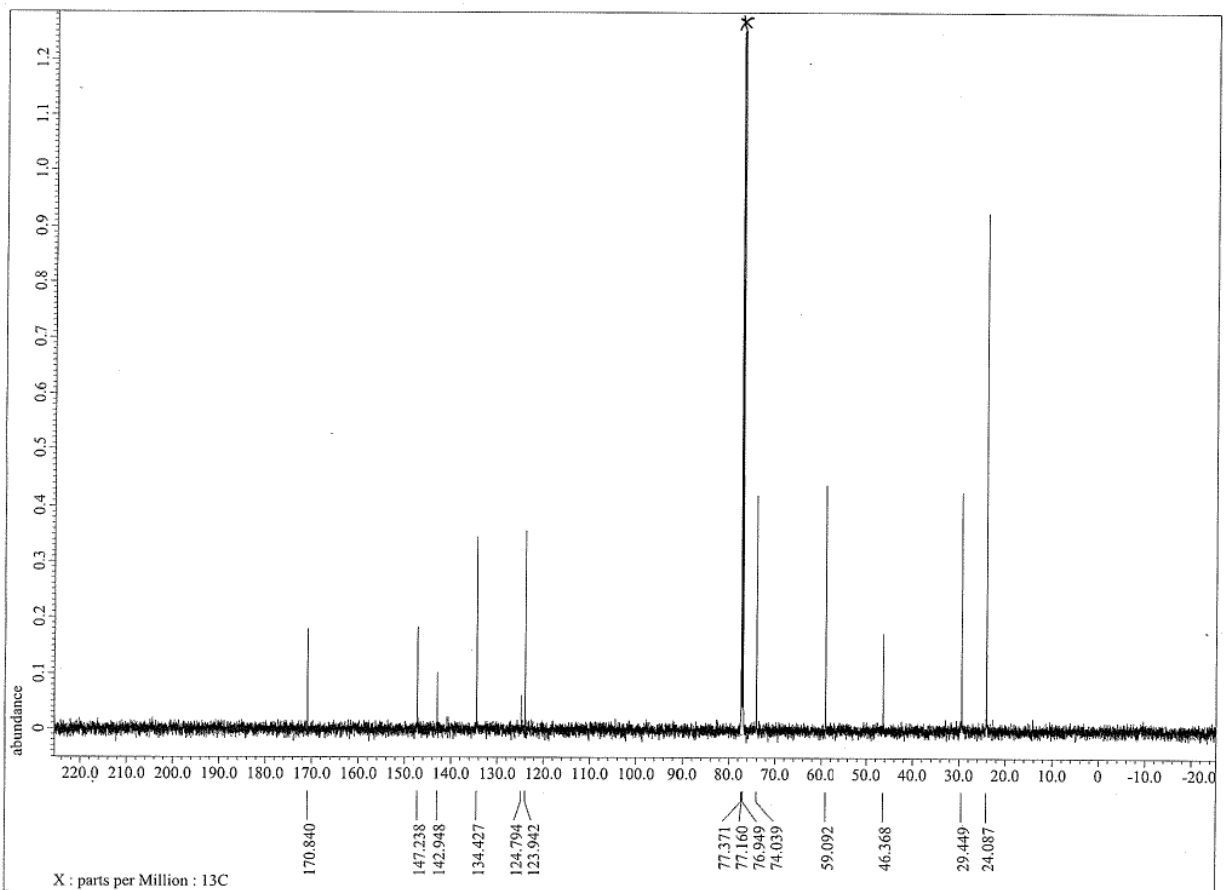

**Supplementary Fig. 53.** $^{13}$C NMR spectrum of **S6** in CDCl$_3$ at 25 °C. The peak marked with * indicates residual solvents.

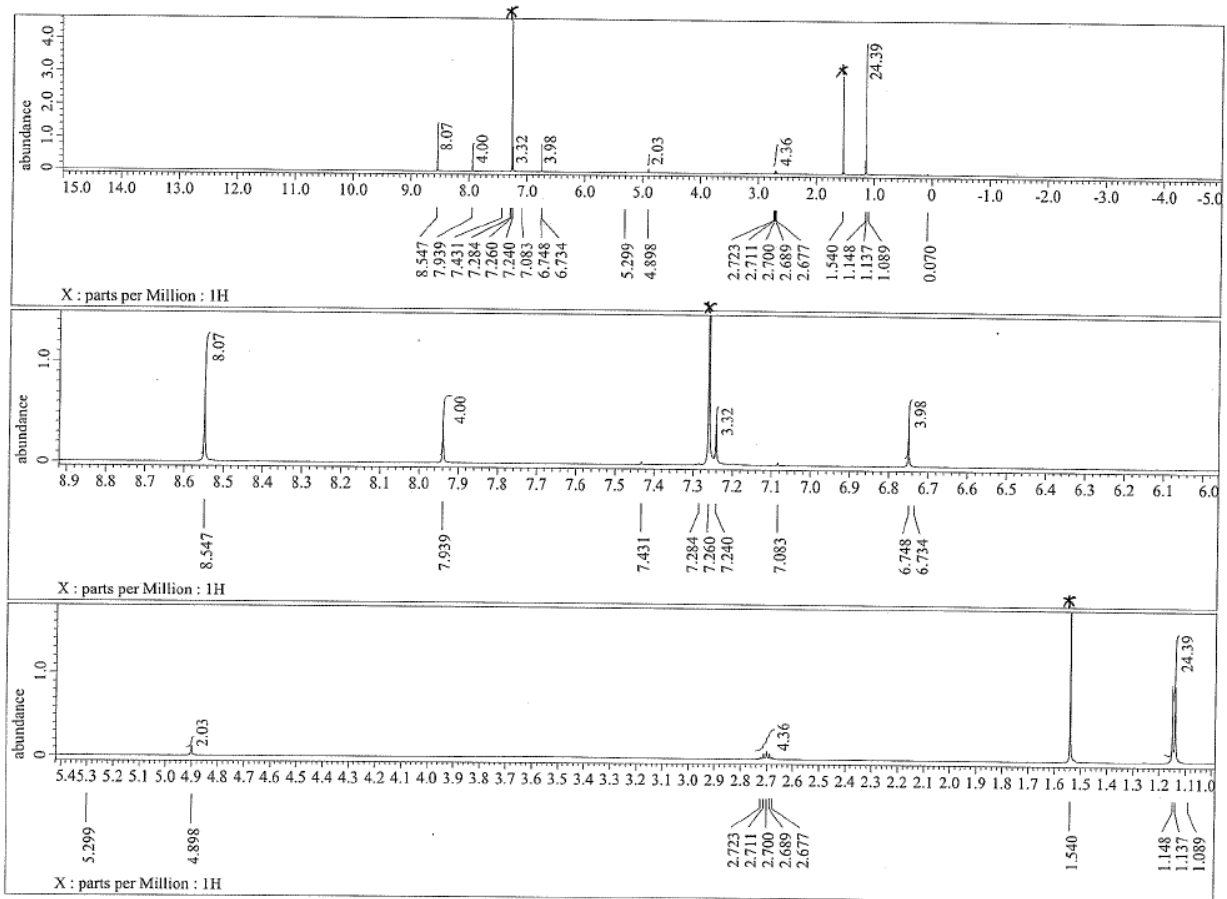

**Supplementary Fig. 54.** $^1$H NMR spectra of **FLAP1** in CDCl$_3$ at 25 °C. The peaks marked with * indicate residual solvents.

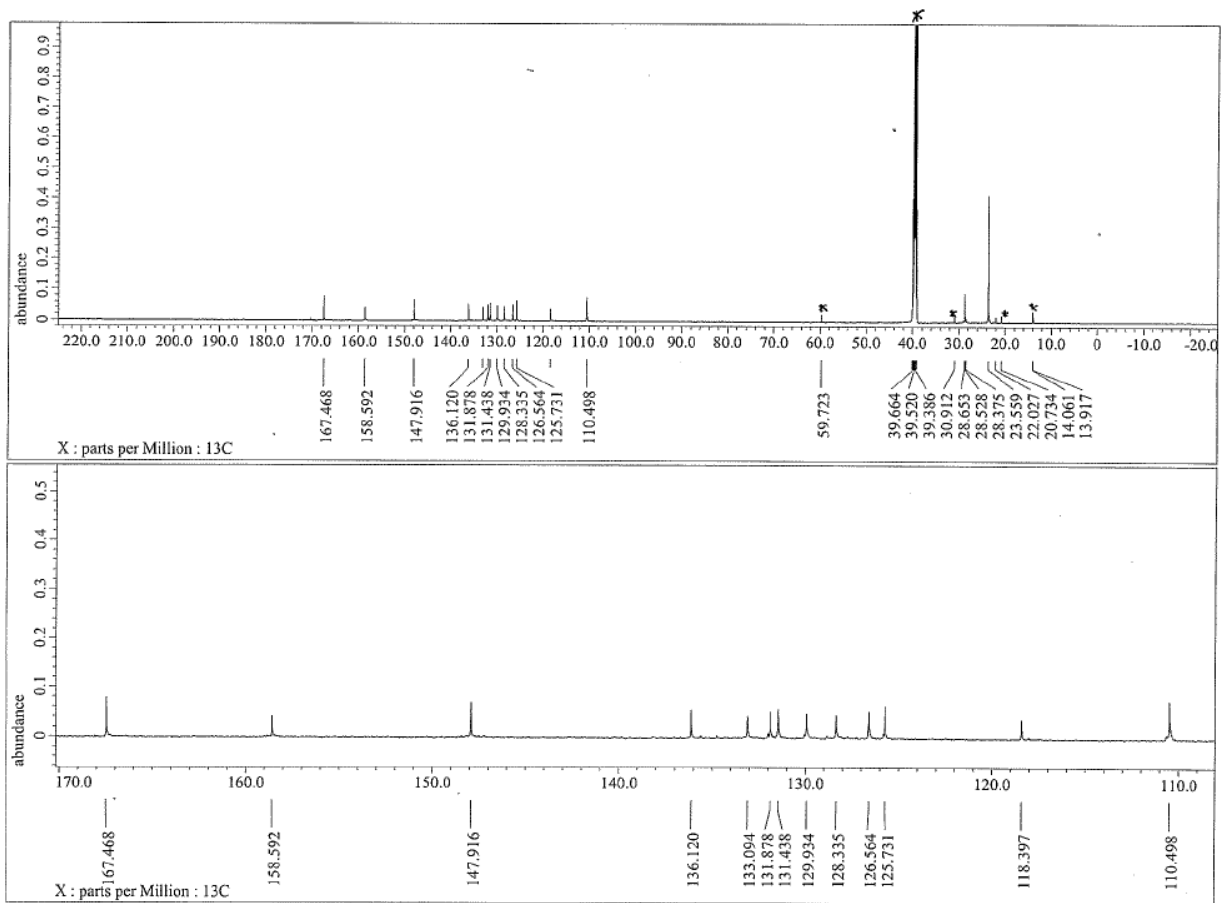

**Supplementary Fig. 55.** $^{13}$C NMR spectra of **FLAP1** in DMSO-$d_6$ at 25 °C. The peaks marked with * indicate residual solvents.

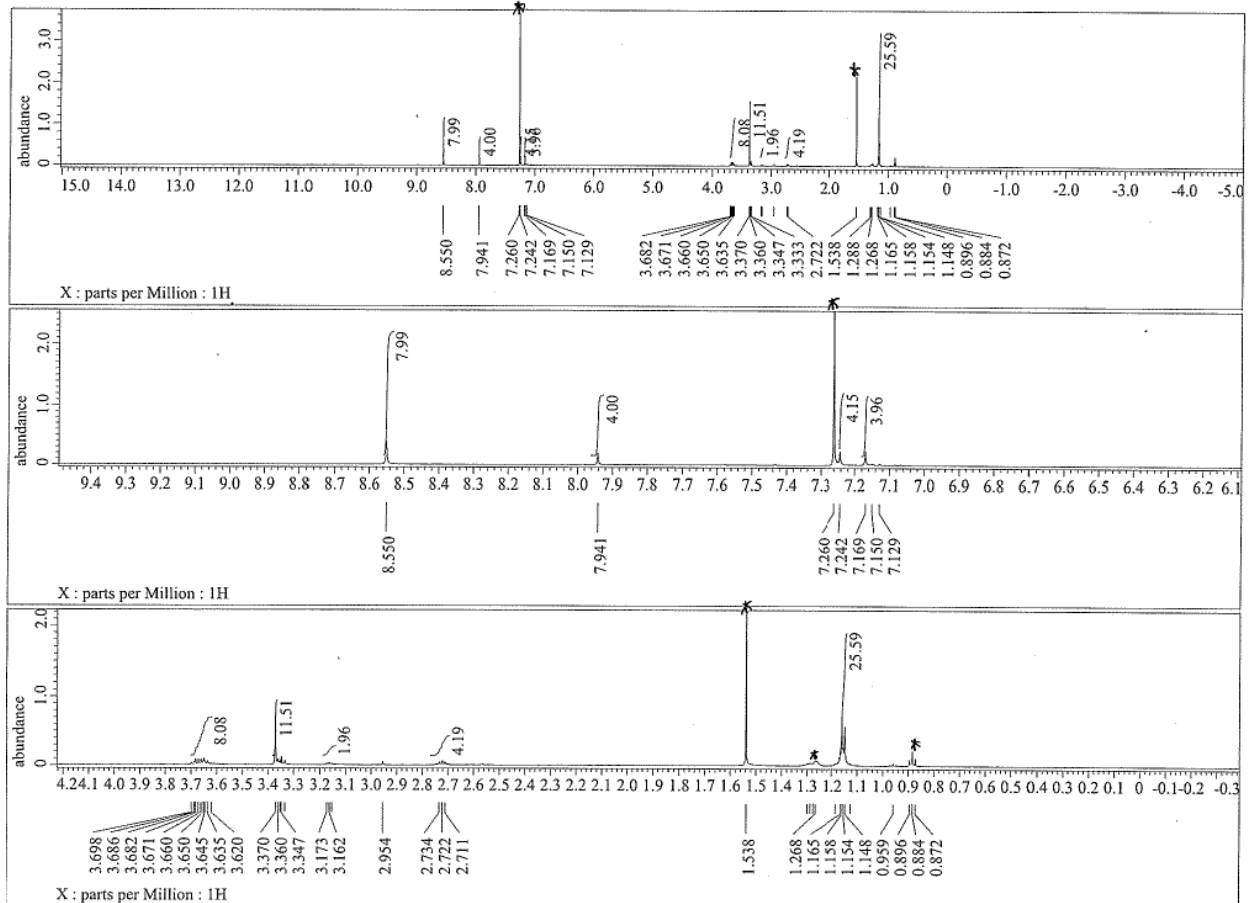

**Supplementary Fig. 56.** $^1$H NMR spectra of **S8** in CDCl$_3$ at 25 °C. The peaks marked with * indicate residual solvents.

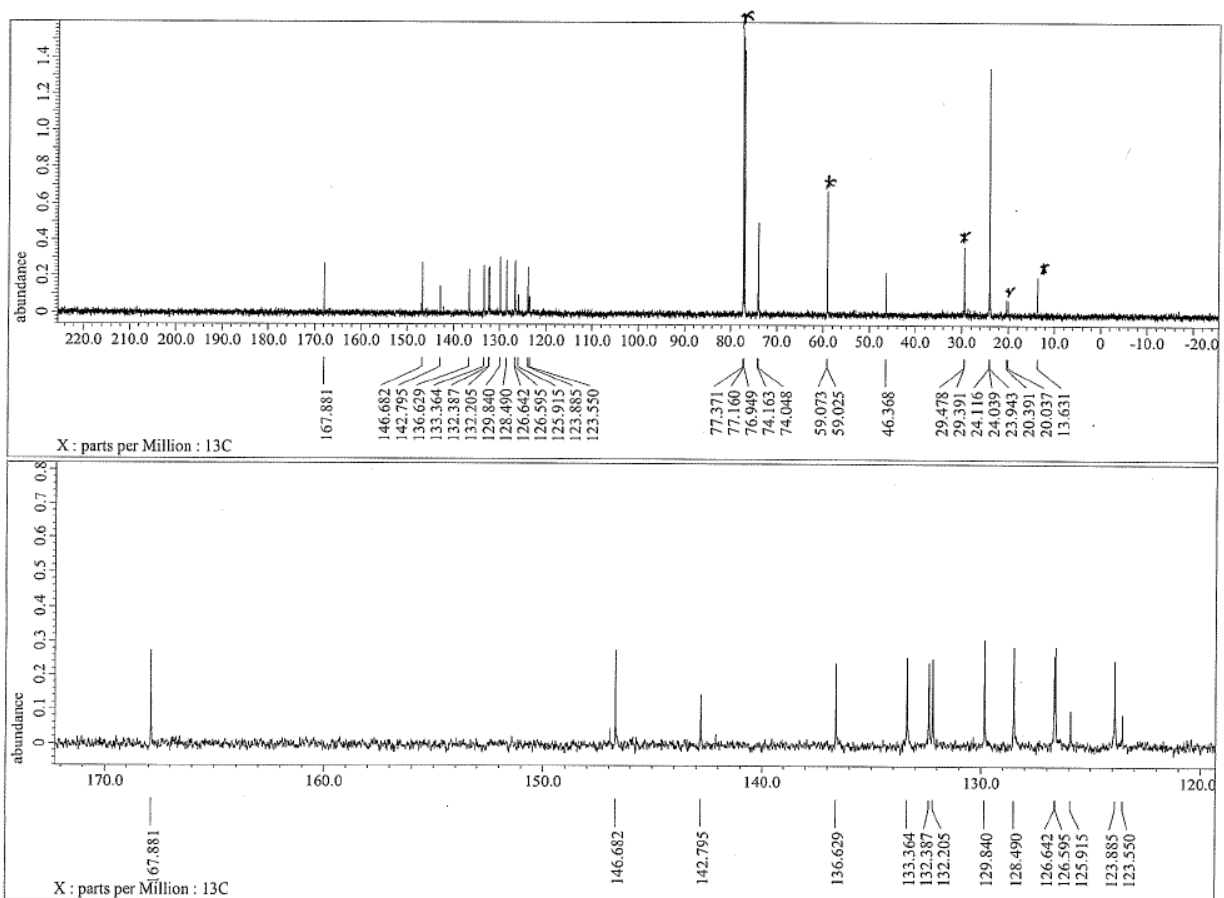

**Supplementary Fig. 57.** $^{13}$C NMR spectra of **S8** in CDCl$_3$ at 25 °C. The peaks marked with * indicate residual solvents.

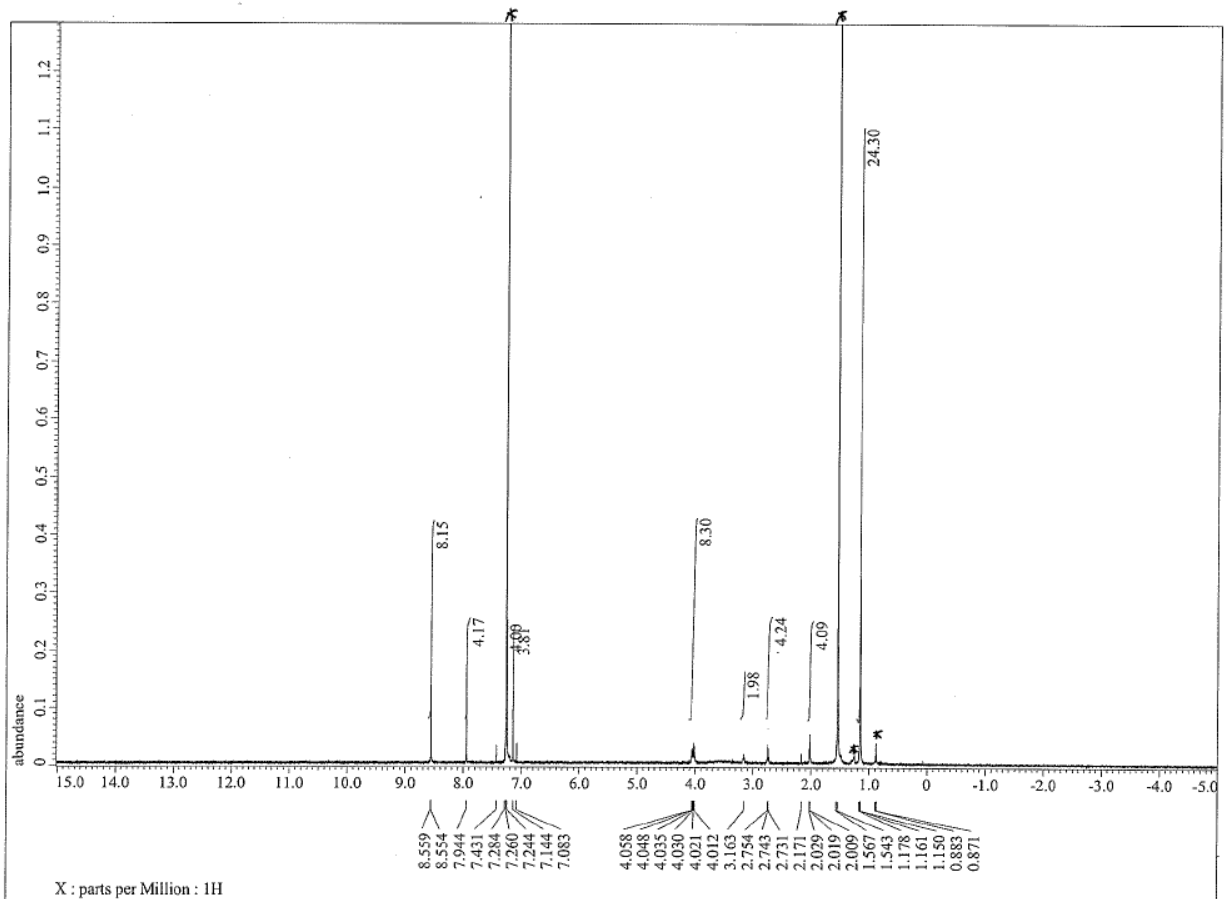

**Supplementary Fig. 58.** $^1$H NMR spectrum of **FLAP2** in CDCl$_3$ at 25 °C. The peaks marked with * indicate residual solvents.

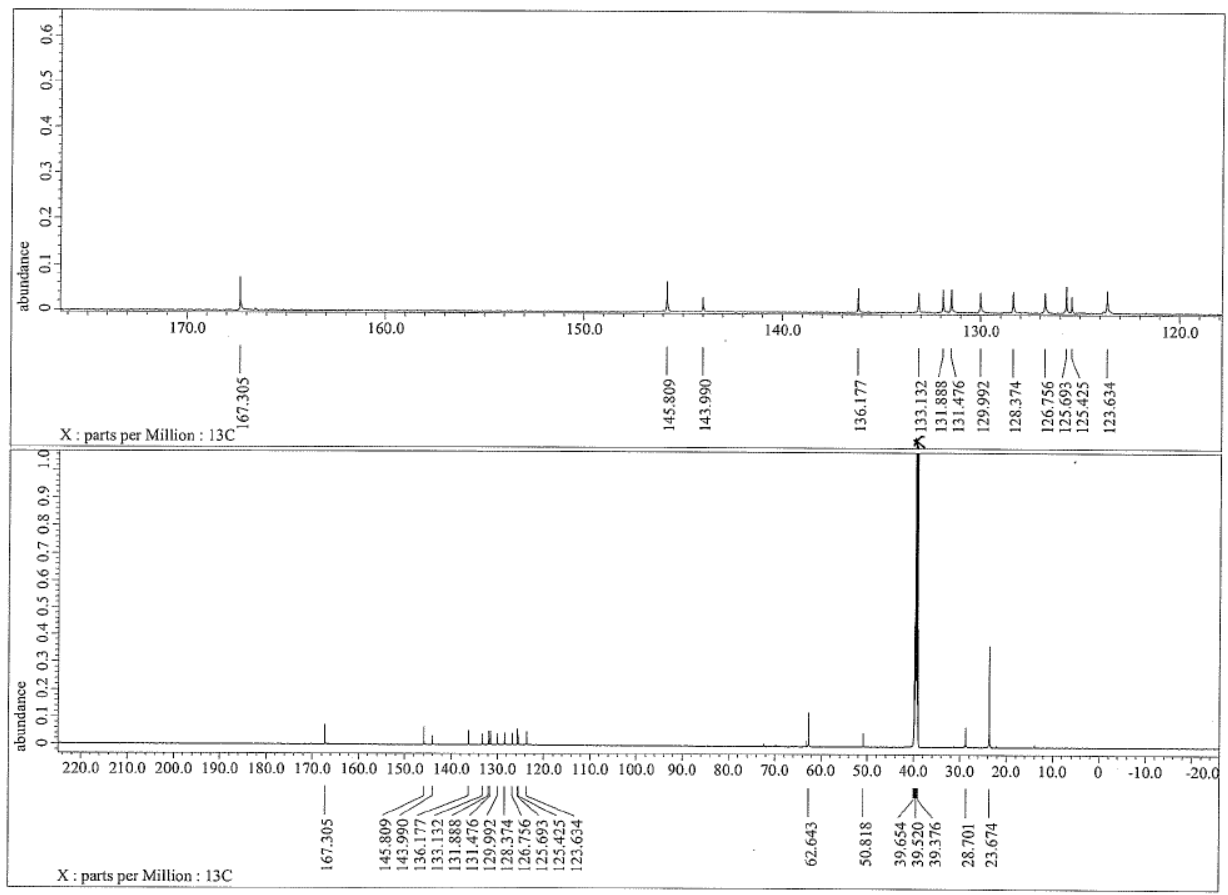

**Supplementary Fig. 59.** $^{13}$C NMR spectra of **FLAP2** in DMSO-$d_6$ at 25 °C. The peak marked with * indicates residual solvents.

## Mass spectra of new compounds

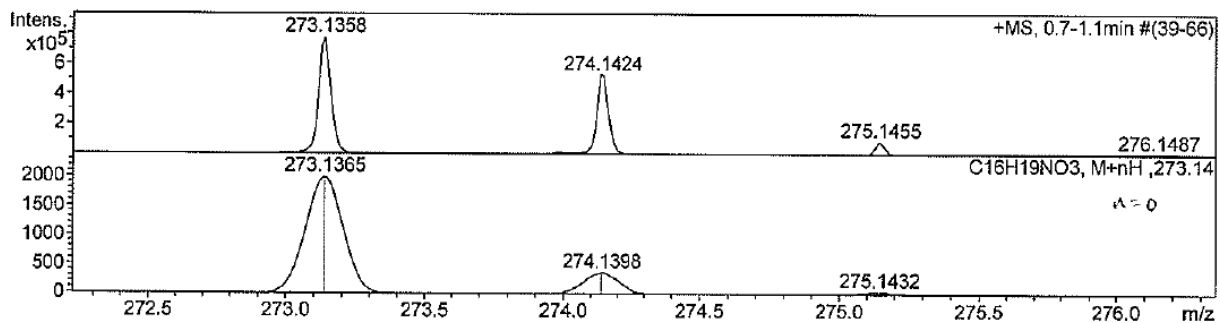

**Supplementary Fig. 60.** HR-APCI-TOF mass spectra of **S1**. Top: observed, bottom: simulated.

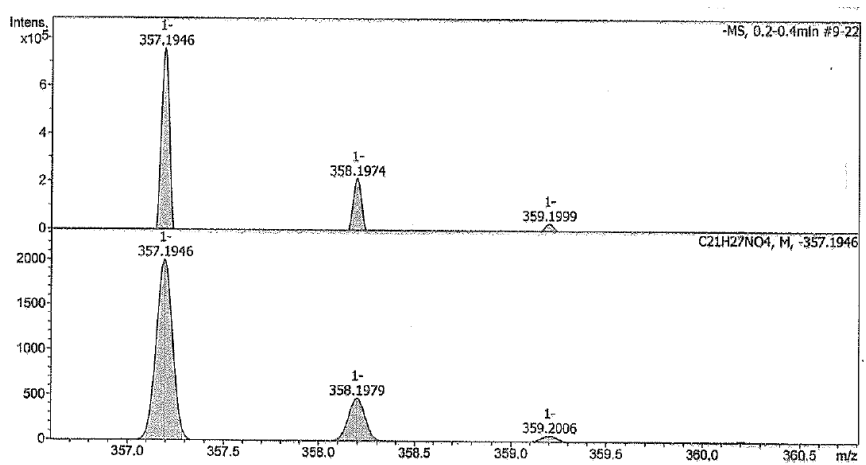

**Supplementary Fig. 61.** HR-APCI-TOF mass spectra of **S2**. Top: observed, bottom: simulated.

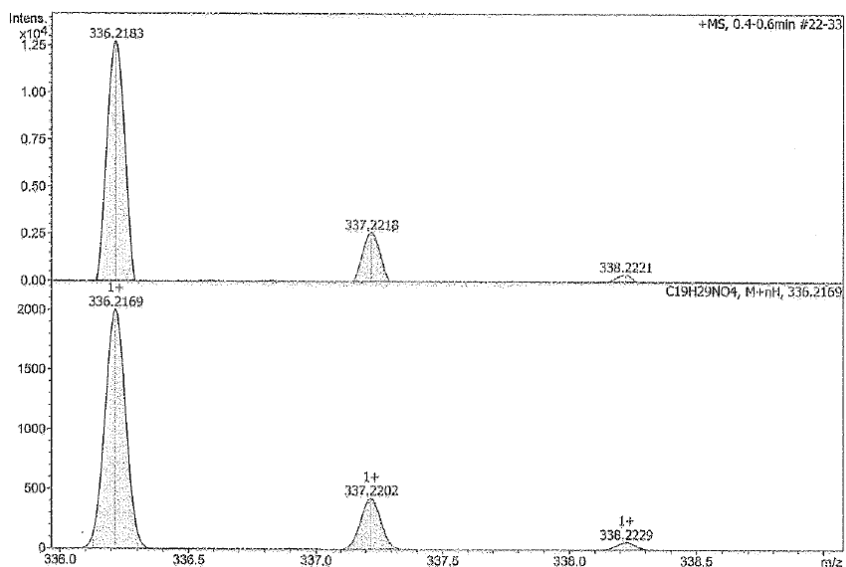

**Supplementary Fig. 62.** HR-APCI-TOF mass spectra of **S3**. Top: observed, bottom: simulated.

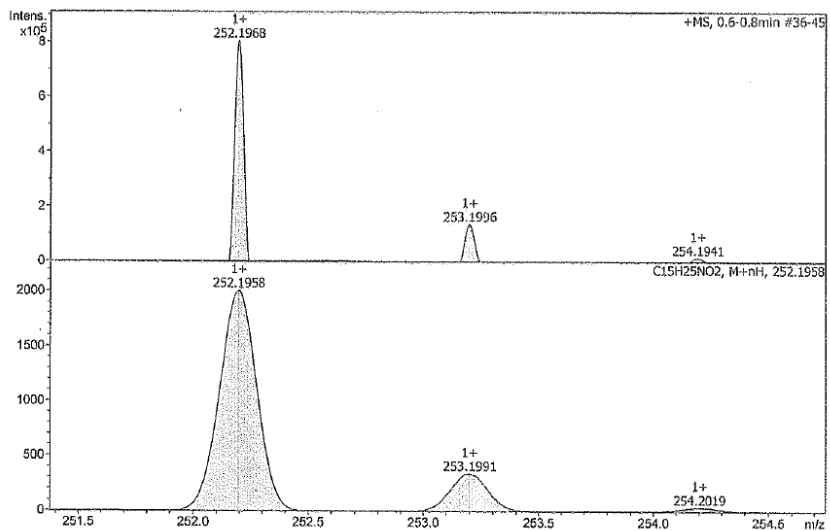

**Supplementary Fig. 63.** HR-APCI-TOF mass spectra of **S4**. Top: observed, bottom: simulated.

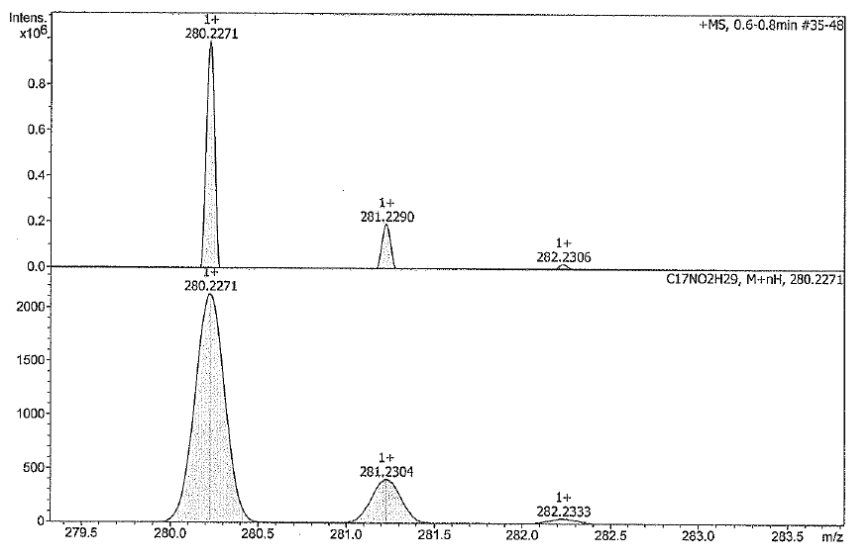

**Supplementary Fig. 64.** HR-APCI-TOF mass spectra of **S5**. Top: observed, bottom: simulated.

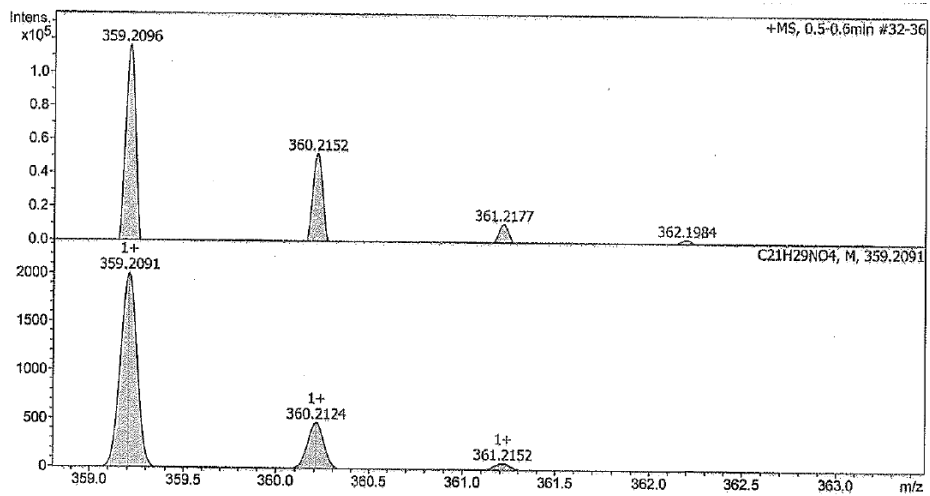

**Supplementary Fig. 65.** HR-TOF mass spectra of **S6**. Top: observed, bottom: simulated.

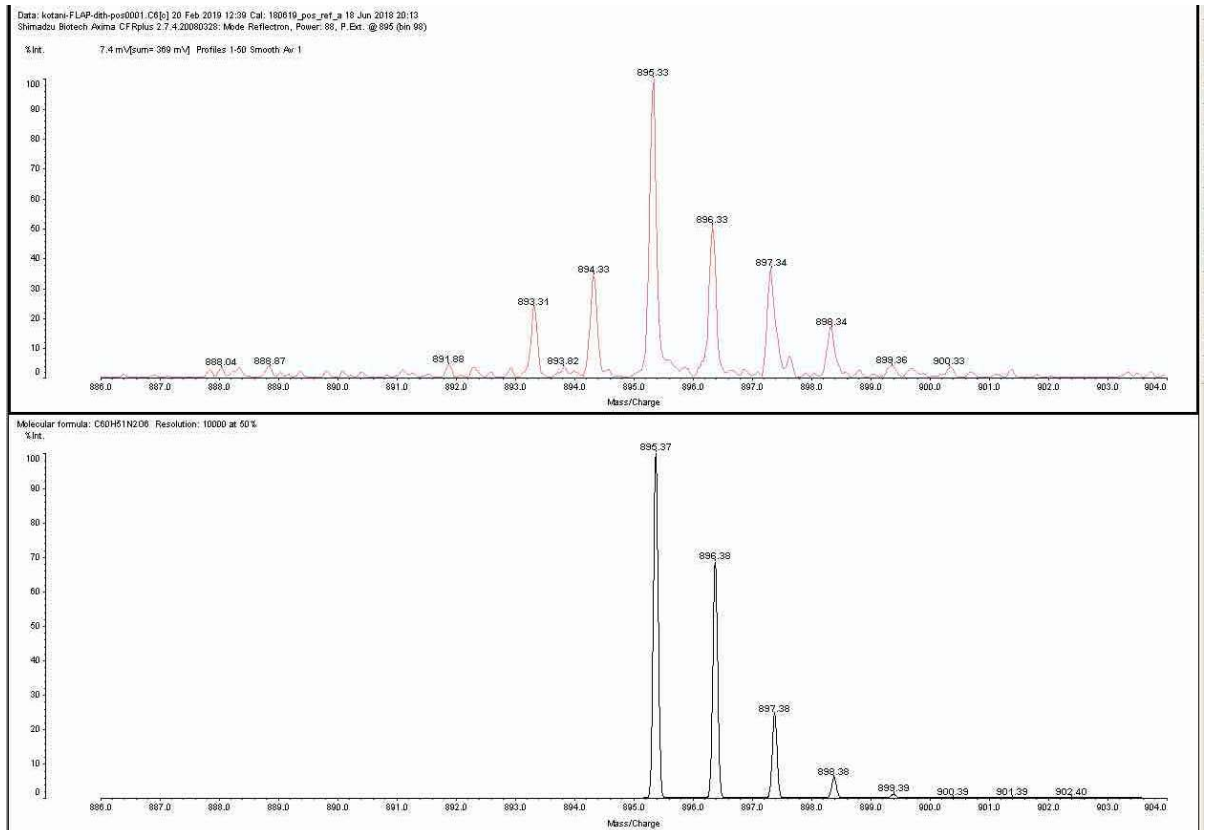

**Supplementary Fig. 66.** HR-MALDI-TOF mass spectra of **FLAP1**. Top: observed, bottom: simulated.

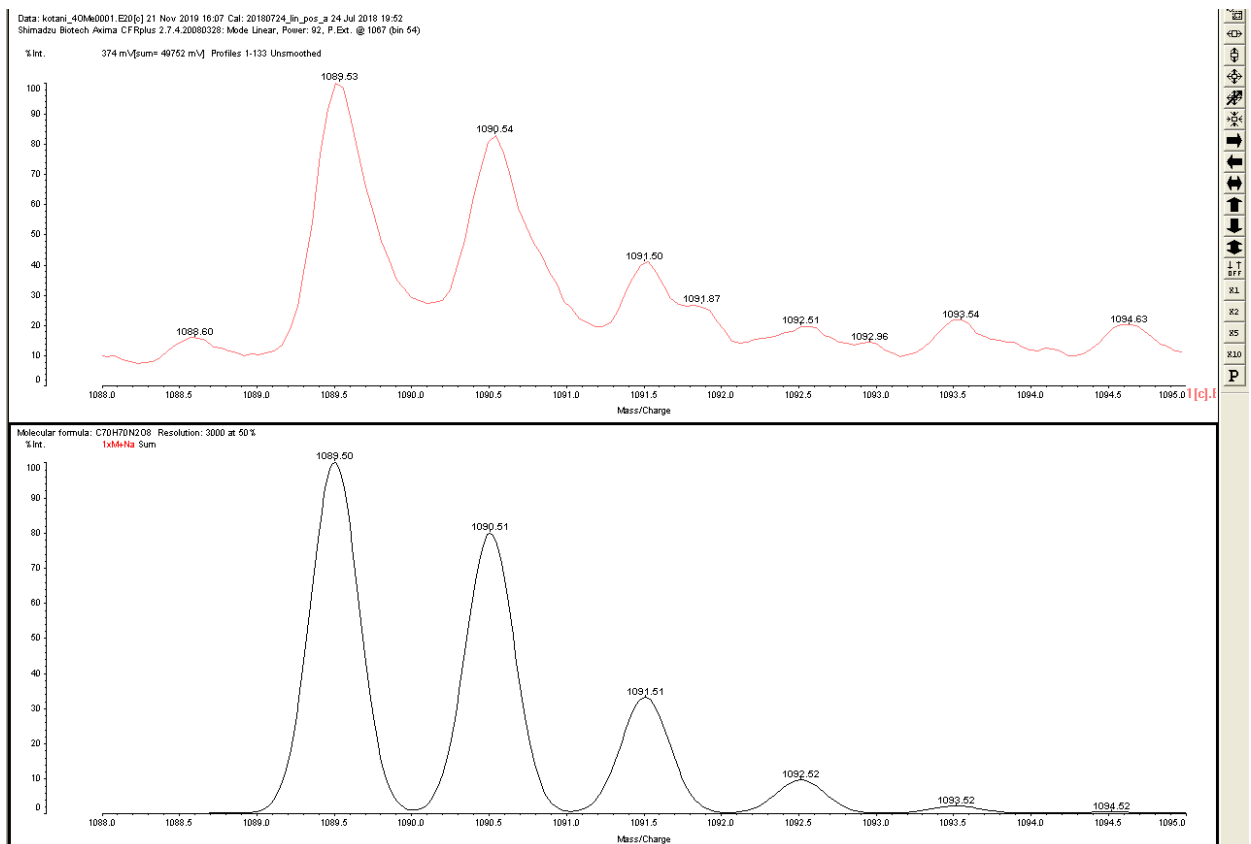

**Supplementary Fig. 67.** HR-MALDI-TOF mass spectra of **S8**. Top: observed, bottom: simulated.

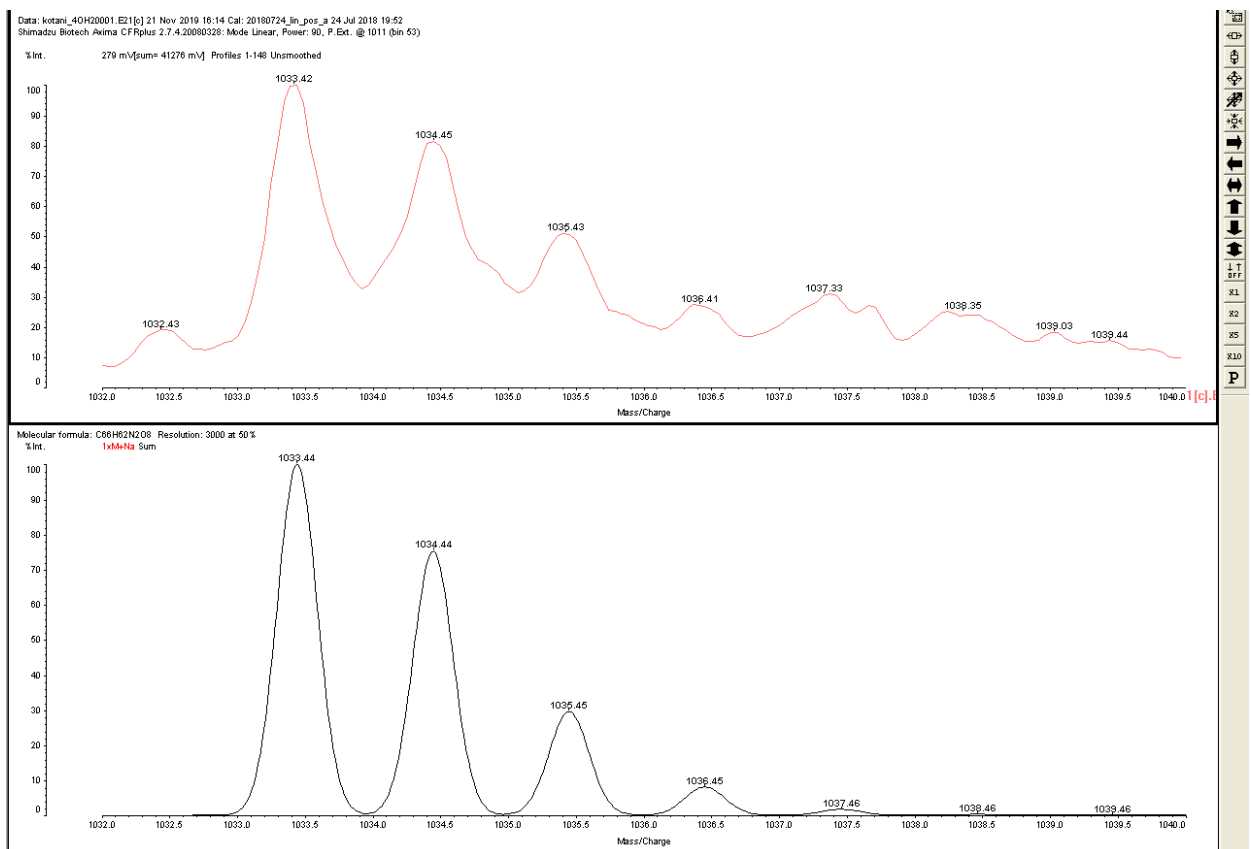

**Supplementary Fig. 68.** HR-MALDI-TOF mass spectra of **FLAP2**. Top: observed, bottom: simulated.